\documentclass[apj]{emulateapj}
\makeatletter\usepackage{graphicx} 

\newcommand{\oii}{[O\thinspace{II}]}
\newcommand{\oiii}{[O\thinspace{III}]}

\begin{document}
\title{Low-Redshift Ly$\alpha$ Selected Galaxies 
from {\em GALEX\/} Spectroscopy:  
A Comparison with Both UV-Continuum Selected Galaxies and
High-Redshift Ly$\alpha$ Emitters\altaffilmark{1,2,3}}
\author{
Lennox~L.~Cowie,$\!$\altaffilmark{4} \email{cowie@ifa.hawaii.edu}
Amy~J.~Barger,$\!$\altaffilmark{5,6,4} \email{barger@astro.wisc.edu}
Esther~M.~Hu$\!$\altaffilmark{4} \email{hu@ifa.hawaii.edu}
}

\altaffiltext{1}{Based in part on data obtained from the Multimission
Archive at the Space Telescope Science Institute (MAST).  STScI is
operated by the Association of Universities for Research in Astronomy,
Inc., under NASA contract NAS5-26555.  Support for MAST for non-HST
data is provided by the NASA Office of Space Science via grant
NAG5-7584 and by other grants and contracts.}
\altaffiltext{2}{Based in part on data obtained at the W. M. Keck
Observatory, which is operated as a scientific partnership among the
the California Institute of Technology, the University of
California, and NASA and was made possible by the generous financial
support of the W. M. Keck Foundation.}
\altaffiltext{3}{This research used the facilities of the Canadian Astronomy 
Data Centre operated by the National Research Council of 
Canada with the support of the Canadian Space Agency.}
\altaffiltext{4}{Institute for Astronomy, University of Hawaii,
2680 Woodlawn Drive, Honolulu, HI 96822.}
\altaffiltext{5}{Department of Astronomy, University of
Wisconsin-Madison, 475 North Charter Street, Madison, WI 53706.}
\altaffiltext{6}{Department of Physics and Astronomy,
University of Hawaii, 2505 Correa Road, Honolulu, HI 96822.}

\shorttitle{Low-Redshift Ly$\alpha$ Emission-Line Galaxies}
\shortauthors{Cowie, Barger, \& Hu}

\slugcomment{Accepted by The Astrophysical Journal}

\begin{abstract}
We construct a sample of low-redshift Ly$\alpha$ 
emission-line selected sources from 
{\em Galaxy Evolution Explorer (GALEX)\/} 
grism spectroscopy of nine deep fields
to study the role of Ly$\alpha$ 
emission in galaxy populations with cosmic time.
Our final sample consists 
of 119 (141) sources selected in the redshift interval 
$z=0.195-0.44$ ($z=0.65-1.25$) from the FUV (NUV) channel.
We classify the Ly$\alpha$ sources as active galactic nuclei 
(AGNs) if high-ionization emission lines are present in 
their UV spectra and as possible star-forming galaxies otherwise.  
We classify additional sources as AGNs using line widths
for our Ly$\alpha$ emitter (LAE) analysis.
These classifications are broadly supported by comparisons 
with X-ray and optical spectroscopic observations, though
the optical spectroscopy identifies a small number of
additional AGNs.
Defining the {\em GALEX\/} LAE sample in the same way
as high-redshift LAE samples, we show that LAEs 
constitute only about 5\% of NUV-continuum 
selected galaxies at $z\sim0.3$.
We also show that they are less common at $z\sim0.3$
than they are at $z\sim3$.  We find that the $z\sim0.3$ 
optically-confirmed Ly$\alpha$ galaxies lie below the 
metallicity-luminosity relation of the $z\sim0.3$ 
NUV-continuum selected galaxies
but have similar H$\alpha$ velocity widths at similar luminosities,
suggesting that they also lie below the metallicity-mass 
relation of the NUV-continuum selected galaxies.  
We show that, on average, the Ly$\alpha$ galaxies 
have bluer colors, lower extinctions as measured from the 
Balmer line ratios, and more compact morphologies than the 
NUV-continuum selected galaxies.  Finally, we confirm that the 
$z\sim2$ Lyman break galaxies (LBGs) have relatively low 
metallicities for their luminosities, and we find that they
lie in the same metallicity range as the $z\sim0.3$ 
Ly$\alpha$ galaxies.  
\end{abstract}


\section{Introduction}
\label{secintro}

High-redshift galaxy studies rely in large part 
on samples selected using color techniques or extreme 
emission-line properties.  However, it is very difficult
to know how such galaxy populations relate to one another
or how they fit into the overall scheme of galaxy
evolution.  Adding complexity to the problem is 
the fact that active galactic nucleus (AGN) contributions 
to the light of different galaxy populations may cause 
different selection biases, often in ways that are hard to 
determine and quantify.  For example, Cowie et al.\ (2009) 
showed that at low redshifts the AGN contribution dominates
the light at wavelengths below the Lyman continuum edge.

A classic example of a highly valued population whose
relationship to other samples is unclear is the 
Ly$\alpha$ emitter (LAE) population, which features 
prominently in the very highest redshift galaxy studies 
(e.g., Hu \& Cowie 2006 and references therein).  
Current estimates of the
fraction of galaxies that exhibit LAE properties
are mostly based on comparisons of the LAE and
Lyman break galaxy (LBG) populations at $z\sim 2-3$
(Shapley et al.\ 2003).  However, it 
would be useful to know how that fraction varies with 
redshift, as well as whether the presence of a strong 
Ly$\alpha$ emission line might be related to other 
galaxy properties.

Of course, the observed properties of a galaxy 
are not enough, as one also needs to know how best
to translate those properties into interesting physical 
quantities, such as star formation rates, masses, and 
metallicities.  
Because Ly$\alpha$ is resonantly scattered by neutral 
hydrogen, such translations are particularly difficult 
to do for LAEs.  Determining the escape path of Ly$\alpha$ 
and hence its dust destruction is an extremely 
complex problem both theoretically (e.g., Neufeld 1991; 
Finkelstein et al.\ 2007) and observationally 
(e.g., Kunth et al.\ 2003; Schaerer \& Verhamme 2008).  
In particular, we do not have a very clear understanding 
of how to translate Ly$\alpha$ luminosities into star 
formation rates in individual galaxies.

Fortunately, new possibilities for making significant 
advances in placing the LAEs in context have become
available through {\em Galaxy Evolution Explorer (GALEX};
Martin et al.\ 2005) grism spectroscopy. 
In a seminal paper Deharveng et al.\ (2008) have shown that low-redshift 
($0.2 < z < 0.35$) galaxies with strong Ly$\alpha$ emission 
lines can be selected from these data. The selection
process is based on finding Ly$\alpha$ in the UV-continuum
selected {\em GALEX\/} sources and thus is most analagous to
locating Ly$\alpha$ emitters in the high-redshift LBG
population via spectroscopy (Shapley et al.\ 2003). However,
the procedure enables the selection of a substantial sample of
sources that can be compared to the high-redshift LAEs.
These low-redshift galaxy samples 
have many advantages.  For one thing, the galaxies are 
bright and can be easily studied at other wavelengths. 
Perhaps even more importantly, however, these 
low-redshift galaxy samples can be integrated into 
comprehensive studies of the galaxy populations at 
these redshifts to understand some of the selection 
biases.  For example, we can determine the 
fraction of NUV-continuum selected galaxies in this redshift 
interval that have strong Ly$\alpha$ emission.
Moreover, by comparing the Ly$\alpha$ properties of
the low-redshift galaxies with their optical properties,
including their H$\alpha$ line strengths, we can 
calibrate the conversion of Ly$\alpha$ luminosity to
star formation rate and answer many of the questions 
on how LAEs are drawn from the 
more general NUV-continuum selected population.  
Finally, we can see whether the presence of strong
Ly$\alpha$ emission in a galaxy is related to the
galaxy's metallicity, extinction, morphology, or kinematics.

The structure of the paper is as follows.
In Section~\ref{secgalex} we use the {\em GALEX\/} 
extracted spectra (Morrissey et al.\ 2007) on nine 
high galactic latitude fields with deep {\em GALEX\/} 
grism exposures to generate a catalog of 261 
Ly$\alpha$ emission-line sources in the redshift 
intervals $z=0.195-0.44$ (low redshift) and $z=0.65-1.25$
(moderate redshift). 
This sample extends to an NUV (AB) magnitude of 21.8.
We divide the Ly$\alpha$ sources into two classes: obvious 
AGNs, due to the presence of high-excitation lines in their
UV spectra, and Galaxies, which do not have such lines. 
It should be emphasized that some of the objects in
the ``Galaxy" class may in fact be AGNs, where either
the high-excitation lines are weak in the UV or
where they lie in missing portions of the UV spectrum.

In Section~\ref{secopt} we present optical spectroscopy
of a subsample of the sources with {\em GALEX\/}
UV spectra. These observations were obtained using
multi-object masks with the DEep Imaging Multi-Object 
Spectrograph (DEIMOS; Faber et al.\ 2003)
on the Keck~II 10~m telescope and cover sources 
both with and without UV spectral identifications. 

In Section~\ref{secagngal} we use the optical data 
and a comparison with X-ray observations to show that 
our AGN and Galaxy classifications from the UV spectra
are broadly robust.  We also analyze to what extent the 
{\em GALEX\/} line widths may be used to discriminate 
further between star-forming galaxies and AGNs.  

In Section~\ref{lalpha} we present the {\em GALEX\/} 
LAE sample defined in the same way as high-redshift
LAE samples, and we measure the LAE number counts 
and luminosity functions.  We compare these with
the number counts and luminosity functions constructed 
for both NUV-continuum selected samples 
and high-redshift LAE samples. 
We also consider the equivalent 
width distributions and kinematics of the 
{\em GALEX\/} LAE sample.

In Section~\ref{lae_lbg} we study the properties
of low-redshift Ly$\alpha$ galaxies using
the Ly$\alpha$ and NUV-continuum selected sources 
that we optically confirmed as galaxies, as well
as an essentially spectroscopically
complete sample of NUV~$<24$ galaxies in the 
Great Observatories Origins Deep Survey-North
(GOODS-N; Giavalisco et al.\ 2004) field
from Barger et al.\ (2008). 
We present the Ly$\alpha$/H$\alpha$ flux ratios and
compare the equivalent widths, metallicities,
line widths, colors, extinctions, and morphologies 
of the galaxies in the two populations.  

Finally, in Section~\ref{seccon} we present our
conclusions and compare the low-redshift Ly$\alpha$
and NUV-continuum selected galaxy populations 
with the $z~\sim 2$ galaxy population of Erb et al.\ (2006).

An analysis of the AGN sample will be given in
a second paper (A. Barger \& L. Cowie 2009, in preparation).
Readers who are primarily interested in the results
may skip directly to Sections~\ref{lalpha},
\ref{lae_lbg}, and \ref{seccon}, which are largely 
self-contained.
We use a standard $H_{o}$ = 70~km~s$^{-1}$~Mpc$^{-1}$, 
$\Omega_{\rm M}$ = 0.3, $\Omega_{\Lambda}$ = 0.7 
cosmology throughout.

\section{The GALEX Sample}
\label{secgalex}

For the present study we use nine blank 
high galactic latitude fields with the deepest {\em GALEX\/} 
grism observations. We summarize the fields in 
Table~\ref{tab1}, where we give the {\em GALEX\/} name,
the J2000 right ascension and declination, the 
exposure time in kiloseconds, the limiting NUV magnitude 
to which the spectra were extracted, the galactic $E(B-V)$ 
in the direction of the field from Schlegel et al.\ (1998),
the galactic latitude, the number of sources with spectra
lying within a radius of $32\farcm5$ from the field center,
and the number of such sources found
to have a Ly$\alpha$ emission line in the redshift intervals
$z=0.195-0.44$ and $z=0.65-1.25$ based on their UV spectra.
The SIRTFFL~00 and SIRTFFL~01 fields have a small overlap.
In this region we used the deeper SIRTFFL~00 observations.

For each field we obtained the one and two dimensional 
spectra of all the sources with {\em GALEX\/} grism 
observations from the Multimission Archive at STScI (MAST).
The extracted sources per field constitute nearly complete
samples to the NUV limiting magnitudes listed in 
Table~\ref{tab1}. Morrissey et al.\ (2007) describe the
spectral extraction techniques used by the {\em GALEX\/} 
team in analyzing the grism data and detail the properties 
of the UV spectra.

\begin{figure}
\hskip -1.5cm
\includegraphics[width=6.0in,angle=0,scale=0.8]{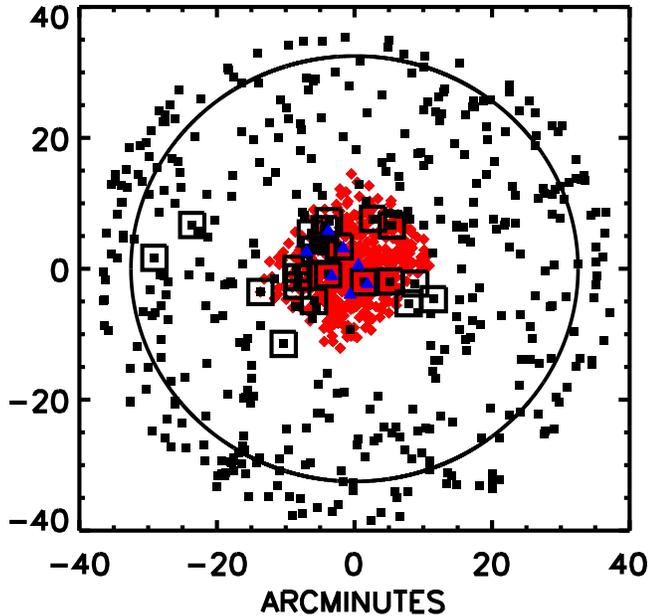}
\caption{The positions of the {\em GALEX\/} sources with 
extracted spectra in the field surrounding the HDF-N 
are shown with small black squares. Those with
known redshifts in the literature are enclosed in
larger open squares. FUV~$<21.5$ sources
in the GOODS-N are shown with blue triangles. The 503
X-ray sources in the CDF-N are shown with red diamonds.
The large black circle shows the $32\farcm5$ radius
field that we use.
\label{pos_image}
} 
\end{figure}

For our analysis we only consider sources within a 
$32\farcm5$ radius of each field center, since in 
the outermost regions of the fields there is a higher 
fraction of poor quality spectra. Even with this 
restriction the {\em GALEX\/} fields are extremely large.
In Figure~\ref{pos_image} we illustrate the size of a 
single {\em GALEX\/} field (the field around the Hubble 
Deep Field-North or HDF-N) by showing the positions of 
the {\em GALEX\/} sources with extracted spectra 
with black squares.  We enclose the {\em GALEX\/} 
sources with known redshifts in the literature from 
Barger et al.\ (2008) and references therein in a larger 
open square.  We also show the 503 sources
found in the  deep 2~Ms X-ray observations 
of the {\em Chandra\/} Deep Field-North (CDF-N)
from Alexander et al.\ (2003) with red diamonds and
the FUV~$<21.5$ sources in the GOODS-N with blue triangles.

With our radius restriction we have from just under 
300 spectra in the shallowest field to just under 1200 
spectra in the deepest field (see Table~\ref{tab1}). 
The FUV spectra cover a wavelength range of approximately
$1300-1800$~\AA\ at a resolution of $\sim 10$~\AA, and the
NUV spectra cover a wavelength range of approximately 
$1850-3000$~\AA\ at a resolution of $\sim 25$~\AA. 
However, the spectra become very noisy at the edges of 
the wavelength ranges. Thus, we only consider sources 
that have a Ly$\alpha$ emission line in the redshift 
interval $z=0.195-0.44$ corresponding to $1452.5-1750$~\AA\ 
in the FUV spectrum or in the redshift interval $z=0.65-1.25$ 
corresponding to $2006-2735$~\AA\ in the NUV spectrum.

\begin{figure}
\includegraphics[width=3.4in,angle=0,scale=1.]{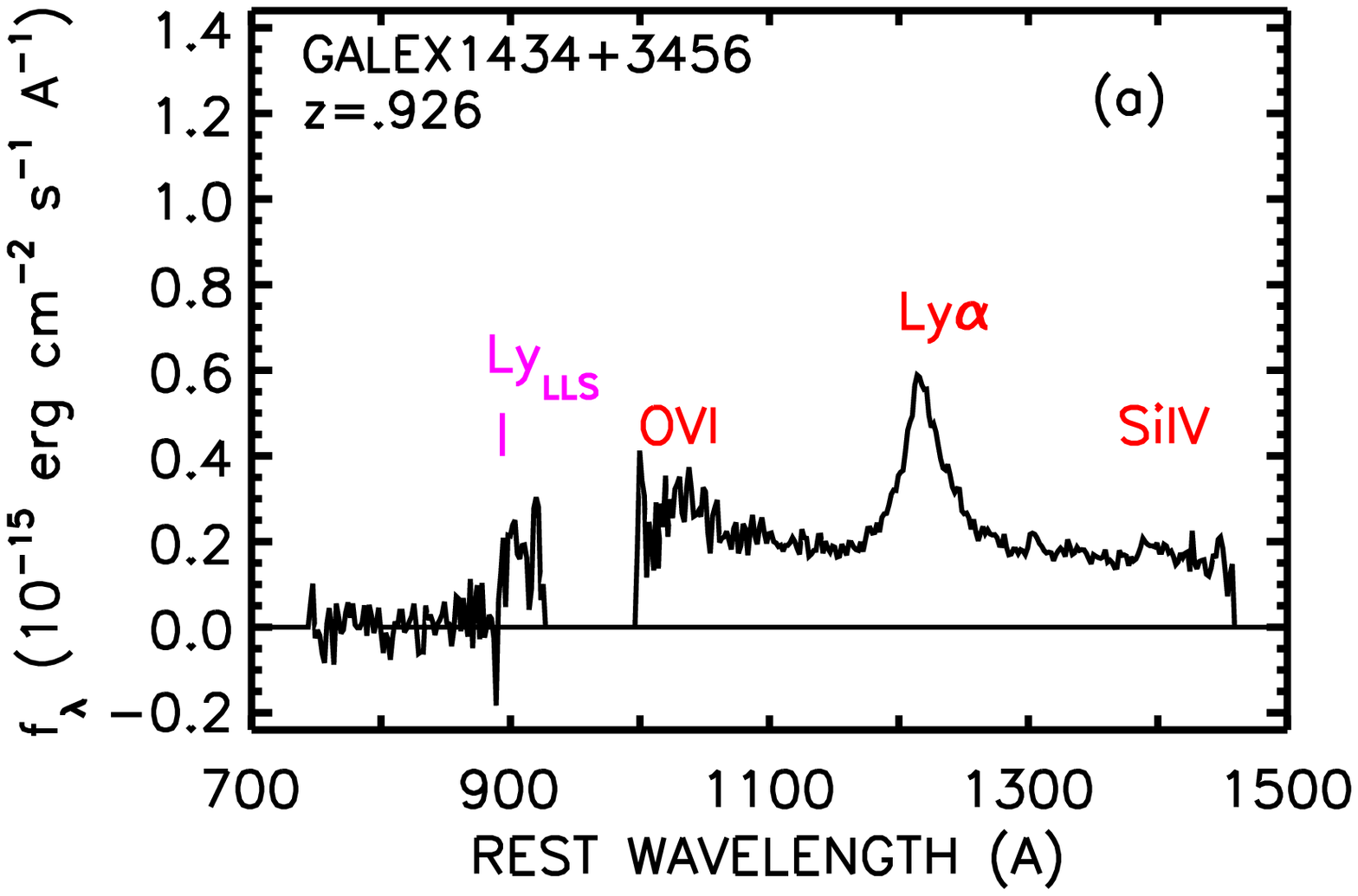}
\includegraphics[width=3.4in,angle=0,scale=1.]{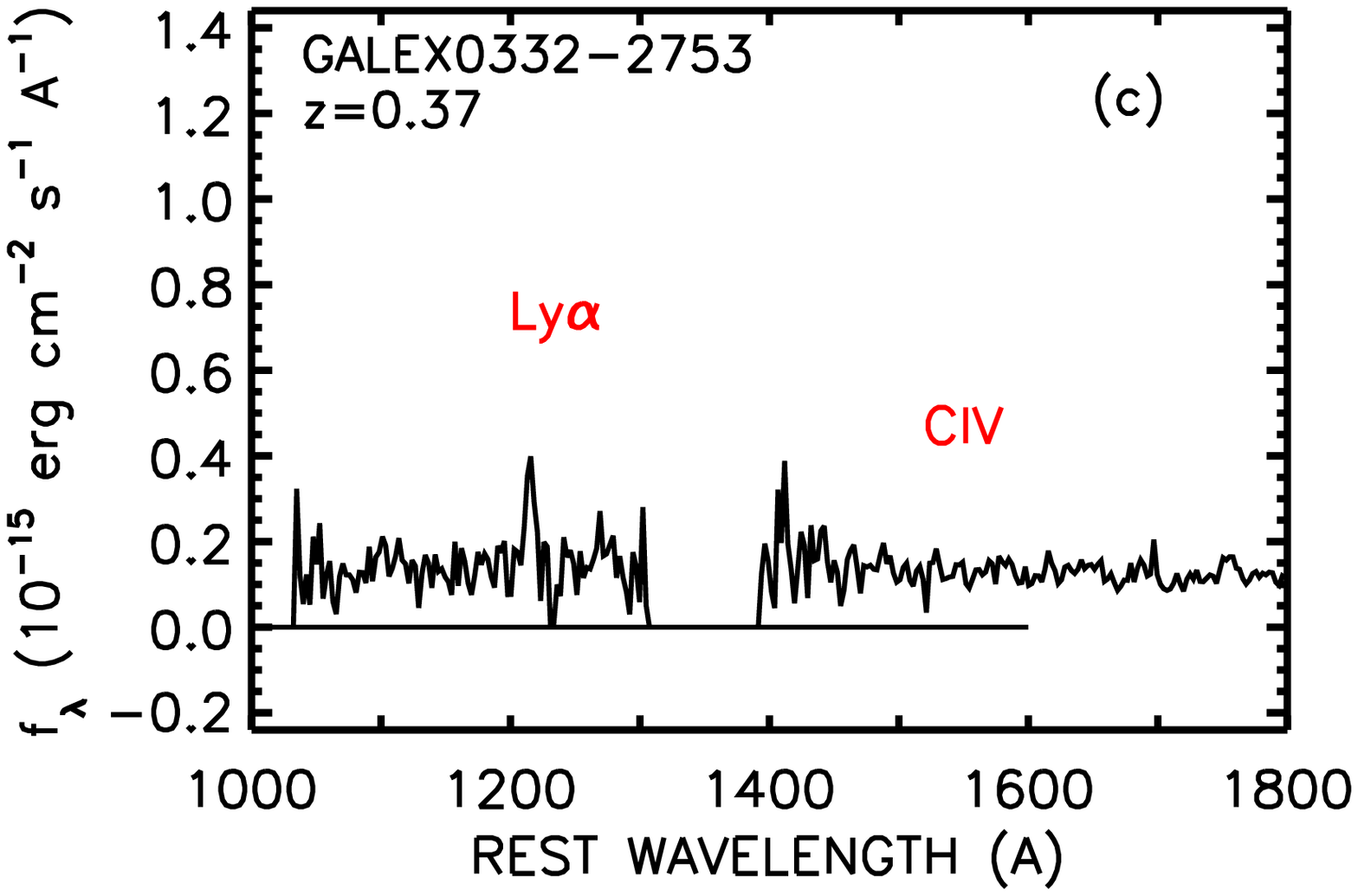}
\includegraphics[width=3.4in,angle=0,scale=1.]{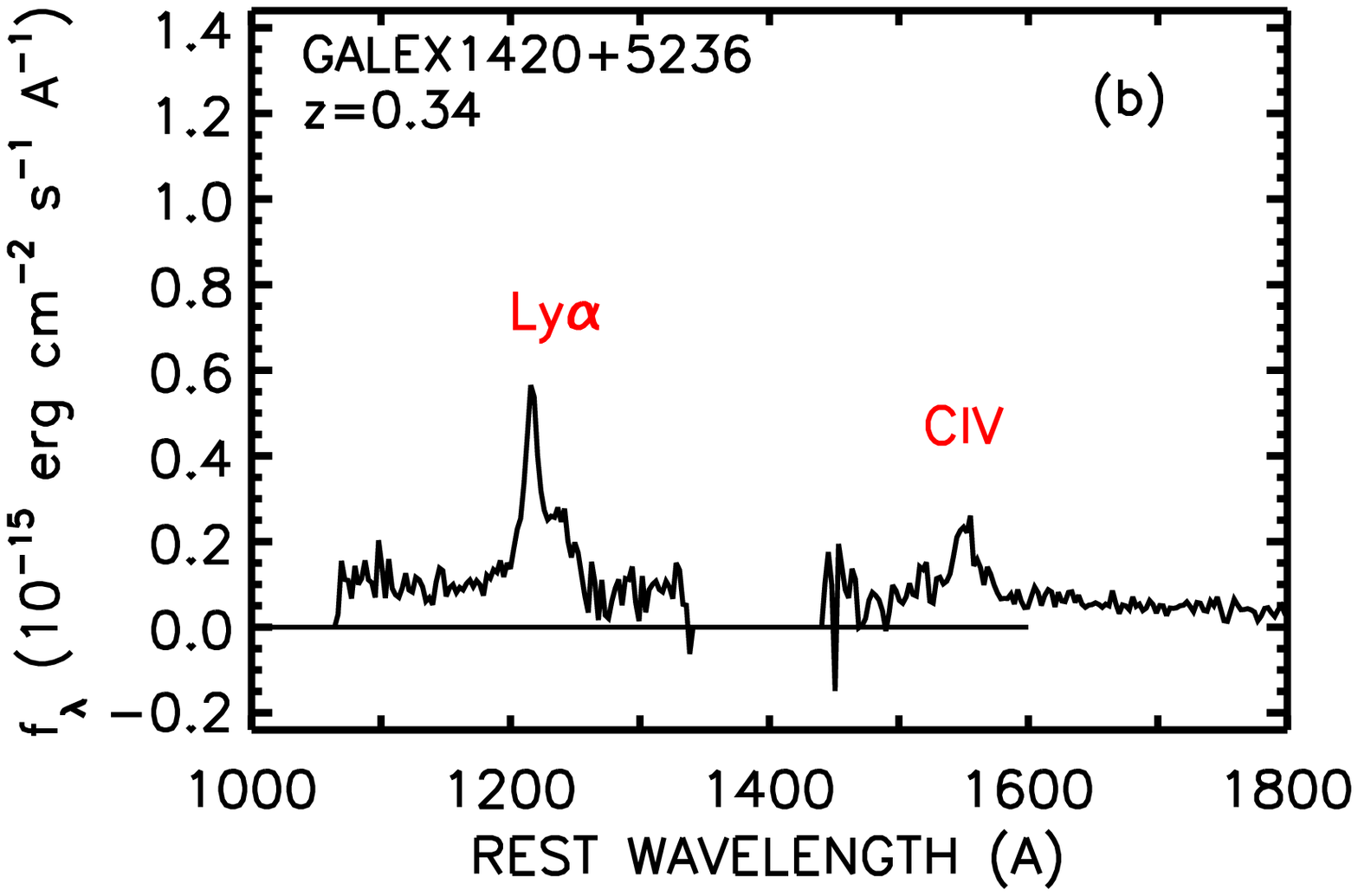}
\caption{
Some sample {\em GALEX\/} spectra illustrating the classifications 
for the Ly$\alpha$ selected sources.
(a) Broad-line AGN. The positions of the various broad emission 
lines are marked, as is the position of a strong intervening 
Lyman limit system. 
(b) Narrower line AGN, where strong CIV
and NV (not marked) are seen in addition to the Ly$\alpha$ line.
(c) A spectrum classified as a candidate Ly$\alpha$ Galaxy. 
There is no emission seen at the CIV position, and the 
single line is narrow.
\label{sample_spectra}
}
\end{figure}

We first ran an automatic search procedure for
emission lines in the spectra. For each source we fitted 
the higher signal-to-noise regions in the FUV and NUV
spectra separately with a second order polynomial continuum 
and a Gaussian. Sources with a significant signal in the 
Gaussian portion were then flagged.
We then visually inspected every spectrum to 
eliminate false emission line detections and to add any 
cases where the automatic procedure had missed an emission 
line. Most of the emission lines were detected by the 
automatic procedure with only a relatively small number of 
corrections required from the visual check. A very small
number of sources were eliminated based on
visual inspection of the images. On average 
4\% of the sources (recall that these are NUV-continuum 
selected and comprise both galaxies and a significant number of 
stars) have detected emission lines. This percentage
is relatively invariant from field to field 
(see Table~\ref{tab1}), though it is slightly higher 
in the deeper fields; in the four deepest fields, 5\%
of the sources have detected emission lines.

\begin{figure}
\hskip -0.6cm
\includegraphics[width=3.7in,angle=0,scale=1.]{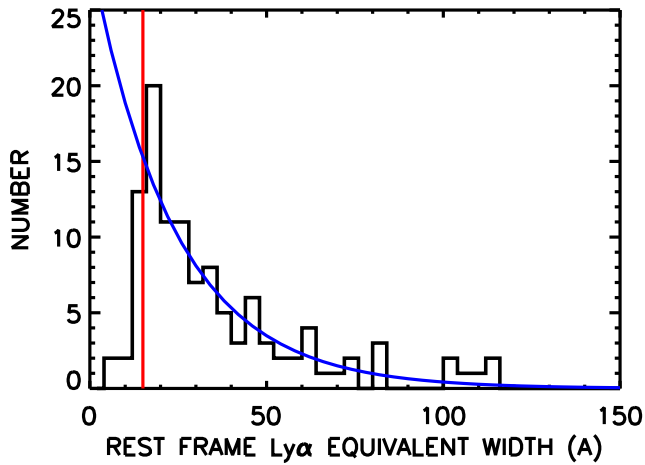}
\caption{
The rest-frame EW(Ly$\alpha$) distribution for our candidate 
Ly$\alpha$ Galaxy sample. The numbers correspond to 4~\AA\ bins. 
The red line shows the rest-frame EW(Ly$\alpha$)$\sim 15$~\AA\ 
limit of the sample. The blue curve shows the exponential
fit to the data described in the text, which has a
scale length of 23~\AA.
\label{ew_dist}
}
\end{figure}

We next measured a redshift for each source with an 
emission line and split the sources into one of two 
classes: an AGN class if there were high-excitation lines 
(usually OVI or CIV) present in addition to the 
Ly$\alpha$ line, or a Galaxy class to denote a potential 
star-forming galaxy if there were only a single line 
visible. In the latter case we assumed the line was 
Ly$\alpha$ in determining the redshift, an 
assumption we will test and find to be extremely 
reliable using optical spectra.  We will frequently
refer to the sources in our Galaxy class as our
candidate Ly$\alpha$ Galaxies, using
the word ``candidate'' to emphasize 
that some of these will turn out to be AGNs.
The relative classifications are illustrated in 
Figure~\ref{sample_spectra}, where we show examples of
(a) a broad-line AGN, (b) a narrower line AGN, and 
(c) a candidate Ly$\alpha$ Galaxy. 

For each source we fitted a 140~\AA\ rest-frame region 
around the Ly$\alpha$ line with a Gaussian 
and a flat continuum to  determine four parameters: 
the Ly$\alpha$ line width (this includes the rather 
wide instrumental resolution), 
the observed-frame equivalent width (EW) of the line,
the central wavelength, and the continuum level. 
We used the IDL MPFIT procedures of Markwardt (2008),
which are based on the More (1978) implementation of
the Levenberg-Marquardt algorithm and are very robust.
We used the {\em GALEX\/} noise vector to determine the 
input error for the spectrum and determined the
statistical errors from the covariance matrix returned
by the program.

All of the sources have detectable continuum at the
Ly$\alpha$ position. This is a direct consequence of the
NUV magnitude selection in the {\em GALEX\/} spectral 
extraction for the NUV selected Ly$\alpha$ lines, and of 
the NUV magnitude selection and the small difference between 
the FUV and NUV magnitudes for the FUV selected Ly$\alpha$ 
lines at lower redshifts. However, changes in the fitting
procedure (such as choosing a different wavelength
range or fitting the baseline with a linear fit rather
than a constant) can change the fitted parameters. In 
general these changes correspond to errors in the EW
of less than 20\%, and, for the most part, the errors are 
smaller than 10\%, which are comparable to the statistical 
errors. However, in a very small number of cases (less 
than a few percent) more substantial errors can occur
as the fitting procedure finds a substantially different 
solution.  Visual inspection suggests that the present 
choice of parameters provides the most reasonable results.

The final Ly$\alpha$ selected samples are 
summarized in Tables~\ref{tab2}$-$\ref{tab10} for 
each of the nine fields individually. For each source 
we give the J2000 right ascension and declination,
the NUV and FUV magnitudes, the 
redshift, the line width in km~s$^{-1}$ together
with the $1\sigma$ error, and the AGN
or Galaxy classification from the UV spectra. 
The NUV and FUV magnitudes are from the deep broadband
{\em GALEX\/} images of the fields with exposure times ranging
from 26~ks to 240~ks in the NUV band and 28~ks to 120~ks
in the FUV band. For the shallowest exposures the 
$1\sigma$ error for the NUV band is approximately
26, and for the FUV band it is approximately 26.4.
The sources with a Galaxy classification can be compared 
with those given in Deharveng et al.\ (2008)
for the subsample of the present fields analyzed in that 
work. In general the agreement is very good. The small 
number of sources omitted from one list but included in 
the other are generally marginal cases where
the inclusion or exclusion is somewhat arbitrary.
We provide a more detailed comparison in
Section~\ref{lae_prop}.
 
We show the distribution of the rest-frame EW(Ly$\alpha$)
for our candidate Ly$\alpha$ Galaxies in 
Figure~\ref{ew_dist}. For the $z=0.195-0.44$ redshift
interval the sample
appears to be relatively complete down to a rest-frame
EW(Ly$\alpha$)~$\sim 15$~\AA, which is shown by the 
red line. (More precisely, the limit of the 
observed-frame EW(Ly$\alpha$) is $\sim 20$~\AA.) 
This limit is about a factor of two times higher
in the $z=0.65-1.25$ redshift interval.
In our comparisons with high-redshift LAE samples in
Section~\ref{lalpha}, we will include only 
sources having a rest-frame EW(Ly$\alpha$)~$>20$~\AA, 
which is normally used as the definition of high-redshift
LAEs (e.g., Hu et al.\ 1998).
Such a sample should be very robust for the $z=0.195-0.44$
interval.  The blue curve shows an exponential fit of the
form $\exp(-$EW/scale length) to the data above a
rest-frame EW of 20~\AA where we believe the sample to be
substantially complete.
The normalization is $29\pm4$, and the EW scale length is 
$23.7\pm2.2$~\AA.  This distribution is nearly identical in 
shape to that found by Shapley et al.\ (2003) for 
the $z\sim3$ LBGs.  

Only seven of our candidate Ly$\alpha$ Galaxies have a 
rest-frame EW(Ly$\alpha$)~$>100$~\AA.  As can be seen 
from Figure~\ref{ew_dist}, five of these lie in the 
$100-120$~\AA\ range.  There are two  further sources 
which lie above the upper bound of the figure,
but these are all almost certainly  broad-line AGN
where we are only seeing the Ly$\alpha$ line.

\begin{figure}
\includegraphics[width=3.4in,angle=0,scale=1.]{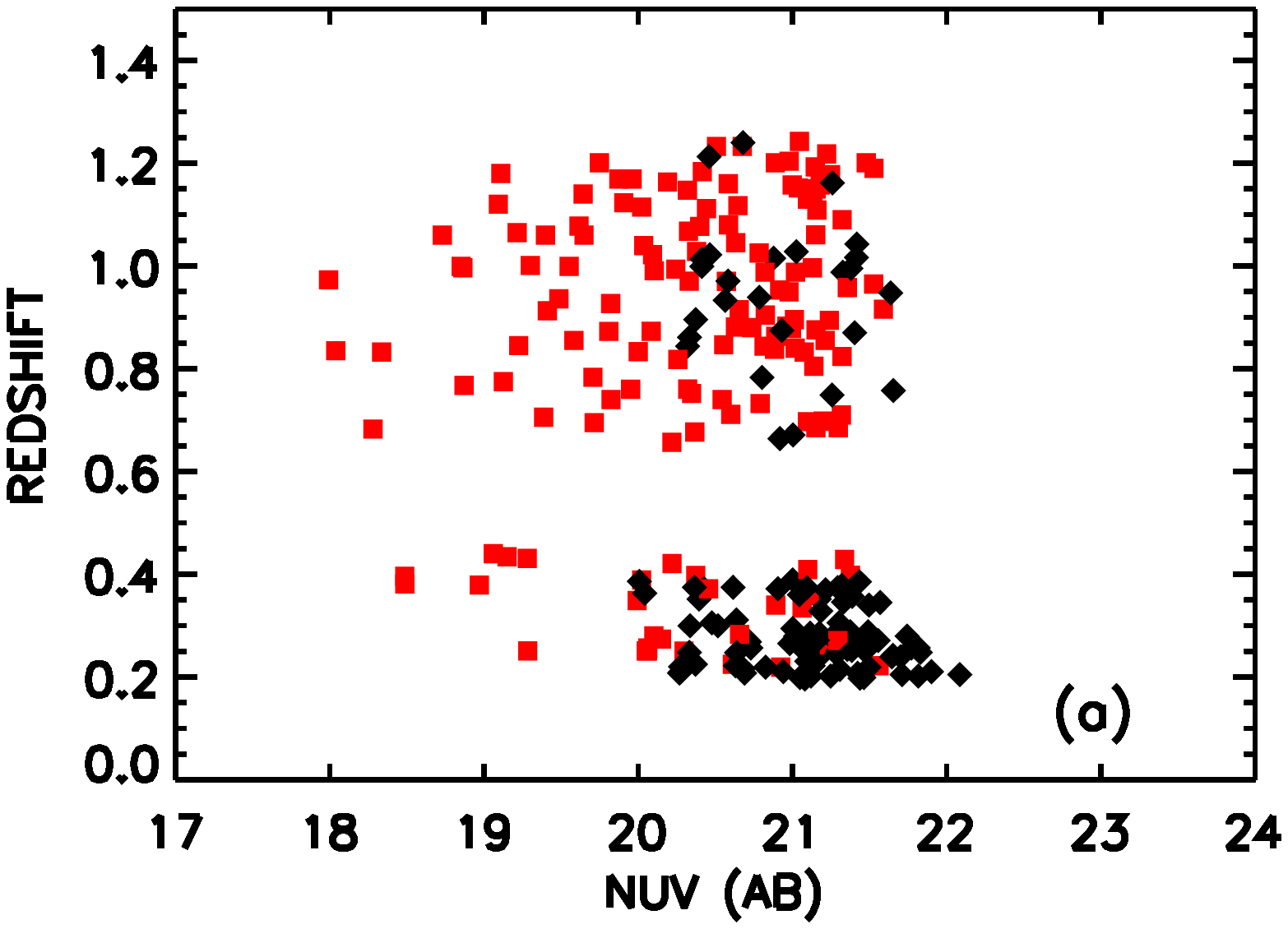}
\includegraphics[width=3.4in,angle=0,scale=1.]{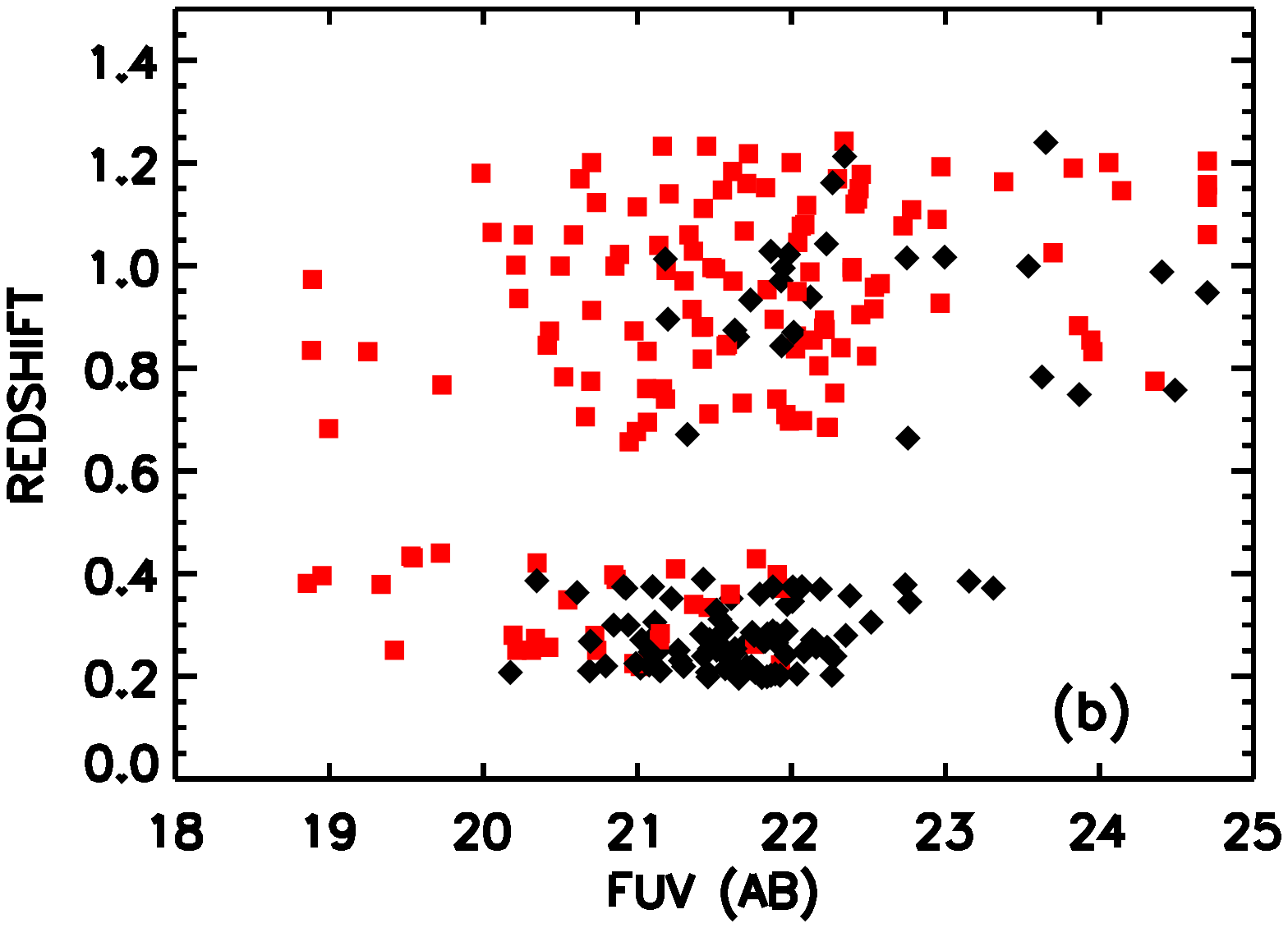}
\includegraphics[width=3.4in,angle=0,scale=1.]{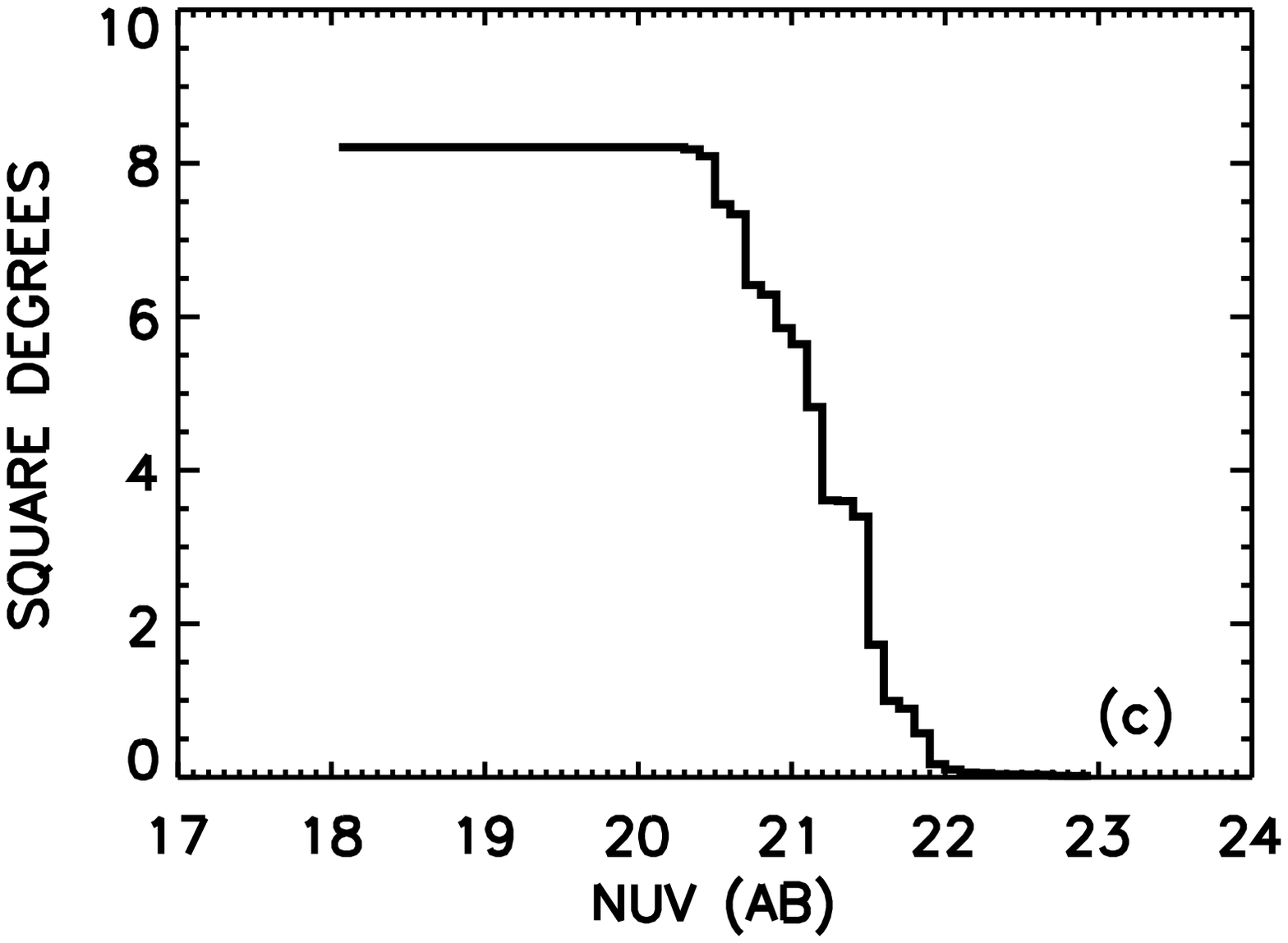}
\caption{
(a) Redshift vs. NUV magnitude for our Ly$\alpha$ 
selected sample from all nine {\em GALEX\/} fields.
Sources classified as AGNs based on high-excitation 
lines in the spectra (otherwise) are shown as red squares
(black diamonds).  The depths of the fields range from
NUV(AB)$=20.5$ in HDFN~00 to 21.8 in GROTH~00.
(b) Redshift vs. FUV magnitude with symbols as in (a).  
Here sources fainter than 24.7 are shown at a nominal 
magnitude of 24.7.  
(c) Total observed area for the nine fields vs. NUV magnitude.
\label{redshift-dist}
}
\end{figure}

In Figure~\ref{redshift-dist} we show the redshift 
distributions of our AGNs (red squares) and our 
candidate Ly$\alpha$ Galaxies (black diamonds) 
versus magnitude.  In (a) we plot the
sources versus NUV magnitude and in
(b) versus FUV magnitude. Nearly all of 
the more luminous sources are AGNs. 
As Deharveng et al.\ (2008) 
note, this results in there being relatively
few candidate Ly$\alpha$ Galaxies in the moderate-redshift 
interval.  In the low-redshift interval the candidate 
Ly$\alpha$ Galaxy population begins to enter 
in substantial numbers around a NUV magnitude of 21. 

Figure~\ref{redshift-dist}(b) suggests immediately
that we are misclassifying some sources as candidate
Ly$\alpha$ Galaxies when they are in fact AGNs, since 
we would expect
$z\sim1$ galaxies to be very faint or undetected in the
FUV band, which lies below the Lyman continuum edge
at these redshifts (e.g., Siana et al.\ 2007; 
Cowie et al.\ 2009). Quantifying this contamination is 
one of the keys to understanding the star-forming population.
We will address this issue in Section~\ref{secagngal}. 

We will require the observed area as a function
of NUV magnitude when we calculate the number
counts and luminosity functions of the LAEs.
We determined this for each field separately by computing the ratio 
of sources with {\em GALEX\/} spectra (identified or not) 
at a given NUV magnitude to sources with that NUV 
magnitude in the continuum catalog.
We then multiplied this ratio for each field
by the area corresponding 
to the $32\farcm5$ selection radius and summed the
results for the nine fields to form the area-magnitude 
relation shown in Figure~\ref{redshift-dist}(c). 
At magnitudes brighter than NUV~$=20.5$ the area is
the 8.2~deg$^2$ area of the nine fields. 
The area then drops as 
we reach the limiting magnitudes of each of the individual 
fields, eventually falling to zero at $\sim21.8$, which is 
the limiting depth of the deepest (GROTH~00) field.

Given the low spatial resolution of the {\em GALEX\/}
data, some sources may be blends. 
While this does not affect the Ly$\alpha$ fluxes or 
luminosities, it will diminish the EWs if the 
measured continuum is too high relative to the line
due to a blend.  It will also 
boost the NUV magnitudes so that we are including 
sources which are fainter than the nominal limits. 

We first re-measured the {\em GALEX\/} coordinates
by visually determining the peak position in the NUV image closest to
the cataloged position in the data archive. 
It is these coordinates which we use in the optical spectroscopy.
The dispersion of the re-measured positions relative to the catalog
positions is $0\farcs5$. However, a small number of sources
have more significant offsets: about 6\% have offsets
greater than $1\farcs5$ and 2\% greater than $3''$.
In Tables~\ref{tab2} through \ref{tab10} we show our
re-measured coordinates, marking those which have a 
significant offset from the cataloged position in the data archive with the 
label (o) in the name column. 

We next tested for the fraction of blends by comparing the 
NUV images  with the ultradeep $u^*$-band images obtained 
on the GROTH~00 and COSMOS~00 fields as part of the 
Canada-France-Hawaii Telescope (CFHT)
Legacy Survey deep observations. 
These are available from the MegaPipe
database at the Canadian Astronomy Data Centre (CADC). 
We corrected for the small absolute
astrometric offsets between the CFHT and {\em GALEX\/} 
coordinates prior to the comparison. We find that 
about 5\% are significant enough blends to seriously affect
the photometry and coordinates. We mark these
in Tables~\ref{tab2} and \ref{tab5} with a (b) in the 
name column.  The blending problem is small enough
to be neglected in the analysis, but the small
number of objects with significant positional 
offsets should be borne in mind when using the 
{\em GALEX\/} coordinates in the data archive.

\section{Optical Spectroscopy}
\label{secopt}

While a number of sources in our {\em GALEX\/} AGN and 
candidate Ly$\alpha$ Galaxy samples have pre-existing optical 
redshifts, these identifications are generally selected 
in some way, often from optical colors or because of the 
presence of X-ray emission or other AGN signatures.  
However, in order to understand the statistical distribution 
of the optical properties of the sources in these
samples, it is critical to observe at optical wavelengths 
sources randomly chosen from these samples.

In addition, in order to understand how the {\em GALEX\/} 
Ly$\alpha$ selected sources relate to the 
more general population of NUV-continuum selected sources, 
it is important to observe at optical wavelengths sources 
randomly chosen from the {\em GALEX\/} spectroscopic sample 
without UV spectral identifications (i.e., without strong 
Ly$\alpha$ emission). 

We therefore made optical spectroscopic follow-up
observations with the DEIMOS spectrograph 
(Faber et al.\ 2003) on the Keck~II 10~m telescope. 
We obtained observations of 39 sources randomly chosen from 
our {\em GALEX\/} candidate Ly$\alpha$ Galaxy sample, 
38 of which lie in the redshift interval $z=0.195-0.44$ and 
one in the redshift interval $z=0.65-1.25$ (Table~\ref{tab11}), 
15 sources randomly chosen from our {\em GALEX\/} AGN 
sample (Table~\ref{tab12}), and 124 sources 
randomly chosen from our {\em GALEX\/} spectroscopic 
sample without UV spectral identifications. For the latter 
we summarize only the 31 sources found to lie in the 
redshift interval $z=0.195-0.44$ (Table~\ref{tab13}).

In Tables~\ref{tab11}$-$\ref{tab13}
we give the {\em GALEX\/} source name, the J2000 right 
ascension and declination, the NUV magnitude, the SDSS
model C $g$ magnitude from the DR6 release
(Adelman-McCarthy et al.\ 2008; these are only given in
Tables~\ref{tab11} and \ref{tab13}), 
the {\em GALEX\/} UV redshift (these are only given in
Tables~\ref{tab11} and \ref{tab12}), the optical 
spectroscopic redshift, the observed-frame EW(H$\alpha$) 
in \AA\ and its $1\sigma$ error (these are only given
in Tables~\ref{tab11} and \ref{tab13}), and the 
spectral type that we determined from the line
widths of the optical spectrum.

\begin{figure}
\includegraphics[width=3.0in,angle=0,scale=1]{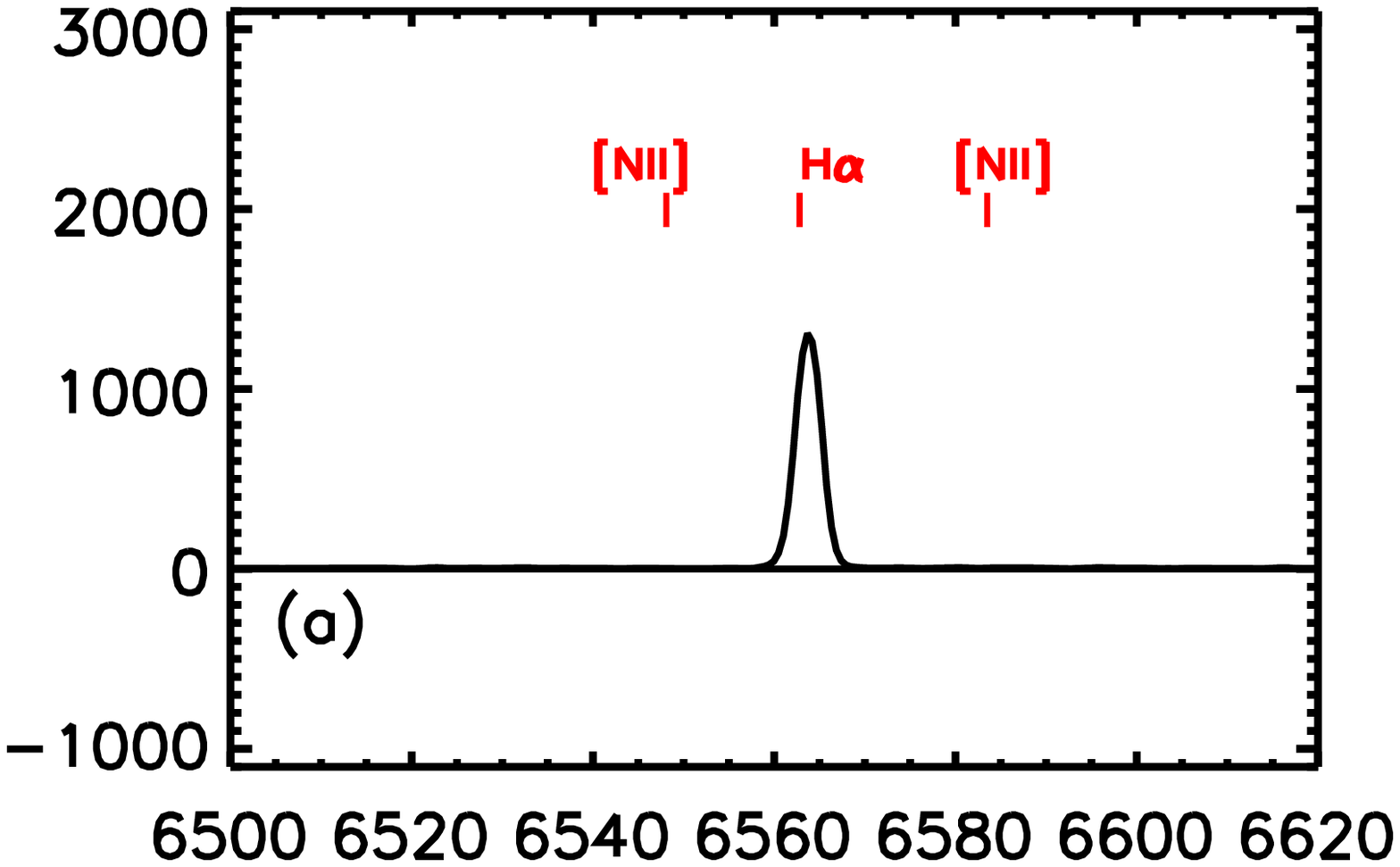}
\includegraphics[width=3.0in,angle=0,scale=1]{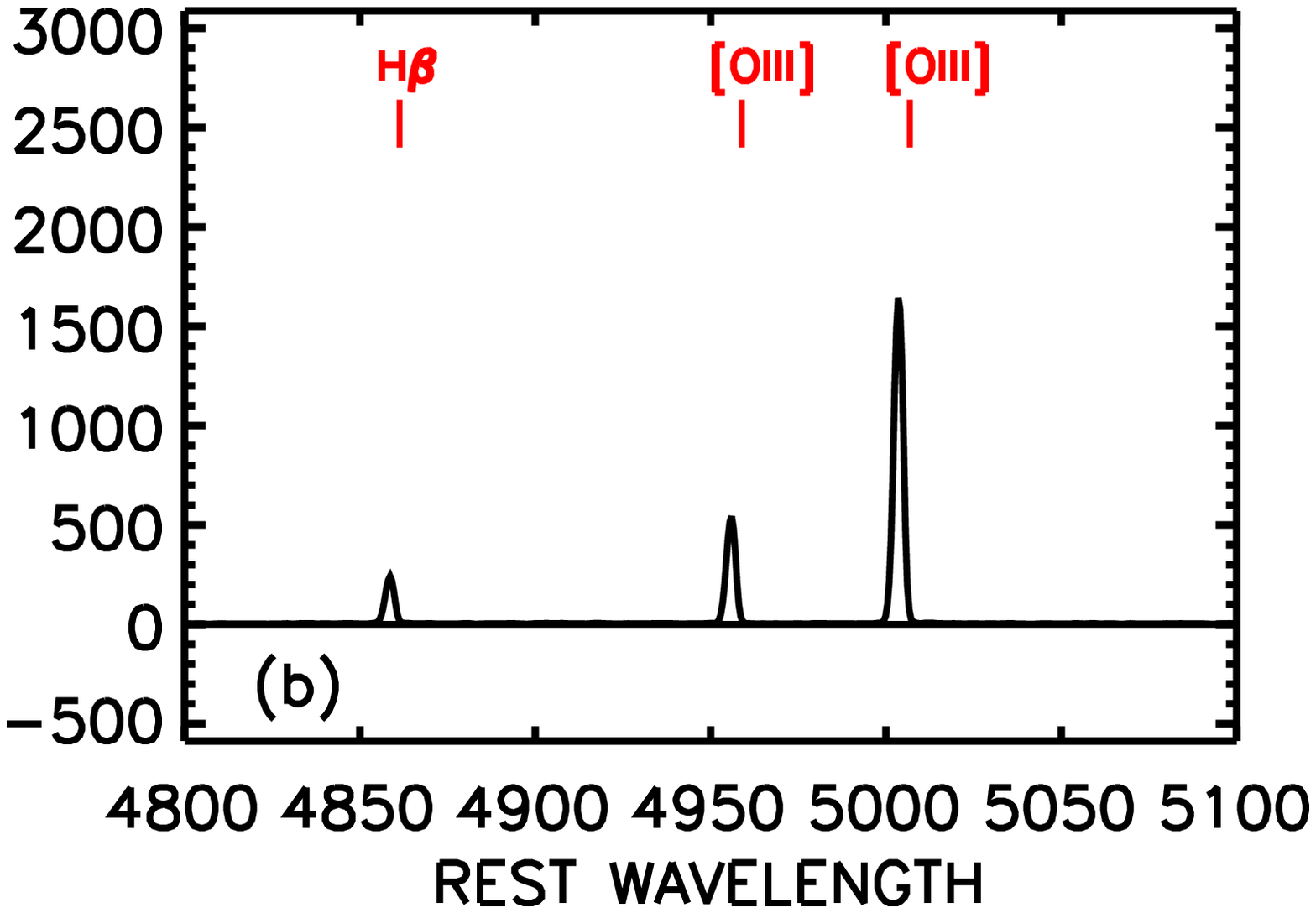}
\includegraphics[width=3.0in,angle=0,scale=1]{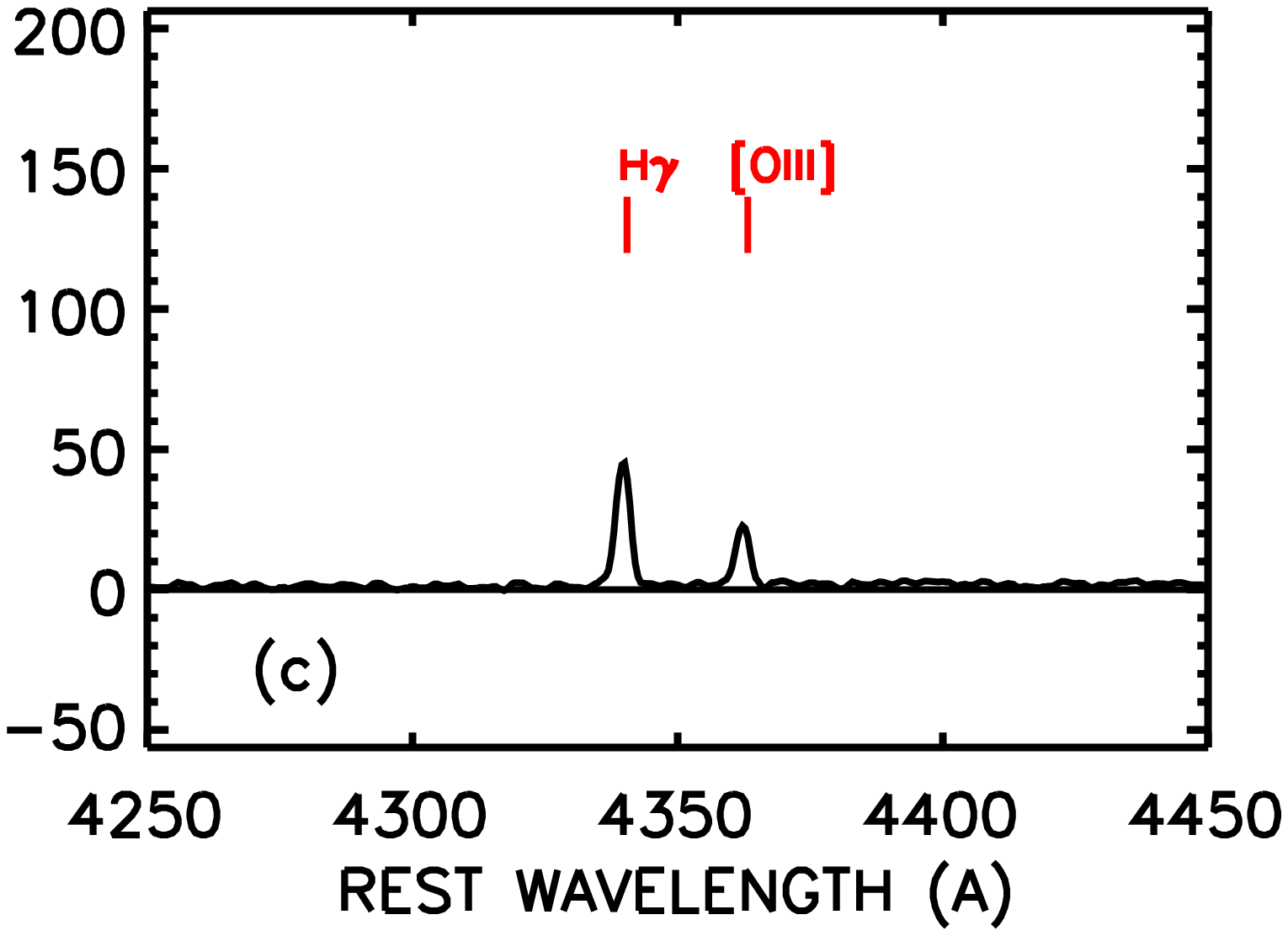}
\includegraphics[width=3.0in,angle=0,scale=1]{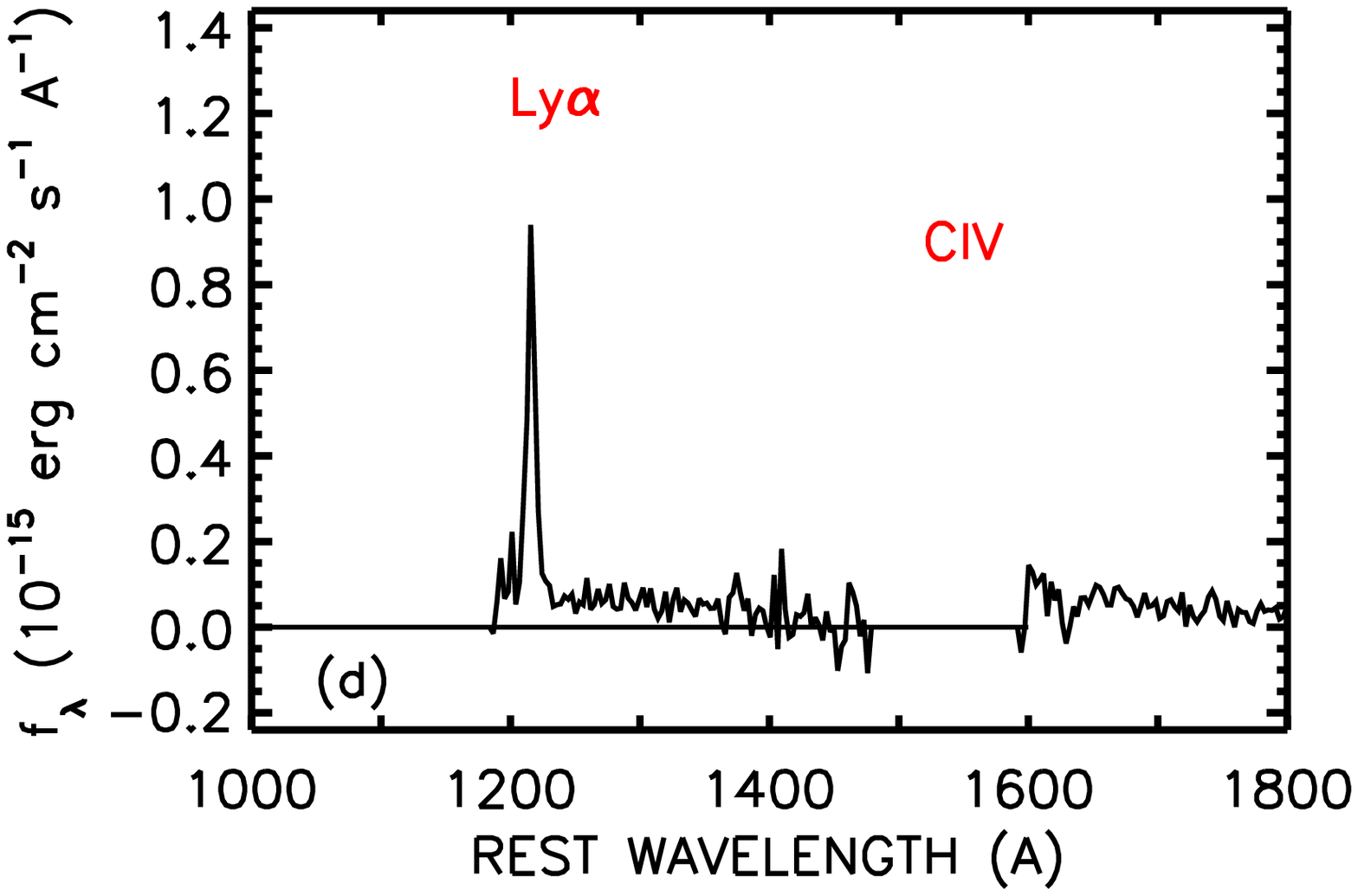}
\caption{
The optical and UV spectrum for the candidate Ly$\alpha$ 
Galaxy GALEX$1417+5228$ in 
the GROTH~00 field at $z=0.206$ (see Table~\ref{tab11}). 
The optical spectrum is dominated by the H$\alpha$
line (in a) and the [OIII] doublet (in b). 
(c) The H$\gamma$ and [OIII]$\lambda4363$ auroral
lines in the spectrum.
(d) The UV spectrum. There are no signs of any 
broad lines in the optical spectrum, and the line ratios 
are consistent with this being an extremely low-metallicity 
emission-line galaxy. The optical spectrum panels are shown 
in $\mu$Jy units, but since the absolute calibration is 
not well determined due to not having photometric conditions, 
we have not labeled them.
\label{whole_spectrum}
}
\end{figure}

In order to provide the widest 
possible wavelength coverage, we used 
the ZD600 $\ell$/mm grating.  We used $1''$ wide 
slitlets, which in this configuration give a resolution 
of 4.5~\AA, sufficient to distinguish the \oii$\lambda$3727 
doublet structure for the  higher redshift sources ($z>0.65$). 
The spectra cover a wavelength range of
approximately 5000~\AA\ and were centered at an average
wavelength of $6800$~\AA, though the exact wavelength 
range for each spectrum depends on the slit position with 
respect to the center of the mask along the dispersion 
direction.  Each $\sim 0.5$~hr exposure was broken 
into three subsets with the objects stepped along the 
slit by $1\farcs5-2''$ in each direction. 
The two-dimensional spectra were reduced following the
procedure described in Cowie et al.\ (1996), and the final
one-dimensional spectra were extracted using a profile 
weighting based on the continuum of the spectrum. 
Calibration stars were observed in the same configuration 
on each of the individual nights.
 
\begin{figure}
\includegraphics[width=3.5in,angle=0,scale=1]{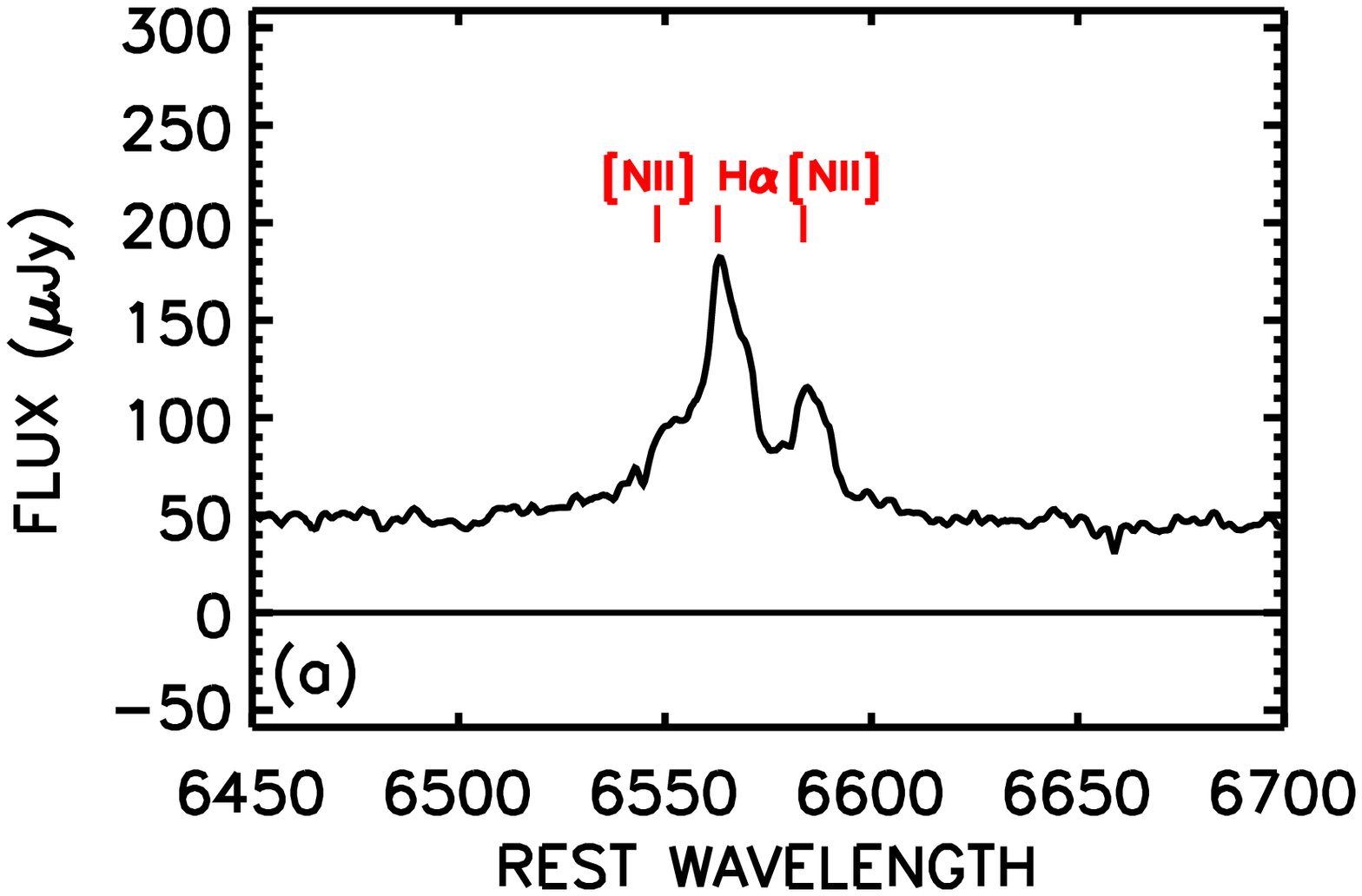}
\includegraphics[width=3.5in,angle=0,scale=1]{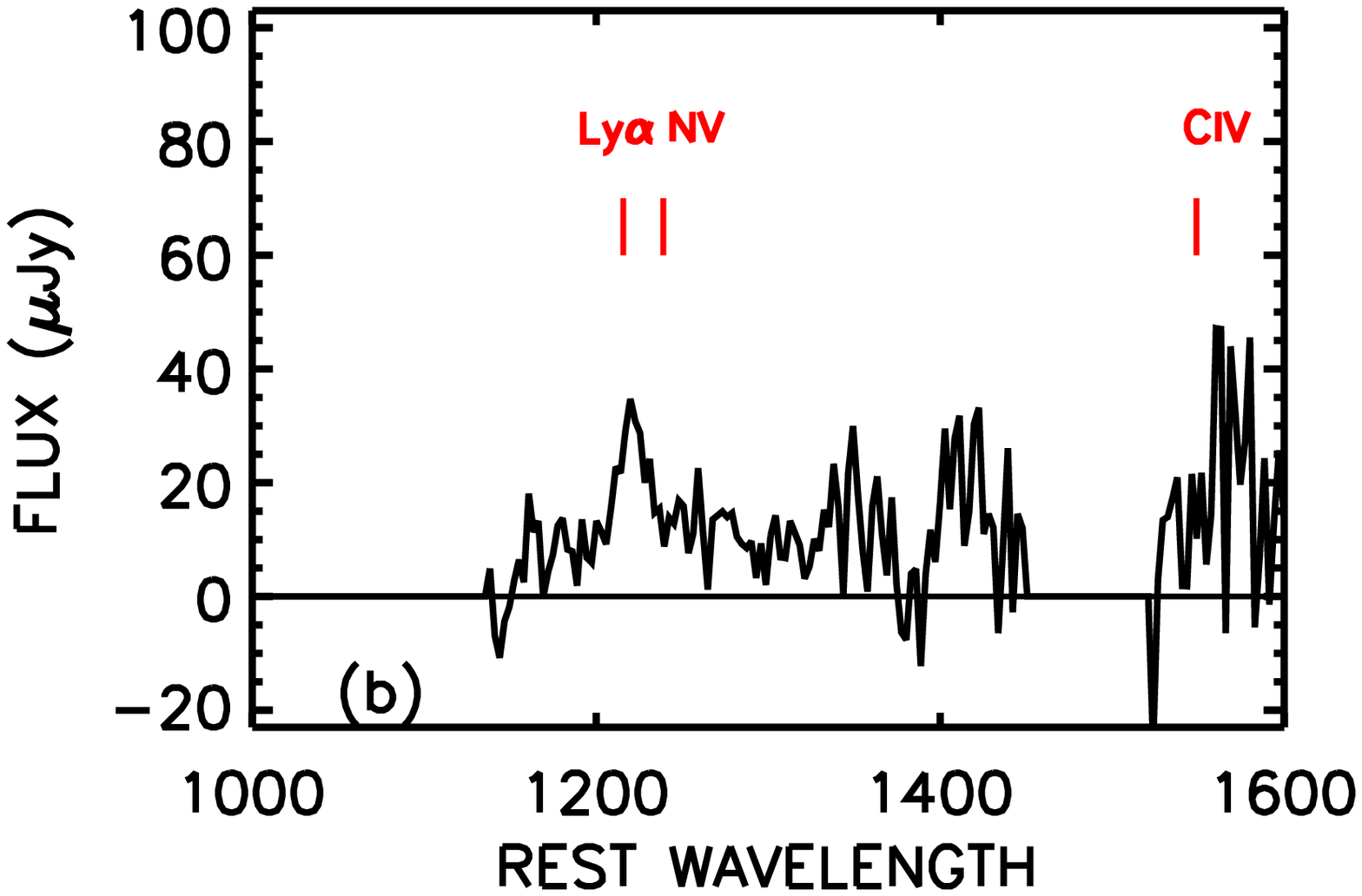}
\caption{
The (a) optical and (b) UV spectrum for the candidate 
Ly$\alpha$ Galaxy GALEX$1047+5827$ 
in the LOCK~00 field at $z=0.2436$ (see Table~\ref{tab11}). 
From the optical data we classify this source as a type~1.8 
Seyfert galaxy based on the broad underlying H$\alpha$ line 
in the optical spectrum and a corresponding 
weak broad H$\beta$ line.  
The optical spectrum was taken in photometric conditions.
While no high-excitation lines 
are seen in the UV spectrum, the Ly$\alpha$ line in the 
UV spectrum is broader than the instrumental resolution.
\label{agn_spectrum}
}
\end{figure}

The spectra were all obtained at a near-parallactic angle
to minimize atmospheric refraction effects.
Each spectrum was initially calibrated using the measured 
response from a calibration star.  
However, since some of the spectra were obtained during a 
night of varying extinction, the absolute flux calibration
using calibration stars is sometimes problematic.
We also note that relative slit losses always pose a 
problem, even for the photometric nights. 
We show pieces of the spectra of two of the 
{\em GALEX\/} Ly$\alpha$ candidate Galaxies
in Figures~\ref{whole_spectrum} and \ref{agn_spectrum}. 
The  absolute fluxes are based on the calibration star, 
but the optical spectrum in Figure~\ref{whole_spectrum} 
was not observed in photometric conditions.

An extensive discussion of the various procedures 
that may be used to determine line fluxes from this type of 
spectrum can be found in Kakazu et al.\ (2007).  However,
here we focus only on the measurements of the relative line 
fluxes, which we need for the metallicity diagnostics, and
on the measurements of the absolute line fluxes,
which we require for the determination of the 
Ly$\alpha$/H$\alpha$ flux ratio.

For each spectrum we fitted a standard set of lines. 
For the stronger lines we used a full Gaussian fit 
together with a linear fit to the continuum baseline.  
For weaker lines we held the full width constant using 
the value measured in the stronger lines and set
the central wavelength to the nominal redshifted value. 
We fitted the [OII]$\lambda3727$ line with two Gaussians 
with the appropriate wavelength separation.  We also 
measured the noise as a function of wavelength by fitting 
Gaussians with the measured line-width from the strong lines
at random positions in the spectrum and computing the 
dispersion in the results. Once again sytematic errors
from the choice of fitting procedure and wavelength
range may exceed the statistical errors and can be
as large as $10-20$\% of the EW.

As long as we restrict the line measurements to short
wavelength ranges where the DEIMOS response is essentially
constant, we can robustly measure the ratios of line fluxes
from the spectra without any flux calibration.
For example, we can assume that the responses 
of neighboring lines (e.g., H$\beta$ and
\oiii$\lambda$5007) are the same and therefore 
measure their flux ratios without calibration. 
In this regard the problems of distinguishing AGNs from 
emission-line galaxies using Baldwin et al.\ (1981)
diagnostic diagrams and of determining metallicities is 
considerably simplified by the presence of H$\alpha$ near 
[NII]$\lambda$6584 (Figure~\ref{whole_spectrum}(a)),
H$\beta$ near \oiii$\lambda$4949 and 
\oiii$\lambda$5007 (Figure~\ref{whole_spectrum}(b)),
and H$\gamma$ near \oiii$\lambda$4363
(Figure~\ref{whole_spectrum}(c)).  For example, by
assuming case~B ratios, we can
take the ratios of the Balmer lines accompanying the
metal lines and the ratios of the metal lines to determine 
the extinction-corrected metal line ratios.  
Unfortunately, we cannot so easily do this 
near \oii$\lambda$3727, where the Balmer lines are weak 
and in some cases contaminated.  Here we must rely on the 
flux calibration made using the measured
response from the calibration star. For this reason
we will use the [NII]$\lambda6584$/H$\alpha$ ratio as our
primary metal diagnostic.

In order to measure the H$\alpha$ line flux, 
we determined the continuum flux from the SDSS model C 
magnitude (DR6 release; Adelman-McCarthy et al.\ 2008) 
at the SDSS bandpass closest to redshifted H$\alpha$, 
and we multiplied this by the observed-frame EW(H$\alpha$). 
This is an approximation, since it assumes that the measured 
EW(H$\alpha$) is representative of the value averaged over 
the total light of the galaxy, including regions outside 
the slit.  However, for the photometric cases we derive 
crudely similar values directly from the calibrated spectra,
suggesting the procedure is relatively robust. In one case
(GALEX1417+5228) there is an H$\alpha$ flux measurement
of $1.3\times10^{-15}$~erg~cm$^{-2}$~s$^{-1}$ from Liang
et al.\ (2004), which may be compared with the presently
derived value of $1.9\times10^{-15}$~erg~cm$^{-2}$~s$^{-1}$.
The Liang et al.\ value is based on calibrated slit spectra
and may be an underestimate, since they measure a corresponding
H$\beta$ flux of $0.61\times10^{-15}$~erg~cm$^{-2}$~s$^{-1}$, 
which for the case B ratio would give an H$\alpha$ flux
of $1.7\times10^{-15}$~erg~cm$^{-2}$~s$^{-1}$ in the absence
of any extinction correction and a higher value if any
extinction were present. However, we will assume, based on
this comparison, that
systematic errors in the absolute H$\alpha$ flux levels could
be as high as 50\%.

\section{AGN-Galaxy Discrimination}
\label{secagngal}

We may expect that sources classified as AGNs based on 
the presence of high-excitation lines in the UV
spectra are truly AGNs. However, the classification of 
the candidate Ly$\alpha$ Galaxies 
is not so clean, and they may contain a substantial amount 
of AGN contamination.  In some cases the high-excitation 
lines of these AGNs may fall at problem wavelengths, and 
in other cases the high-excitation lines may be intrinsically weak.  
The optical spectroscopic data support these points.
All but one of the 15 AGNs classified as such from the UV 
spectra and then observed in the optical are 
straightforwardly identified as intermediate-type Seyfert 
galaxies (see Table~\ref{tab12}; since only one source 
appears as an emission-line galaxy, this suggests that
for the most part one would know that these sources
are AGNs based solely on their optical spectra). 
However, three of the 39 candidate Ly$\alpha$
Galaxies classified as such from the UV spectra 
and then observed in the optical
are also identified as intermediate-type Seyfert galaxies
(see Table~\ref{tab11} and, e.g., Figure~\ref{agn_spectrum}). 
The remaining 36 do not have visible broad lines 
(see, e.g., Figure~\ref{whole_spectrum}).  This
suggests that we have about an 8\%  broad-line AGN contamination rate 
in this population.  Finally, we note that the 31 sources 
chosen randomly from the {\em GALEX\/} spectroscopic sample 
without UV spectral identifications that happen to lie in 
the redshift interval $z=0.195-0.44$ all appear 
to be star-forming galaxies with H$\alpha$ luminosities 
comparable to the candidate Ly$\alpha$ Galaxies
(see Table~\ref{tab13}).

\begin{figure}
\includegraphics[width=3.4in,angle=0,scale=1]{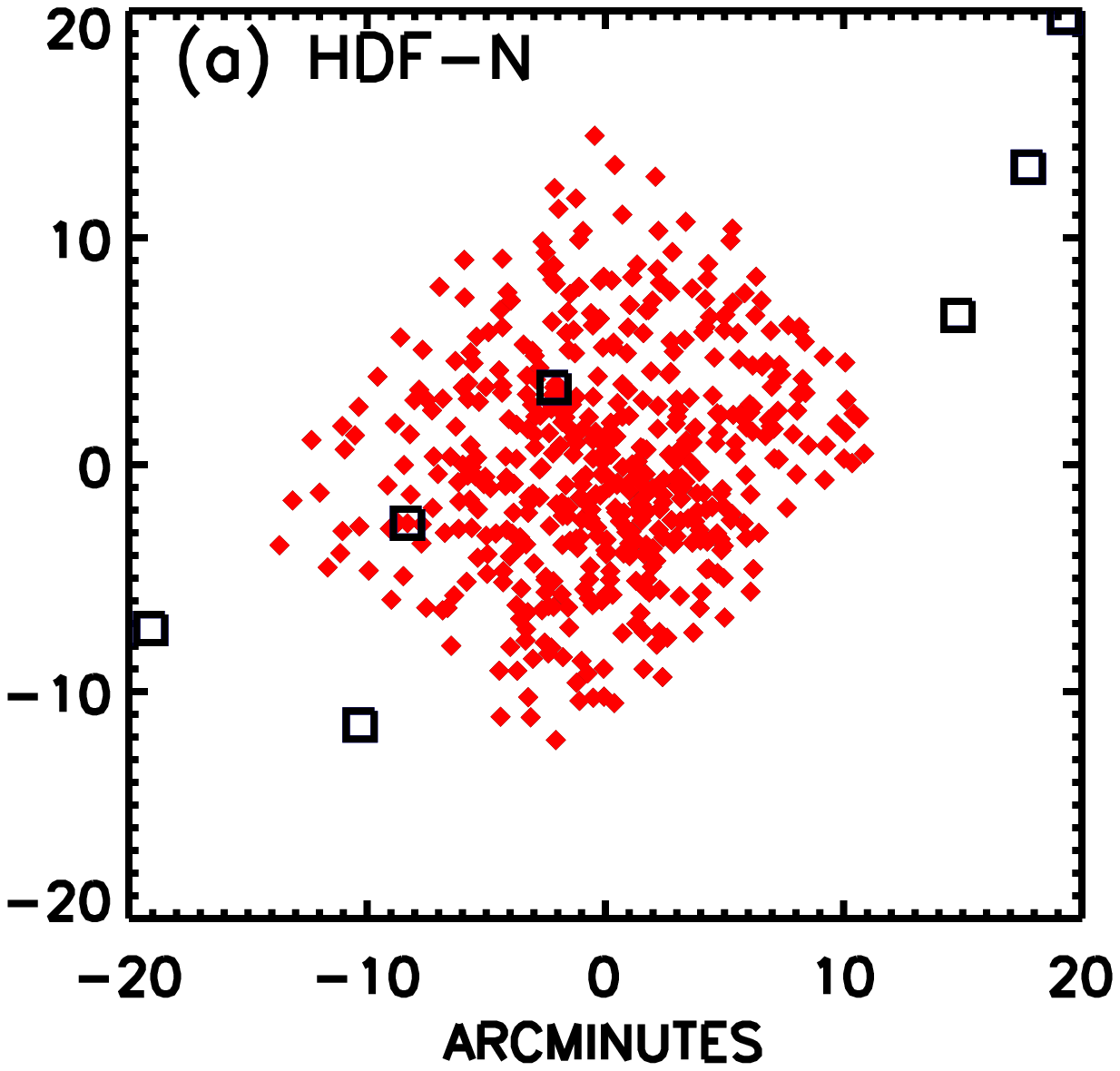}
\includegraphics[width=3.4in,angle=0,scale=1]{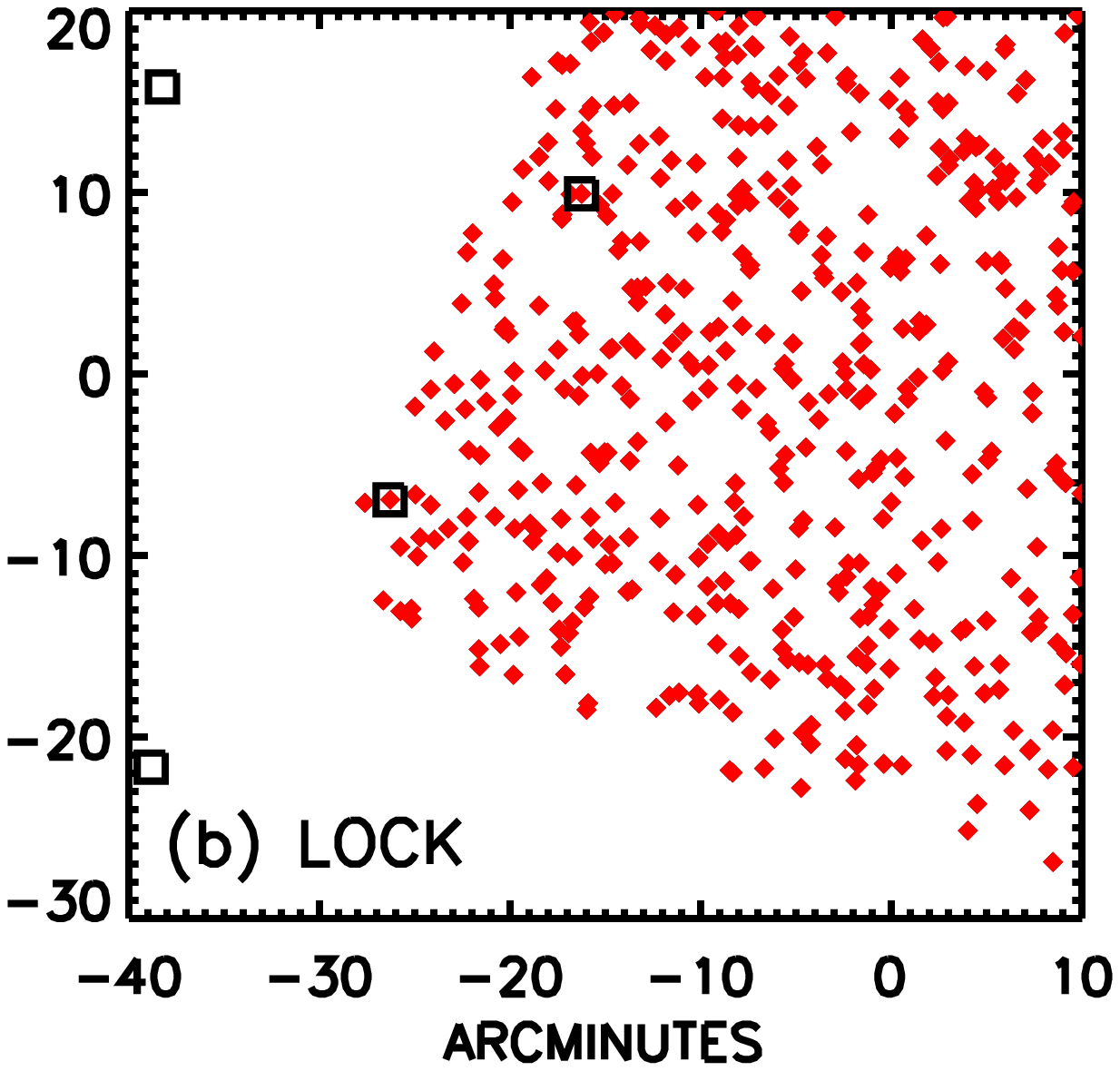}
\includegraphics[width=3.4in,angle=0,scale=1]{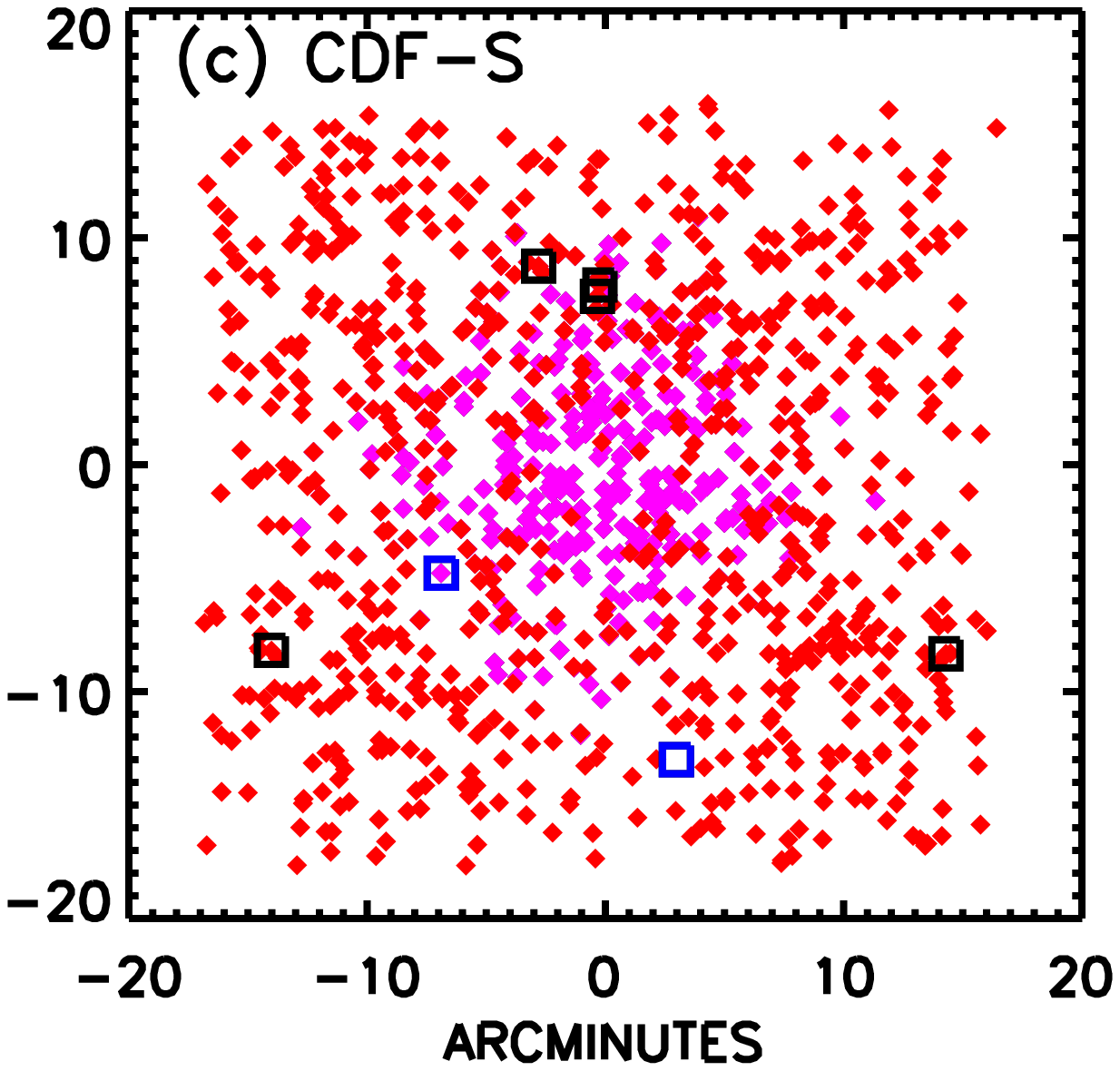}
\caption{
Comparison of the {\em GALEX\/} Ly$\alpha$ selected
sources with the {\em Chandra\/} X-ray sources
in the (a) HDFN~00 (CDF-N), (b) LOCK~00 (CLANS),
and (c) CDFS~00 (CDF-S and Extended CDF-S) fields. 
The X-ray sources are shown with red solid diamonds,
except in the CDFS~00 field where the 2~Ms CDF-S 
(Extended CDF-S) X-ray sources are shown with purple 
(red) solid diamonds.  The {\em GALEX\/} sources are shown 
with black (blue) open squares for AGNs based on the
presence of high-excitation lines in their UV spectra 
(otherwise).
\label{xray_comparison}
}
\end{figure}

There are deep X-ray observations of portions of the 
{\em GALEX\/} areas for several of the fields used in
our study.  In Figure~\ref{xray_comparison} we illustrate 
the overlap of the X-ray sources (red solid diamonds)
with the {\em GALEX\/} Ly$\alpha$ selected sources 
(black open squares for AGNs based on their UV spectra; 
blue open squares otherwise).  
In (a) we compare the CDF-N sources 
(Alexander et al.\ 2003) with the HDFN~00 sources;
in (b) we compare the CLANS sources (Trouille et al.\ 2008, 2009) 
with the LOCK~00 sources; and in (c) we compare the 
Extended CDF-S sources (Lehmer et al.\ 2005; Virani et al.\ 2006) 
and the 2~Ms CDF-S sources (purple solid diamonds;
Luo et al.\ 2008) with the CDFS~00 sources.
In total eleven sources in the {\em GALEX\/} 
Ly$\alpha$ selected sample lie in the X-ray fields.
We summarize their X-ray properties in Table~\ref{tab14}, 
where we give the {\em GALEX\/}
name of the source, the J2000
right ascension and declination, the $2-8$~keV and 
$0.5-2$~keV fluxes, the $2-8$~keV luminosity computed 
using the procedure given in Barger et al.\ (2005),
the redshift from the UV spectra, and the classification 
of the source as an AGN or a candidate Ly$\alpha$ 
Galaxy from the UV spectrum.

All nine of the sources classified as AGNs based on their 
UV spectra are also X-ray sources and have
logarithmic $2-8$~keV luminosities in the range 
$\log L_X=43.24-44.55$~erg~s$^{-1}$ (see Table~\ref{tab14}), 
placing them at or near quasar luminosities ($\log L_X > 44$) 
and near the break in the AGN luminosity 
function at $z\lesssim1.2$ (e.g., Yencho et al.\ 2009). 
In contrast, of the two candidate Ly$\alpha$ Galaxies, 
one is only weakly detected in the $0.5-2$~keV band in 
the extremely deep 2~Ms CDF-S image, and the other is 
not detected in the Extended CDF-S image.
This places both of their $\log L_X$ values 
below 41~erg~s$^{-1}$ (see Table~\ref{tab14}), suggesting that they 
are strong star-forming galaxies rather than AGNs.
Thus, in all of the overlapped sources, the UV 
classification is robustly supported by the X-ray data.

\begin{figure}
\includegraphics[width=3.5in,angle=0,scale=1.]{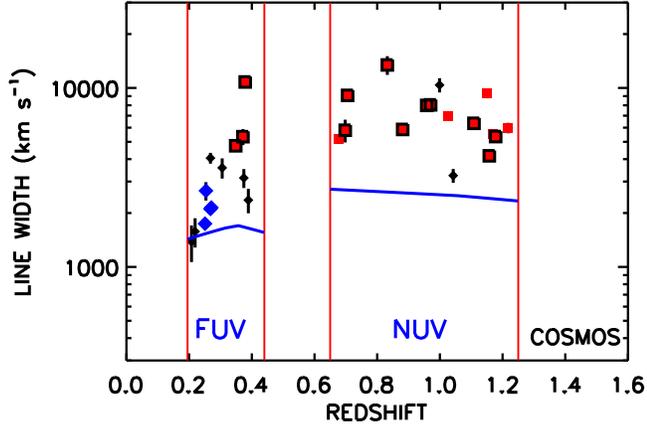}
\caption{
Comparison of the rest-frame line widths of the {\em GALEX\/} 
Ly$\alpha$ selected sample in the COSMOS~00 field
with those of optically selected AGNs 
in the COSMOS field from Prescott et al.\ (2006)
vs. redshift. The {\em GALEX\/}
sources are plotted with red solid squares (AGNs based on the
presence of high-excitation lines in their UV spectra) and black 
small diamonds (otherwise).
The error bars show the $\pm 1\sigma$ range in the line widths.
The {\em GALEX\/} sources with matches to 
the optically selected AGNs in Prescott et al.\ 
are surrounded by black 
large open squares. The {\em GALEX\/} sources with matches 
to the emission-line galaxies in Prescott et al.\ 
are shown with blue solid diamonds.  The blue curves show the 
varying instrumental resolution for a point source.
\label{cosmos_comparison}
} 
\end{figure}

We can also make a comparison with a previous optical 
spectroscopic sample, since Prescott et al.\ (2006) 
observed a large sample of candidate AGNs in a
1.39~deg$^2$ region of the COSMOS {\em HST\/} Treasury 
field (Scoville et al.\ 2007), which overlaps 
significantly with the {\em GALEX\/} COSMOS~00 field.
In Figure~\ref{cosmos_comparison} we plot the
Ly$\alpha$ line widths for the
{\em GALEX\/} Ly$\alpha$ selected sample 
(red solid squares for AGNs based on their UV spectra; 
black small diamonds otherwise) that lie in 
the overlap region with the Prescott et al.\ (2006) 
observations versus redshift. Where Prescott et al.\ (2006)
classifications exist for these sources, we show them
on the figure (black large open squares for 
optically selected AGNs; blue solid diamonds for 
optical emission-line galaxies). 
All eleven of the {\em GALEX\/} AGNs with a 
Prescott et al.\ (2006) optical identification were 
classified by those authors as AGNs based on their 
optical spectra, and all three of the {\em GALEX\/} 
candidate Ly$\alpha$ Galaxies with a Prescott et al.\ (2006) 
optical identification were classified by those authors as 
emission-line galaxies based on their optical spectra. 
All of the redshifts in the two samples are fully consistent.

\begin{figure}
\includegraphics[width=3.5in,angle=0,scale=1.]{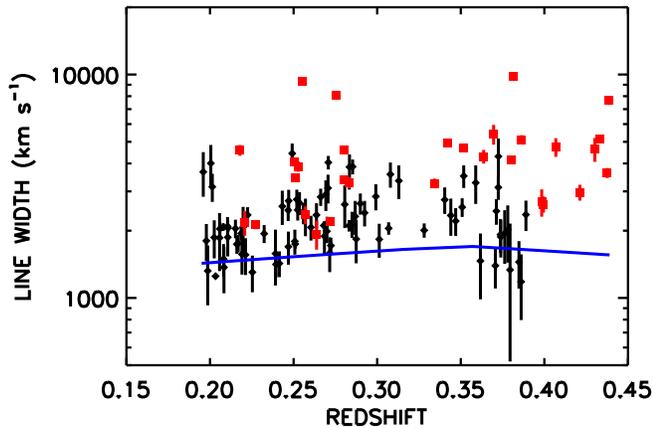}
\caption{
Rest-frame line widths of the {\em GALEX\/} Ly$\alpha$ selected 
sample in the low-redshift interval.
The sources are plotted with red solid
squares (AGNs based on the presence of high-excitation
lines in their UV spectra) and black diamonds (otherwise). 
The error bars show the $\pm1\sigma$ range in the line widths.
The blue curve shows the varying instrumental resolution for a 
point source.
\label{line_widths}}
\end{figure}

As Figure~\ref{cosmos_comparison} illustrates, 
the Ly$\alpha$ line widths provide a second diagnostic 
that may be used to separate AGNs from emission-line 
Galaxies. In general, at the low resolution of the 
{\em GALEX\/} grisms the emission lines from
galaxies should be essentially unresolved, and only broad 
AGN lines ($\sim$~several thousand km~s$^{-1}$) should 
produce significant broadening in the {\em GALEX\/}
spectra. (In the present paper we only determine
the intrinsic galaxy line widths from the 
optical spectroscopy.) This problem is further
complicated by the nature of the grism data where
the final resolution depends on the image size.
Deharveng et al.\ (2008) 
used the line width as a second criterion to separate AGNs from 
emission-line Galaxies in their {\em GALEX\/} sample.  
However, as we illustrate 
further in Figure~\ref{line_widths}, a good deal of scatter 
in the measured line widths and a substantial overlap 
between AGNs and candidate Ly$\alpha$ Galaxies make the 
choice of a dividing line difficult. In particular, the
average line width of the candidate Ly$\alpha$ Galaxies
appears to be larger than the nominal resolution of
a point source in the {\em GALEX\/} data.

\begin{figure}
\includegraphics[width=3.5in,angle=0,scale=1.]{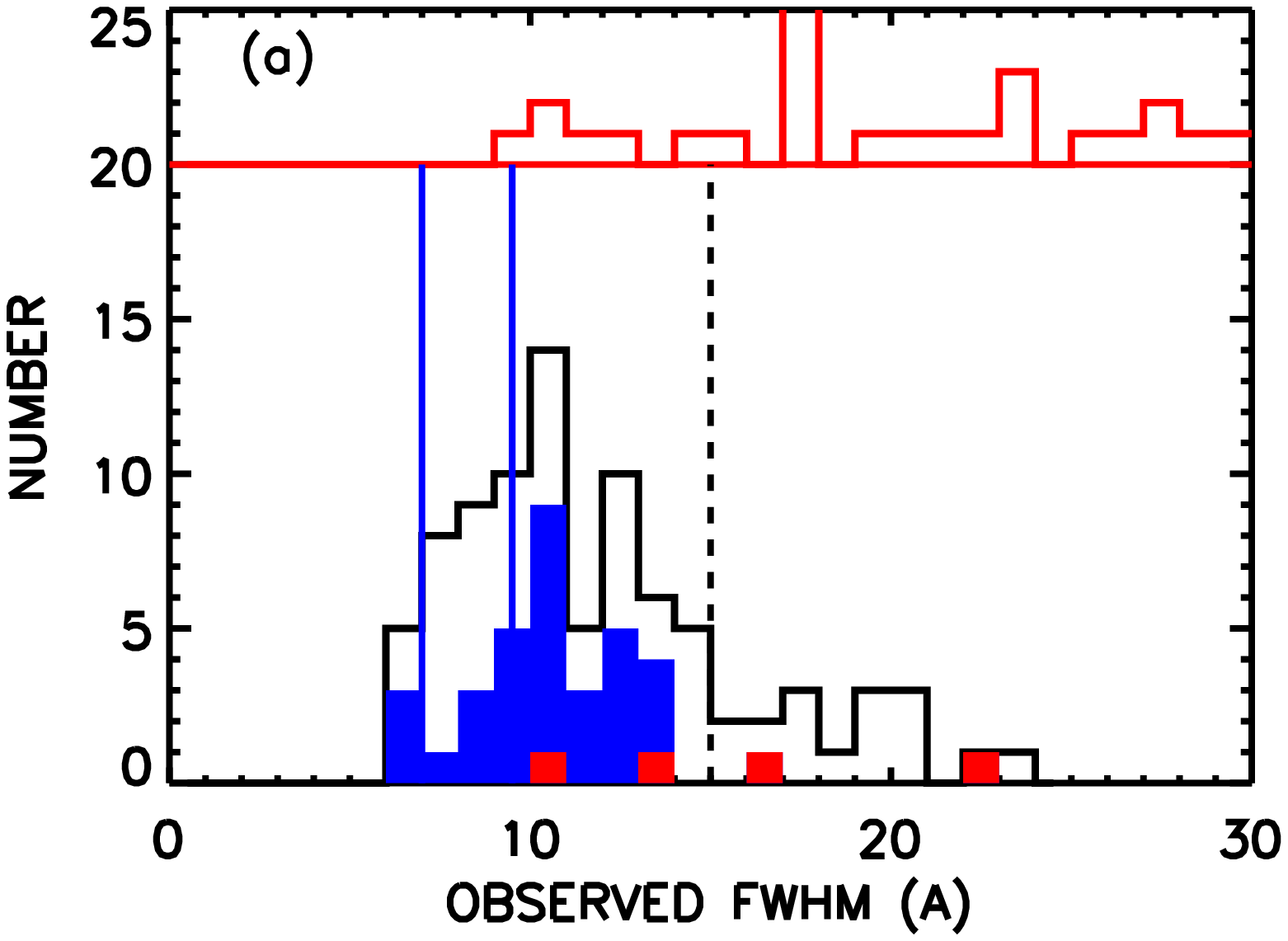}
\includegraphics[width=3.5in,angle=0,scale=1.]{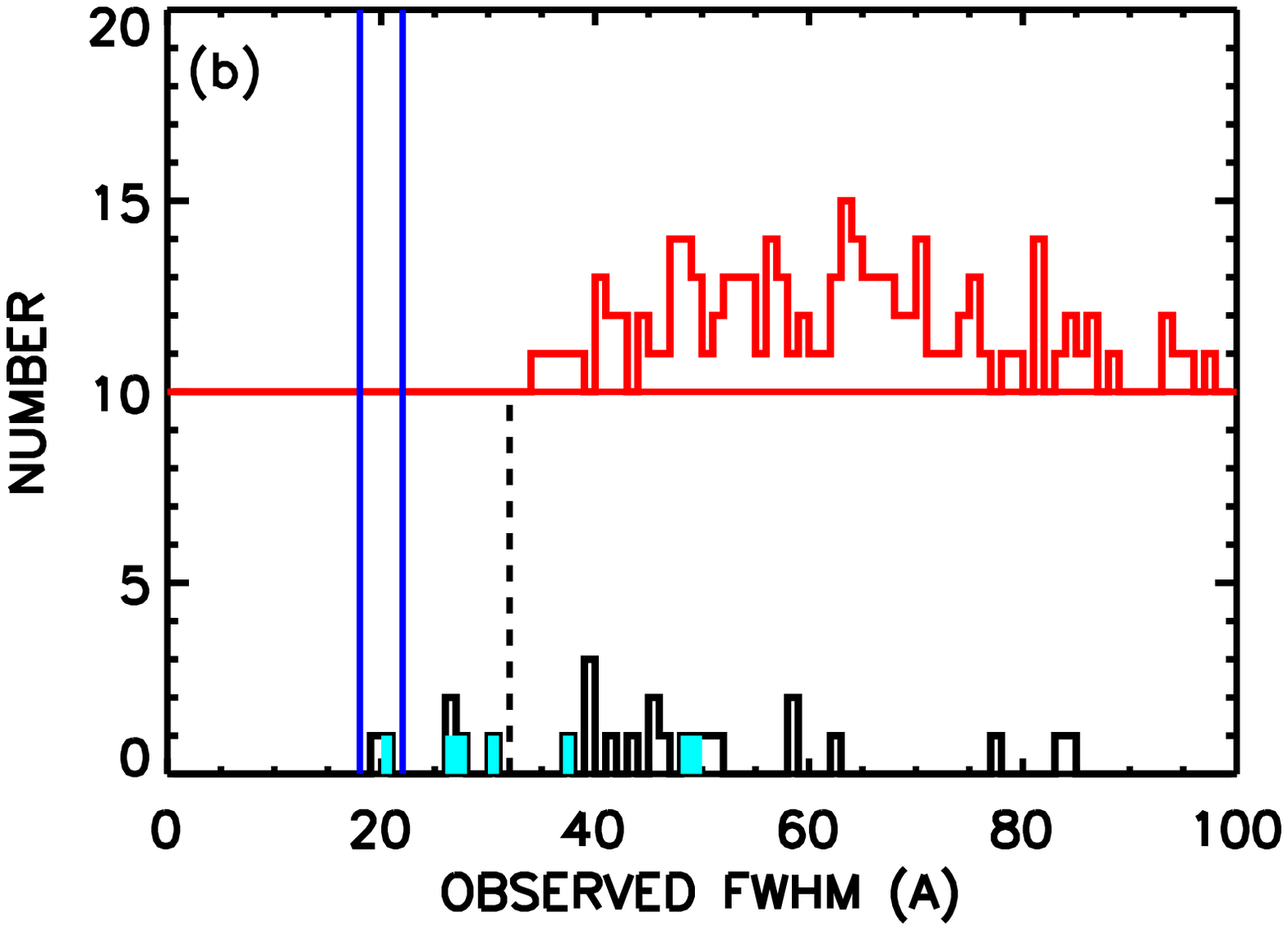}
\caption{
Distribution of observed-frame
line widths in the (a) $z=0.195-0.44$
and (b) $z=0.65-1.25$ {\em GALEX\/} Ly$\alpha$ selected
samples.  The black open histogram shows the candidate 
Ly$\alpha$ Galaxies, and the red open histogram shows the 
AGNs (vertically displaced by 20 in (a) and 10 in (b) 
for clarity). In each panel the blue vertical lines show
the range of the instrumental resolution in the wavelength
interval,
and the black vertical dashed line shows the proposed 
split above which we consider that the FWHM line width 
alone may show a source to be an AGN 
(i.e., 15~\AA\ in the FUV
spectra and 30~\AA\ in the NUV spectra). In (a)
the blue (red) shaded histogram shows sources in the
candidate Ly$\alpha$ Galaxy population confirmed 
to be Galaxies (broad-line AGNs) from optical spectroscopy.
In (b) the cyan shaded histogram
shows sources in the candidate Ly$\alpha$
Galaxy population with strong Lyman continuum breaks. 
\label{wid_dist}}
\end{figure}

We consider this more quantitatively in 
Figure~\ref{wid_dist}(a), where we show the 
distribution of line widths (now in the observed frame 
so we may compare with the instrumental resolution) 
in the {\em GALEX\/} $z=0.195-0.44$ Ly$\alpha$ selected 
sample.  We show the distribution for the candidate 
Ly$\alpha$ Galaxies with the black open histogram, and 
we show the distribution for the sources classified
as AGNs based on their UV spectra with the red open
histogram (offset upwards for clarity).
There is a substantial spread of the Ly$\alpha$ 
Galaxies about the instrumental resolution 
(whose range with wavelength is shown by the 
blue vertical lines) resulting from the
effects of statistical and systematic
noise in the Gaussian fitting. As we have discussed
above, the lines are also wider on average than
the nominal instrumental resolution for a point-like object.

We show all 33 candidate Ly$\alpha$ Galaxies that we 
have confirmed as star-forming  emission-line galaxies using 
our DEIMOS optical spectroscopy (see Table~\ref{tab11}; hereafter, 
we refer to these as our optically-confirmed Ly$\alpha$ 
Galaxies) with the blue shaded histogram.  
This excludes the four galaxies with broad lines in
the optical spectra labeled as AGNs in Table~\ref{tab11},
as well as one additional object classified as an
AGN based on its optical line ratios (see Section~\ref{la_metal}).
All of the optically-confirmed Ly$\alpha$
Galaxies lie near the instrumental 
resolution, as do two of the four candidate Ly$\alpha$
Galaxies that we found to be broad-line AGNs using our DEIMOS optical 
spectroscopy (see Table~\ref{tab11}; red shaded histogram).  
The remaining two AGNs (including the one in the 
LOCK~00 field shown in Figure~\ref{agn_spectrum}) have 
substantial Ly$\alpha$
 line widths.  We have chosen an 
observed-frame FWHM line width of 15~\AA, shown as
the black vertical dashed line in Figure~\ref{wid_dist}(a), 
to roughly split the spread of true galaxy line
widths from the extended tail, which may be
predominantly AGN-dominated.  However, further
work is clearly needed to better define this
separation, and, even with this elimination,
there will be AGNs where the lines are not
resolved (as is evident from the red open histogram).  
There are 72 candidate Ly$\alpha$ Galaxies in Figure~\ref{wid_dist}(a)
with widths less than 15~\AA, of which 33 are 
optically-confirmed Ly$\alpha$ Galaxies and 3 are 
optically-confirmed AGNs (this includes the AGN
identified from the line ratios).  This suggests that even 
with a line width criterion, we have about a 10\% AGN 
contamination in the candidate Ly$\alpha$ Galaxy sample.

\begin{figure}
\hskip -0.6cm
\includegraphics[width=3.7in,angle=0,scale=1.]{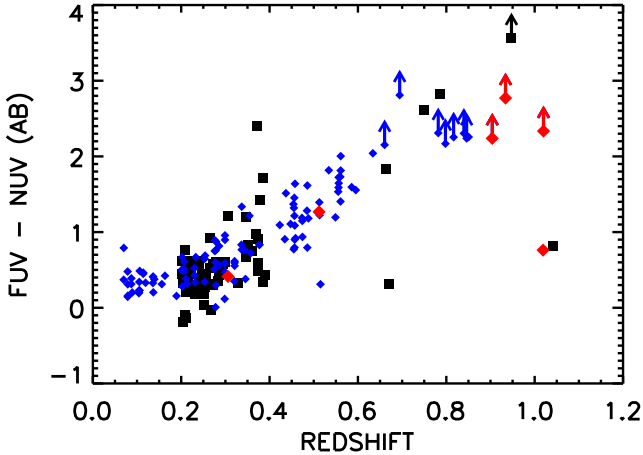}
\caption{
Comparison of the broadband FUV$-$NUV colors of the
{\em GALEX\/} candidate Ly$\alpha$ Galaxies that 
satisfy the line width cuts (black squares) with the 
colors of a spectroscopically complete sample of 
NUV~$<23$ galaxies in the GOODS-N from 
Barger et al.\ (2008; red diamonds when
identified as AGNs based on their X-ray luminosities
and blue diamonds otherwise).
Note that one of the {\em GALEX\/} $z=0.65-1.25$ 
candidate Ly$\alpha$ Galaxies has a color redder than 
the $y$-axis upper limit of the plot.  Sources with 
FUV~$>25.2$ are shown at that value 
with upward-pointing arrows. For most of the data points
the errors in the colors are small compared to the symbol
sizes.
\label{galaxy_color}
}
\end{figure}

In Figure~\ref{wid_dist}(b) we show the distribution 
of Ly$\alpha$ line widths in the {\em GALEX\/} $z=0.65-1.25$ 
Ly$\alpha$ selected sample.  Here most of the candidate
Ly$\alpha$ Galaxies are clearly too broad relative to 
the instrumental resolution (whose range with wavelength 
is shown by the blue vertical lines), even when allowance
is made for the line-width noise and the uncertainty
in the instrumental resolution,
and must be AGNs. We consider only the 6 sources with
observed-frame FWHM line widths less than 30~\AA\ 
(black vertical dashed line) to be 
plausible candidate Ly$\alpha$ Galaxies. However, in 
contrast to the lower redshift sources, we can test this
classification further by looking for
the presence of the Lyman break at the Lyman continuum
edge.  The nine sources with possible Lyman continuum 
breaks are shown by the cyan shaded histogram in
Figure~\ref{wid_dist}(b).  Four of the six
sources in the low-velocity width range (i.e., $<30$~\AA) 
have breaks, suggesting that at least for these sources 
we have confirmation of the line width classification by 
the Lyman break criterion.  Alternatively, however, they 
may just be type~2 AGNs with Lyman continuum breaks,
as the $>30$~\AA\ sources with Lyman continuum breaks 
likely are. Only one of these sources (Table~11) has
been observed and confirmed as a star former with optical 
spectroscopy.

We test our line width cuts in Figure~\ref{galaxy_color}, 
where we compare the colors of the candidate 
Ly$\alpha$ Galaxies that satisfy our line width cuts 
(black squares) with the colors of a spectroscopically
complete sample of NUV~$<23$ galaxies in the GOODS-N 
from Barger et al.\ (2008;
red diamonds for sources classified as AGNs based on 
having X-ray luminosities $>10^{42}$~erg~s$^{-1}$ and
blue diamonds otherwise).  Sources fainter than FUV~$=25.2$ 
are shown at this value with upward-pointing arrows. 
The {\em GALEX\/} candidate Ly$\alpha$ Galaxies generally 
lie along the color track of the GOODS-N galaxies, 
except for the two $z=0.65-1.25$ sources without Lyman 
breaks whose colors are well below the track, placing
them in the AGN category.
As can also be seen from Figure~\ref{galaxy_color}, 
many of the AGNs in the GOODS-N follow the galaxy
color track, so the {\em GALEX\/} Ly$\alpha$ Galaxy
sample might have further AGN contamination. 
Thus, the four sources with a Lyman continuum
break and an observed-frame 
line width less than 30~\AA\ in the $z=0.65-1.25$ interval 
should be considered as an upper limit on the Ly$\alpha$
selected star-forming galaxy population in this redshift interval.

\section{Comparison of the GALEX LAE Sample
with High-Redshift LAE Samples}
\label{lalpha}

\subsection{LAE Selection}
\label{lae_prop}

In order to make valid comparisons with the high-redshift 
LAE samples, we need to construct our
{\em GALEX\/} LAE sample carefully.  We start with our candidate 
Ly$\alpha$ Galaxy sample, which already
excludes sources with detectable high-excitation lines in the 
UV spectra.  We now also exclude sources with observed-frame
FWHM line widths greater than 15~\AA\ (30~\AA) in the redshift 
interval $z=0.195-0.44$ ($z=0.65-1.25$). 
For the moderate-redshift interval we further exclude the
two $<30$~\AA\ sources without a strong Lyman continuum break.

\begin{figure}
\hskip -0.6cm
\includegraphics[width=3.7in,angle=0,scale=1.]{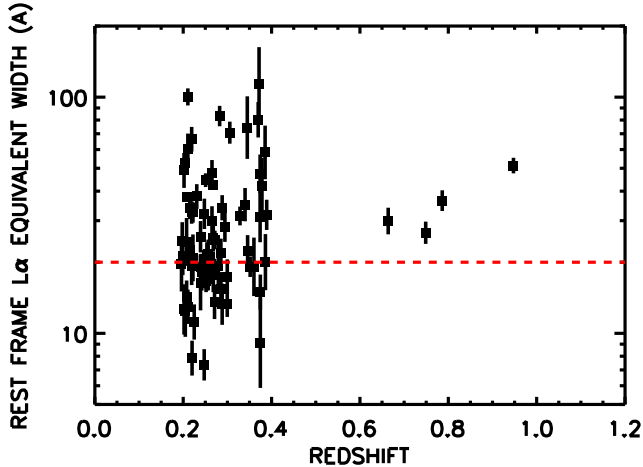}
\caption{
Rest-frame EW(Ly$\alpha$) for the candidate Ly$\alpha$
Galaxies vs. redshift. This consists of
objects with no high-excitation lines in
the UV and with widths
less than 15~\AA\ in the low-redshift interval
and less than 30~\AA\ in the moderate-redshift
interval. Objects with FUV-NUV~$<1.8$
are also excluded in the moderate-redshift
interval. The final LAE sample is taken 
to be those sources with a rest-frame 
EW(Ly$\alpha$)~$>20$~\AA\ (red dashed line).
The error bars show the $\pm 1\sigma$ errors from
the formal statistical fit of the Gaussian and
baseline. As discussed in the text, there may
be comparable systematic errors resulting
from the choice of fitting procedure.
\label{lae_sample}
}
\end{figure}

In Figure~\ref{lae_sample} we plot rest-frame EW(Ly$\alpha$)
versus redshift for the above sample.
A substantial fraction of the sources have an EW(Ly$\alpha$) less 
than or equal to the 20~\AA\ value normally used to define the 
high-redshift LAE population (e.g., Hu et al.\ 1998). If we follow the 
EW(Ly$\alpha)>20$~\AA\ definition, we are left with 41 LAEs in the 
low-redshift interval and 4 in the moderate-redshift interval. This LAE 
sample is tabulated in Table~\ref{tab15} (low-redshift interval)
and Table~\ref{tab16} (moderate-redshift interval), where we give 
the name, the J2000 right ascension and declination, 
the NUV and FUV magnitudes, the {\em GALEX\/} redshift,
the logarithm of the Ly$\alpha$ luminosity, 
$L=4\pi d_{L}^2 f({\rm Ly}\alpha)$, and its $1\sigma$ error, where
$f({\rm Ly}\alpha)$ is the observed line flux and $d_L$ 
is the luminosity distance, the rest-frame EW(Ly$\alpha$)
and its $1\sigma$ error, and the ground-based redshift, if available.
The tables are sorted by rest-frame EW(Ly$\alpha$), with
the highest EW first. We also include in the table
sources with 15~\AA$<$EW(Ly$\alpha)\le 20$~\AA, though 
these are not part of our final LAE sample.

The ground-based redshifts are based on this paper's
observations, except for two sources in the COSMOS~00 
field, where the redshifts are from Prescott et al.\ (2006), 
and two sources in the GROTH~00 field, where the redshifts 
are from Finkelstein et al.\ (2009).  The redshifts for 
the Prescott et al.\ and Finkelstein et al.\ sources are 
enclosed in parentheses and annotated with a ``pr'' or ``f''.
Hereafter, we concentrate on the low-redshift LAE sample,
since the moderate-redshift LAE sample is so small.

The low-redshift sample may be expected to be relatively 
complete, but, as we have discussed, we expect it to have
$\sim20$\% AGN contamination.  For our LAE analysis
we do not eliminate any sources that we know to be 
AGNs from their optical spectra, since we do not have 
this information for a substantial part of the sample. 
However, we do note in the table where a source 
is clearly an AGN based on its optical spectrum, 
either because it has broad lines or based
on the Baldwin et al.\ (1981) diagnostic
diagram of [OIII]$\lambda5007$/H$\beta$ versus
[NII]$\lambda6584$/H$\alpha$. We denote these latter
sources with BPT AGN.
Finkelstein et al.\ (2009) classify the two sources 
in the GROTH~00 field where we have used their redshifts
as AGNs.  In one case this is based on broad lines, and 
in the other it is based on the Baldwin et al.\ (1981) diagnostic. 
They also classified
GALEX1417+5228 as an AGN, but we believe this source to 
be a very high-excitation star former.  We discuss this 
very interesting source  in Section~\ref{la_metal}.  

Considering only the sources
for which we have obtained optical spectroscopic data
and classifying sources as AGNs only if they show
AGN signatures in their optical spectra, 
we find the AGN contamination in our final LAE 
sample to be 4 out of 23 or $9-31$\%, where the range 
is $\pm1\sigma$.  
Finkelstein et al.\ (2009) give $17-61$\% for this range 
based on observations of a subsample of the 
Deharvang et al.\ (2008) 
sources in the GROTH~00 field.  The small sample sizes 
leave the exact value somewhat statistically uncertain, 
but it is important to note that the degree of 
remaining AGN contamination is a function of the 
NUV-continuum selection procedure 
and may vary with different samples, so it is not a
particularly interesting or physical quantity in itself.
The more interesting quantity is the total fraction 
of AGNs in all the NUV-continuum selected sources in 
a given redshift and magnitude range, including both
those identified in the UV and the additional objects 
identified with the optical spectra.  We shall 
return to analyze this further in 
A. Barger \& L. Cowie (2010, in preparation).

Our low-redshift LAE sample of 41 sources is substantially 
smaller than the sample of 96 sources used by 
Deharveng et al.\ (2008), 
despite the larger number of fields used here. This partly
reflects our smaller redshift interval ($z=0.195-0.44$), 
together with our geometric restriction ($R<32\farcm5$).
Only 67 of the Deharveng et al.\ (2008) sources lie within 
these bounds:  59 of these are contained in our initial sample
of LAEs, while the remaining 8 are marginal lines that fell out of 
our selection in both the automatic and the visual search.
Of the 59 sources, we classified 12 as AGNs based on the UV spectra 
(mostly on the basis of the line width rather than on
the presence of high-excitation lines), 
and of the remaining 47 sources, only 28 have 
EW(Ly$\alpha)>20$~\AA.  Thus, 28 of our low-redshift 
LAE sample 
overlaps with Deharveng et al.\ (2008)'s sample, and 
the remaining sources are either drawn from the other 
three fields not used in their analysis or are 
additional sources in the fields in common that were 
included in our sample but not in their sample.
All 28 overlapping sources agree in the {\em GALEX\/} 
redshift, but the Deharveng et al.\ (2008) Ly$\alpha$ 
luminosities are about 10\% higher than ours.
This may follow from the slight differences in methodology; 
in particular, we have calculated the fluxes directly from 
the spectra rather than recalibrating to the broadband 
fluxes. Thus, 10\% is probably a reasonable measure of the 
systematic uncertainty in the luminosities. 

We note 
that one of the sources in our moderate-redshift LAE sample
(GALEX1437+3541) is included in the Deharveng et al.\ (2008)
sample at a redshift of $z=0.468$. The present redshift 
of $z=0.664$ is based on a much stronger emission line 
seen in the NUV spectrum of this object. The line identified
in the FUV by Deharvang et al.\ is weak and appears to be an 
artifact.  (Note that Deharveng et al.\ only searched the FUV spectra 
for Ly$\alpha$ emission lines.)

\subsection{LAE Number Counts}
\label{lae_num}

\begin{figure}
\includegraphics[width=3.5in,angle=0,scale=1.]{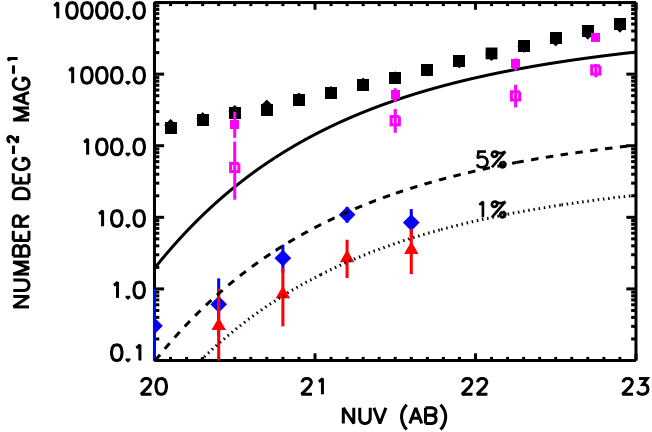}
\caption{Comparison of the number counts as a function of 
NUV magnitude of the $z=0.195-0.44$ LAEs from our 
{\em GALEX\/} fields with rest-frame EW(Ly$\alpha$)~$>20$~\AA\ 
(blue diamonds) and $>45$~\AA\ (red triangles) and
the number counts of all the NUV continuum sources 
in our {\em GALEX\/} fields (black squares)
and in the GOODS-N field (purple solid squares), 
as well as all the $z=0.195-0.44$ galaxies in the
GOODS-N field (purple open squares).
The $z=0.195-0.44$ galaxies 
constitute about 30\% of the sources at NUV~$>21$. 
The solid curve shows the number counts obtained
from the translated $z=0.2-0.4$ NUV-continuum
LF of Arnouts et al.\ (2005).  
It matches the $z=0.195-0.44$ GOODS-N points reasonably 
well. The dashed curve shows 5\% of 
the Arnout et al.\ (2005) counts, which 
matches the EW(Ly$\alpha$)~$>20$~\AA\ LAE counts, and the
dotted curve shows 1\% of the Arnouts et al.\ (2005) 
counts, which matches the EW(Ly$\alpha$)~$>45$~\AA\ LAE counts. 
\label{fract_ew}
}
\end{figure}

In Figure~\ref{fract_ew} we show the number counts 
per unit magnitude of the LAEs in the redshift
interval $z=0.195-0.44$ versus NUV magnitude. 
This is simply the sum of the inverse areas over all 
the sources in the magnitude and redshift interval
divided by the magnitude interval.  We have divided 
this into two rest-frame EW(Ly$\alpha$) intervals: 
$>20$~\AA\ (blue diamonds) 
and $>45$~\AA\ (red triangles) to show that the EW 
distribution does not change rapidly with NUV magnitude. 
We compare the number counts of the LAEs with the number 
counts of all the NUV continuum sources in our {\em GALEX\/} 
fields (black squares) and in the GOODS-N field 
(purple solid squares).  These points include stars, 
galaxies, and AGNs.  The purple open squares show the 
number counts of all the $z=0.195-0.44$ galaxies in the 
GOODS-N sample using the Barger at al.\ (2008) 
redshift information and after removing all sources
classified as AGNs based on having X-ray luminosities
$>10^{42}$~erg~s$^{-1}$.

Arnouts et al.\ (2005) produced a $z=0.2-0.4$ 
rest-frame $1500$~\AA\ luminosity function (LF) from 
{\em GALEX\/} VIMOS-VLT Deep Survey data, which
we translated to number counts in the same redshift 
interval versus what corresponds to approximately
observed-frame 1900~\AA. We then applied a small 
$-0.2$~mag differential $K-$correction to move 
that to an NUV magnitude.  The result is the solid 
curve, which agrees reasonably well with the 
$z=0.195-0.44$ GOODS-N points.

Both the EW(Ly$\alpha$)~$>20$~\AA\ (blue diamonds) and 
EW(Ly$\alpha$)~$>45$~\AA\ (red triangles) LAE number counts 
show a substantial rise to NUV~$=21$ where, as we have 
discussed previously (e.g., Figure~\ref{redshift-dist}(a)), 
the LAEs appear to onset in substantial numbers in the
redshift interval $z=0.195-0.44$.
As can be seen from Figure~\ref{fract_ew}, this trend mirrors 
the overall number counts in the redshift interval $z=0.195-0.44$, 
which enter the population in significant numbers at magnitudes
fainter than 21. We have scaled the Arnouts et al.\ (2005) 
curve by 0.05 (dashed) and 0.01 (dotted) to show the fraction 
of LAEs in a particular EW(Ly$\alpha$) range.
Thus, $\sim5$\% of the NUV-continuum selected 
galaxies in this low-redshift interval are LAEs. 
The curves provide a reasonable approximation to the shape
of the LAE counts in both EW(Ly$\alpha$) ranges,
showing that the EW(Ly$\alpha$) distribution is not changing
rapidly over the observed magnitude range.

\subsection{LAE Luminosity Function}
\label{lae_lumfun_sect}

We may quantify this further by computing the LAE LF.
As Deharveng et al.\ (2008) stress in their derivation
of the LAE LF, the procedure is complicated
because we are using a NUV-continuum selected sample
and the line flux limits depend on the EW(Ly$\alpha$) 
distribution.  This issue is somewhat alleviated by 
the relatively invariant rest-frame UV colors above the 
Lyman continuum break (see the low-redshift sources in
Figure~\ref{galaxy_color}) so that the NUV
limiting magnitude corresponds to an approximate
Ly$\alpha$ flux for a given observed EW(Ly$\alpha$).
In principle, by choosing a high enough flux limit in each field
we could construct a roughly complete flux-limited sample. 
This would require us to set the flux limit to correspond
to a source with the maximum observed EW(Ly$\alpha$)
(about 100~\AA; e.g., Figure~\ref{ew_dist}) at the magnitude
limit of the particular field. 
However, in practice, the present sample is too small to 
allow such a procedure. An alternative procedure, following 
Deharveng et al.\ (2008), is to include all the sources above 
a lower flux limit than this. We can then attempt to correct
for the incompleteness by assuming the EW(Ly$\alpha$)
distribution of Figure~\ref{ew_dist} is invariant as a 
function of NUV magnitude.
This allows us to correct for the missing high-EW(Ly$\alpha$) 
sources with NUV magnitudes fainter than the magnitude
limit of the field. This assumption of invariance in the 
EW(Ly$\alpha$) distribution may well fail as we move to 
fainter magnitudes, where we may see higher EW(Ly$\alpha$) 
sources, so it is important to minimize the extrapolation.
We use the latter procedure here.

We first set the Ly$\alpha$ flux limits high 
enough---corresponding
to a rest-frame EW(Ly$\alpha)$~$>45$~\AA\ at the limiting 
NUV magnitude of each field---both to include a significant 
fraction of the sources in Figure~\ref{ew_dist} and to 
minimize the incompleteness corrections.  The downside of 
this high flux cut is that we reduce the already small 
sample significantly and are restricted to high Ly$\alpha$ 
luminosities.  However, the incompleteness corrections are 
small.  We also try a low flux cut with a flux limit 
corresponding to a rest-frame EW(Ly$\alpha$)~$>25$~\AA\ 
at the limiting NUV magnitude of each field, which allows 
us to probe to lower luminosities at the expense of a 
higher incompleteness correction.

Specifically, for each field we determined a Ly$\alpha$ flux 
limit corresponding to a source with the limiting NUV magnitude 
of Table~\ref{tab1} corrected to an FUV magnitude by adding
an offset of 0.37~mag (see Figure~\ref{galaxy_color}) and with 
an observed EW(Ly$\alpha$) corresponding to the chosen 
rest-frame limit
placed at the center of the redshift interval. For a 
rest-frame EW(Ly$\alpha$) of 45~\AA, the adopted flux limit is 
$f({\rm Ly}\alpha) = 3.4\times10^{-15}$~erg~cm$^{-2}$~s$^{-1}$ 
for the GROTH~00 field; it is proportionally higher for the 
remaining shallower fields. Only galaxies with fluxes above 
each field's adopted flux limit were included in the final 
sample.  The observed area at a given flux is then the sum 
of all the field areas where the limiting flux is lower
than this flux.

\begin{figure}
\hskip -0.3cm
\includegraphics[width=3.6in,angle=0,scale=1.]{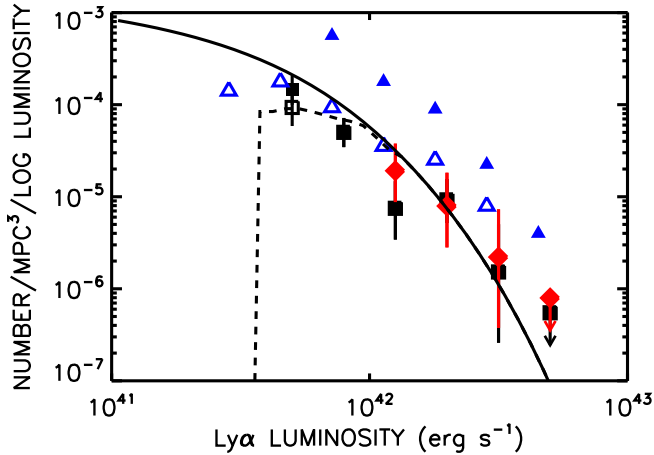}
\caption{LAE (rest-frame EW(Ly$\alpha$)~$>20$~\AA) LF 
in the redshift interval $z=0.195-0.44$ using two flux 
limits (see text). 
The red diamonds show the LF using the high flux cut 
where incompleteness corrections are minimal, and the
black squares show the LF using the low flux cut 
that extends to lower luminosities, but where the 
incompleteness corrections are larger. In each case 
the raw (incompleteness-corrected) LAE LF is shown 
with open (solid) symbols.
The error bars are $\pm1\sigma$ based on the
Poisson errors for the number of sources in each
luminosity bin. The solid curve shows the
LAE LF obtained by convolving the EW(Ly$\alpha$) 
distribution of Figure~\ref{ew_dist} with the 
translated $z=0.2-0.4$ Arnouts et al.\ (2005) 
NUV-continuum LF.  The dashed curve 
shows the same but using the EW(Ly$\alpha$) distribution 
of the NUV~$<21.8$ galaxies only, which should roughly 
match to the raw LAE LFs.  The blue triangles show 
Deharveng et al.\ (2008)'s determination of the LAE 
LF before (open) and after (solid) their incompleteness 
correction.  We have reduced Deharveng et al.\ (2008)'s 
luminosity scale by 10\% to make the Ly$\alpha$ luminosity 
measurements consistent (see Section~\ref{lae_prop}).
\label{lae_lumfun}
}
\end{figure}

We next constructed the LAE LF in the redshift interval 
$z=0.195-0.44$ using the $1/V$ technique (Felten 1976). 
In Figure~\ref{lae_lumfun} we use red open diamonds (black 
open squares) to show the LF for the high flux (low flux) cut. 
The $\pm1\sigma$ errors are based on the Poisson errors 
corresponding to the number of sources in each bin. 
Since the line width split that we used to separate
out AGNs is somewhat subjective, as a check 
we have also calculated the LF including all the sources 
in the redshift interval that do not have high-excitation lines. 
This slightly increases the LF but by an amount which is 
small compared to the uncertainties.

These are the raw LFs, without the incompleteness corrections 
for the missing high-EW(Ly$\alpha$) sources in the flux-limited 
samples.  In order to compute the incompleteness corrections, 
we used the form of the Arnouts et al.\ (2005) LF to obtain the
expected number of continuum sources in the redshift 
interval at fainter UV magnitudes. We then
drew the correct number of sources from the EW(Ly$\alpha$)
distribution of Figure~\ref{ew_dist} to simulate the 
missing high-EW(Ly$\alpha$) sources and recomputed the LFs.
The incompleteness-corrected LFs are
shown with the solid symbols in Figure~\ref{lae_lumfun}. In
general these corrections are small, except in the faintest
bin of the low flux cut sample (black squares).

Given our assumption of an invariant EW(Ly$\alpha$) 
distribution, it is also possible to simply convolve 
this distribution with the observed NUV-continuum LF in the 
redshift interval to determine the LAE LF.  This allows us
to construct a LF to fainter luminosities but at the expense
of assuming the same EW(Ly$\alpha$) distribution applies at 
substantially fainter NUV magnitudes than where it was measured. 
As we have stressed above, the invariant EW(Ly$\alpha$) 
distribution assumption may not be valid if fainter sources 
have different Ly$\alpha$ emission-line properties. 
We did the calculation using the EW(Ly$\alpha$) distribution of 
Figure~\ref{ew_dist} and the translated $z=0.2-0.4$
Arnouts et al.\ (2005) NUV-continuum LF.  This result is 
shown with the solid curve in Figure~\ref{lae_lumfun}.
We repeated the calculation using the EW(Ly$\alpha$) 
distribution of the NUV~$<21.8$ galaxies only 
to illustrate the luminosity range corresponding to the 
actual measurements.  This result is shown with the dashed 
curve.  The incompleteness corrections can then be measured 
from the ratio of the two curves (this more closely mirrors 
the procedure used by Deharveng et al.\ 2008), and they 
agree well with what we found previously.

All of our LAE LF measurements are comparable to the raw 
LAE LF determined by Deharveng et al.\ (2008; blue open 
triangles in Figure~\ref{lae_lumfun}). However,
we do not find the large incompleteness corrections
that they found (the blue solid triangles show their 
incompleteness-corrected LAE LF). Their corrections appear 
remarkably large, particularly at the high-luminosity end.
There may be differences reflecting the selection in
the EW(Ly$\alpha$) (here we are using a rigid rest-frame 
EW(Ly$\alpha$)~$>20$~\AA\ definition of the LAEs) and 
the more stringent exclusion of potential AGNs in the present 
analysis, but it does not appear that these can account 
for the differences. The problem may be caused by a
missing color correction in the Deharvang analysis
(J.-M.~Deharvang, priv. comm.).

\begin{figure}
\hskip -0.6cm
\includegraphics[width=3.7in,angle=0,scale=1.]{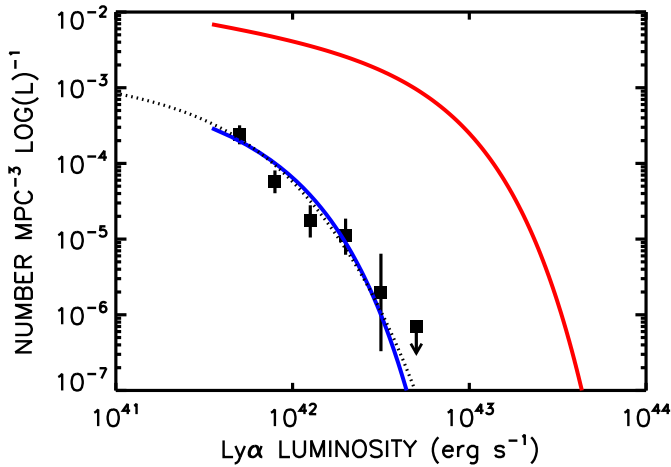}
\caption{
Comparison of our $z=0.195-0.44$ low flux cut 
incompleteness-corrected LAE LF (black squares) 
with the $z=3.1$ LAE LF of Gronwall et al.\ (2007; red curve).
Other determinations of the $z\sim3$ LAE LF
are extremely similar (see text). 
The blue curve shows the maximum likelihood Schechter 
(1976) function
fit to our data for a fixed slope of $\alpha=-1.36$ 
(that of Gronwall et al.\ 2007). There is a drop of 
$\sim8$ in $L_\star$ and $\sim6.5$ in the normalization 
$\phi_\star$ relative to the $z=3.1$ LAE LF. 
The dotted black curve shows the LAE LF obtained by 
convolving the EW(Ly$\alpha$) distribution of Figure~\ref{ew_dist} 
with the translated $z=0.2-0.4$ Arnouts et al.\ (2005) 
NUV-continuum LF.  
\label{lae_lumfun_comparison}
}
\end{figure}

In Figure~\ref{lae_lumfun_comparison} we compare our 
$z=0.195-0.44$ LAE LF (here we adopt the low flux cut
incompleteness-corrected result; black squares), which was 
chosen with the same rest-frame EW(Ly$\alpha$) 
selection (i.e., $>20$~\AA)
as the high-redshift samples, with the $z=3.1$ LAE LF 
of Gronwall et al.\ (2007; red curve). Other determinations 
of the $z\sim3$ LAE LF are extremely similar 
(e.g., Cowie \& Hu 1998; van Breukelen et al.\ 2005; 
Ouchi et al.\ 2008).  We have made a maximum likelihood 
analysis of our data to obtain a Schechter (1976) function 
fit.  Given the limited dynamic range
of our data, we have held the slope $\alpha$ fixed
and only measured $L_\star$. For the Gronwall et al.\ (2007)
slope of $\alpha=-1.36$, we obtain $\log L_\star=41.76\pm0.08$.
Normalizing to the observed number of sources gives 
$\phi_\star = (1.95\pm0.35)\times10^{-4}$~Mpc$^{-3}$.
The errors are $\pm1\sigma$. This fit is shown as the 
blue curve in Figure~\ref{lae_lumfun_comparison}.
For the van Breukelen et al.\ (2005) slope of
$\alpha=-1.6$, we obtain $\log L_\star =41.81\pm0.09$.
Normalizing to the observed number of sources gives
$\phi_\star = (1.68\pm0.30)\times10^{-4}$~Mpc$^{-3}$.

Using the Gronwall et al.\ (2007) slope and comparing
to the Gronwall et al.\ (2007) LF gives a drop
of $\sim8$ in $L_\star$ and $\sim6.5$ in $\phi_\star$
for a drop in the Ly$\alpha$ luminosity density of $\sim50$. 
Using the van Breuken  et al.\ (2005) slope and comparing 
to the van Breukelen et al.\ (2005) LF gives a drop
in the Ly$\alpha$ luminosity density of 55. 

As Deharveng et al.\ (2008) have pointed out, this 
substantial drop in the LF from higher redshift values
is considerably in excess of the corresponding 
continuum UV light density drop.  Thus, it appears that 
LAEs are far less common now than they were in the past 
and that they have lower luminosities.

\subsection{LAE Equivalent Widths}
\label{la_ew_dist_sect}

\begin{figure}
\hskip -0.6cm
\includegraphics[width=3.7in,angle=0,scale=1.]{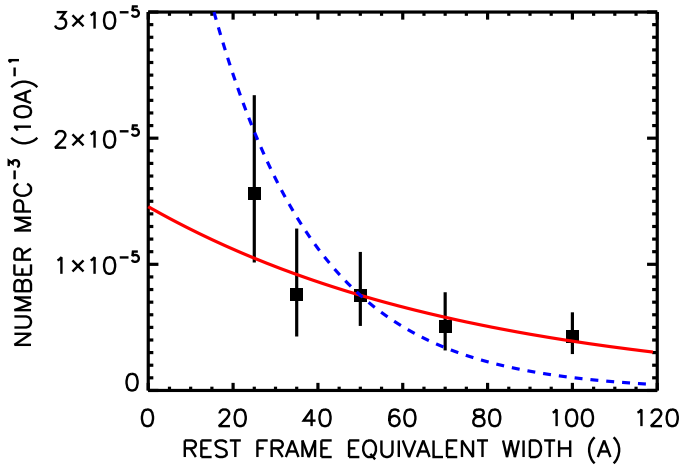}
\caption{Number density of sources above 
$\log L$(Ly$\alpha)=41.6$~erg~s$^{-1}$
vs. rest-frame EW(Ly$\alpha$). The number
densities correspond to a 10~\AA\ bin size.
The red solid curve shows an exponential fit to the data 
with a 75~\AA\ scale length equal to that in the $z=3.1$ 
population, which provides a good description of the data. 
The blue dashed curve shows the steeper fit to the continuum
selected data of Figure~3. 
\label{ew_dist_plot}
}
\end{figure}

In order to compare the LAE EW(Ly$\alpha$) 
distributions at low and high redshifts, we need to
translate our measured EW(Ly$\alpha$) distribution as a 
function of NUV magnitude (Figure~\ref{ew_dist}) into one 
which is a function of LAE luminosity. To do this we 
computed the number density of LAEs in the redshift 
interval $z=0.195-0.44$ as a function of their rest-frame 
EW(Ly$\alpha$) using the same methodology that we used to 
compute the incompleteness-corrected LAE LFs.  
We used a low flux cut corresponding to a rest-frame 
EW(Ly$\alpha$) of 25~\AA\ to probe to low luminosities 
and computed the number density of sources above a 
limiting luminosity of $4\times10^{41}$~erg~s$^{-1}$ by 
summing the inverse volumes of all the sources in the 
EW(Ly$\alpha$) interval lying above this luminosity. 
We then divided by the width of the EW(Ly$\alpha$) bin. 
The results are shown in Figure~\ref{ew_dist_plot}.

The low-redshift LAE EW(Ly$\alpha$) distribution 
is well described by the same exponential
with a scale length of 75~\AA\ that provides a good fit 
to the $z\sim3$ LAE EW(Ly$\alpha$) distribution 
(Gronwall et al.\ 2007).  This is shown by the red 
solid curve in Figure~\ref{ew_dist_plot}.  Thus, the 
form of the EW(Ly$\alpha$) distribution for the LAEs is not 
changing with redshift, even though the number of sources
satisfying the LAE criterion is much smaller at low redshifts.
We do not see 
any extreme EWs(Ly$\alpha$) (greater than 120~\AA) 
in the present sample, but this may be a simple consequence 
of the continuum selection, which is biased against finding 
such objects.

\subsection{Ly$\alpha$ Velocities Relative to H$\alpha$}
\label{la_ha_prop}

A comparison of the redshifts of the Ly$\alpha$ lines 
relative to the redshifts of the H$\alpha$ lines is of 
considerable interest since it may relate to the kinematical 
structure of the galaxy and the escape process of the Ly$\alpha$ 
photons.  However, this comparison is difficult because of the 
wavelength calibration uncertainties in the {\em GALEX\/} grism 
data.  Morrissey et al.\ (2007) give
a calibration error for the {\em GALEX\/} wavelengths of 
about 3\% in the body of each spectral order and about 10\% 
near the edges of each order. 

\begin{figure}
\hskip -0.6cm
\includegraphics[width=3.7in,angle=0,scale=1.]{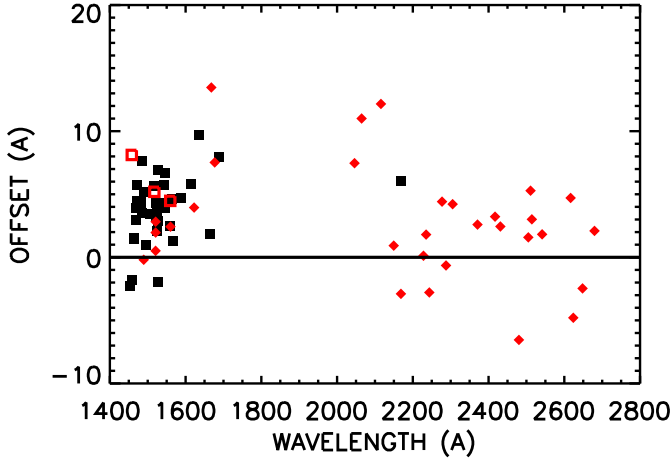}
\caption{The offset of the observed Ly$\alpha$ wavelength
in the {\em GALEX\/} spectra from that expected from
the optical redshift vs. wavelength. 
Black solid squares show the
LAEs, red solid diamonds show AGNs with high-excitation lines 
in the UV spectra, and red open squares show sources without 
high-excitation lines but with line widths greater than 15~\AA.
The optical redshifts come from our observations and from
NED.
\label{wave_scale}
}
\end{figure}

In Figure~\ref{wave_scale} we show the wavelength offsets 
between the Ly$\alpha$ wavelength that would be measured 
from {\em GALEX\/} and the Ly$\alpha$ wavelength 
that would be inferred from the optical redshifts. 
For this figure we have augmented the {\em GALEX\/} LAE 
sample (black solid squares; this includes any source 
where we have measured the redshift from our optical 
data, as well as any source where the optical redshift could 
be obtained from the NED database\footnote{The NASA/IPAC 
Extragalactic Database (NED) is operated by the Jet Propulsion 
Laboratory, California Institute of Technology, under contract 
with the National Aeronautics and Space Administration})
in order to obtain a larger number 
of sources to maximize our understanding of the effect.  
The additional sources are AGNs (red solid diamonds are 
based on the presence of high-excitation lines; red 
open squares are based on the line widths)
in our {\em GALEX\/} fields with optical redshifts, 
again either from our own observations or from NED.

Nearly all of the sources show a {\em GALEX\/} Ly$\alpha$ 
wavelength that is redder than would be expected from the 
optical redshift. The median offset is 3.9~\AA\ for the 
sources (both AGNs and galaxies) in the redshift
interval $z=0.195-0.44$.  Given that both the AGNs and the 
LAEs show this offset, it is probable that much of it is indeed 
due to the absolute calibration uncertainty.
The median wavelength offset is 3.9~\AA\ for just the LAEs and 
3.8~\AA\ for just the AGNs.   
The median redshifting seen in $z\sim2$ galaxies, which is around 
550~km~s$^{-1}$ (Pettini et al.\ 2001), would produce an offset 
of about 2.2~\AA, which is well within the uncertainties.


\section{Properties of the Low-Redshift Ly$\alpha$ Galaxy Sample}
\label{lae_lbg}

\subsection{Ly$\alpha$ Versus H$\alpha$ Fluxes}
\label{la_ha_fluxes}

Focusing on our optical spectroscopic observations of 
the {\em GALEX\/} spectroscopic sample, we now turn 
to analyzing the properties of the optically-confirmed 
Ly$\alpha$ Galaxies in the redshift interval $z=0.195-0.44$ 
(Table~\ref{tab11}, excluding the sources classified 
as AGNs). We compare them to the 
properties of both the {\em GALEX\/} spectroscopic sample
without UV spectral identifications that are optically 
classified as galaxies and lie in the same 
redshift interval (Table~\ref{tab13}; hereafter, we refer to this
as the optically-confirmed NUV-continuum selected galaxy 
sample) and the essentially completely spectroscopically 
identified GOODS-N NUV~$<24$ galaxy sample in the redshift 
interval $z=0.15-0.48$ (hereafter, we refer to this as the 
GOODS-N NUV-continuum selected galaxy sample). In the
latter sample we have removed all the sources classified 
as AGNs based on having X-ray luminosities 
$>10^{42}$~erg~s$^{-1}$. There are no
sources in the GOODS-N sample that appear in the GALEX Ly$\alpha$ Galaxy 
sample.  However, we caution that there may still be some 
Ly$\alpha$ emission-line galaxies in the GOODS-N sample, since the
{\em GALEX\/} grism data for this field do not go as 
faint in NUV as the optical spectroscopy.

\begin{figure}
\hskip -0.6cm
\includegraphics[width=3.7in,angle=0,scale=1.]{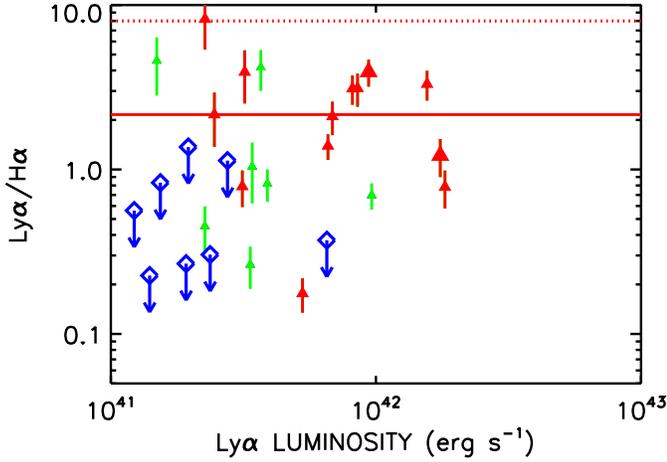}
\caption{
Ly$\alpha$/H$\alpha$ flux ratio vs. Ly$\alpha$ luminosity
for the {\em GALEX\/} sources that appear to be star formers 
based on their UV and optical spectra, lie in the redshift 
interval $z=0.195-0.44$, and have SDSS magnitudes.
The red small solid triangles show the measured ratios for the 
optically-confirmed Ly$\alpha$ Galaxies with rest-frame
EW(Ly$\alpha)>20$~\AA.  
The red solid line shows the median ratio of 2.1 for this sample. 
The red large solid triangles show the sources with EW(Ly$\alpha)>40$~\AA,
and the green very small solid triangles show the sources with detected 
Ly$\alpha$ and EW(Ly$\alpha)\le 20$~\AA.
All error bars are $\pm 1\sigma$.
The red dotted line marks the lower limit of the ratios 
($8-12$) expected for case~B (Ferland \& Osterbrock 1985). 
The blue open diamonds with the downward pointing arrows denote
the upper limits on the Ly$\alpha$/H$\alpha$ flux ratios for the
optically-confirmed NUV-continuum selected galaxies.  
Here we have assumed an observed-frame EW(Ly$\alpha)~=10$~\AA\ 
in calculating the upper limits for the flux ratios
and the Ly$\alpha$ luminosities.
\label{lumra-ha}
}
\end{figure}

In Figure~\ref{lumra-ha} we show the ratio of 
the Ly$\alpha$ flux to the H$\alpha$ flux versus Ly$\alpha$
luminosity.  The optically-confirmed Ly$\alpha$
Galaxies with rest-frame EW(Ly$\alpha)>20$~\AA\ 
(red solid triangles) mostly lie in a fairly narrow region  
from just below 1 to a maximum ratio of 8. 
The median value of $2.2 \pm 1.0$ is shown by the red solid line,
while the average ratio is 2.6.
This value is about 4 times smaller than the case~B 
ratio (which, depending on the electron density, is $8-12$; 
Ferland \& Osterbrock 1985) often used to translate
Ly$\alpha$ luminosity to star formation rate, and
it is consistent with the similar reduction inferred 
by comparing star formation
rates measured from the UV continuum with those measured 
from Ly$\alpha$ in the $z\sim3$ emitters (Gronwall et al.\ 2007). 
It is also consistent with the range of values seen in 
other optical spectroscopic follow-up observations of the 
Deharvang et al.\ (2008) sample (Atek et al.\ 2009; 
Mallery 2009).  The galaxies with detected Ly$\alpha$
lines but rest-frame EW(Ly$\alpha$) weaker than 20~\AA\
are shown with green very small solid triangles.
These generally have lower ratios of Ly$\alpha$/H$\alpha$, with
a median value of 1.03, though the number of objects is
too small to derive a median error. There appears to
be no obvious dependence on other parameters such
as metallicity though a large sample is clearly
required to explore this in depth.

If we use Ly$\alpha$/H$\alpha=2.6$ and 
adopt the widely used Kennicutt (1998) 
conversion of the H$\alpha$ luminosity to star formation 
rate for the Salpeter (1955) initial mass function (IMF) 
extending to 0.1~M$_\sun$, we obtain a conversion of the 
Ly$\alpha$ luminosity to star formation rate of
\begin{equation}
\log {\rm SFR} = -40.67 + \log L({\rm Ly}\alpha) \,.
\label{eqlaSFR}
\end{equation}
The true value for the H$\alpha$ calibration depends on
the time history of the star formation. The value
derived in Cowie \& Barger (2008), which is appropriate
for the average of galaxies at these redshifts, would 
reduce this by 0.2~dex to $-40.47 + \log L({\rm Ly}\alpha$).

While the above equation can be used to get the star
formation rates for Ly$\alpha$ galaxies selected as
in this sample---a selection which corresponds fairly 
closely to the formal LAE definition---it must be noted 
that the observed values of Ly$\alpha$/H$\alpha$ in this 
sample lie above a threshold set by the EW(Ly$\alpha$) 
selection (see Figure~\ref{ew_dist}).
In other words, if the EW(Ly$\alpha$) is too small,
we will not see a Ly$\alpha$ line.  Thus, the narrow
range of observed values in Figure~\ref{lumra-ha}
is merely a selection bias, as we illustrate by putting
upper bounds on the optically-confirmed NUV-continuum 
selected galaxies (blue diamonds).  These stretch up to 
overlap the lower boundary of the optically-confirmed 
Ly$\alpha$ Galaxy sample.

The reason for the selection bias is as follows.  
A given UV continuum luminosity corresponds to an approximate
H$\alpha$ luminosity in the absence of continuum extinction,
and for a given EW(Ly$\alpha$), it also specifies 
the Ly$\alpha$ luminosity.  
For this sample we can use the empirical star formation 
rate calibrations given in Cowie \& Barger (2008),
\begin{eqnarray}
\log {\rm SFR} &=& -42.63 + \log L({\rm FUV}) \,, \\
\log {\rm SFR }&=& -40.90 + \log L({\rm H}\alpha) \,,
\end{eqnarray}
and the conversion from FUV fluxes to NUV fluxes 
based on Figure~\ref{galaxy_color} (NUV=FUV$-0.37$~mag)
to roughly obtain 
a relation between the NUV continuum luminosity and 
the H$\alpha$ luminosity.
Then we use the NUV continuum 
flux and the EW(Ly$\alpha$) to determine the
Ly$\alpha$ luminosity and obtain the relation
\begin{equation}
L({\rm Ly}\alpha)/L({\rm H}\alpha)=2.4 ({\rm EW(Ly}\alpha)/20~{\rm \AA}) \,,
\end{equation}
where EW(Ly$\alpha$) is the rest-frame EW(Ly$\alpha$). Thus,
the effect of the EW(Ly$\alpha$) selection is to place
a lower bound on $L$(Ly$\alpha)/L$(H$\alpha$), while
case~B places an upper bound.  The exact lower bound 
does depend on the details of the star formation history, 
which determines the exact UV continuum to H$\alpha$ 
conversion, and on the extinction, which can lower 
the $L$(Ly$\alpha)/L$(H$\alpha$) ratios.  However, 
the lower limit bias will still be present.
Thus, we cannot simply estimate general escape fractions from
Ly$\alpha$ galaxy samples alone, because we also need to 
deal with all of the Ly$\alpha$ undetected sources, which
still produce some Ly$\alpha$ emission.

\begin{figure}
\includegraphics[width=3.5in,angle=0,scale=1.]{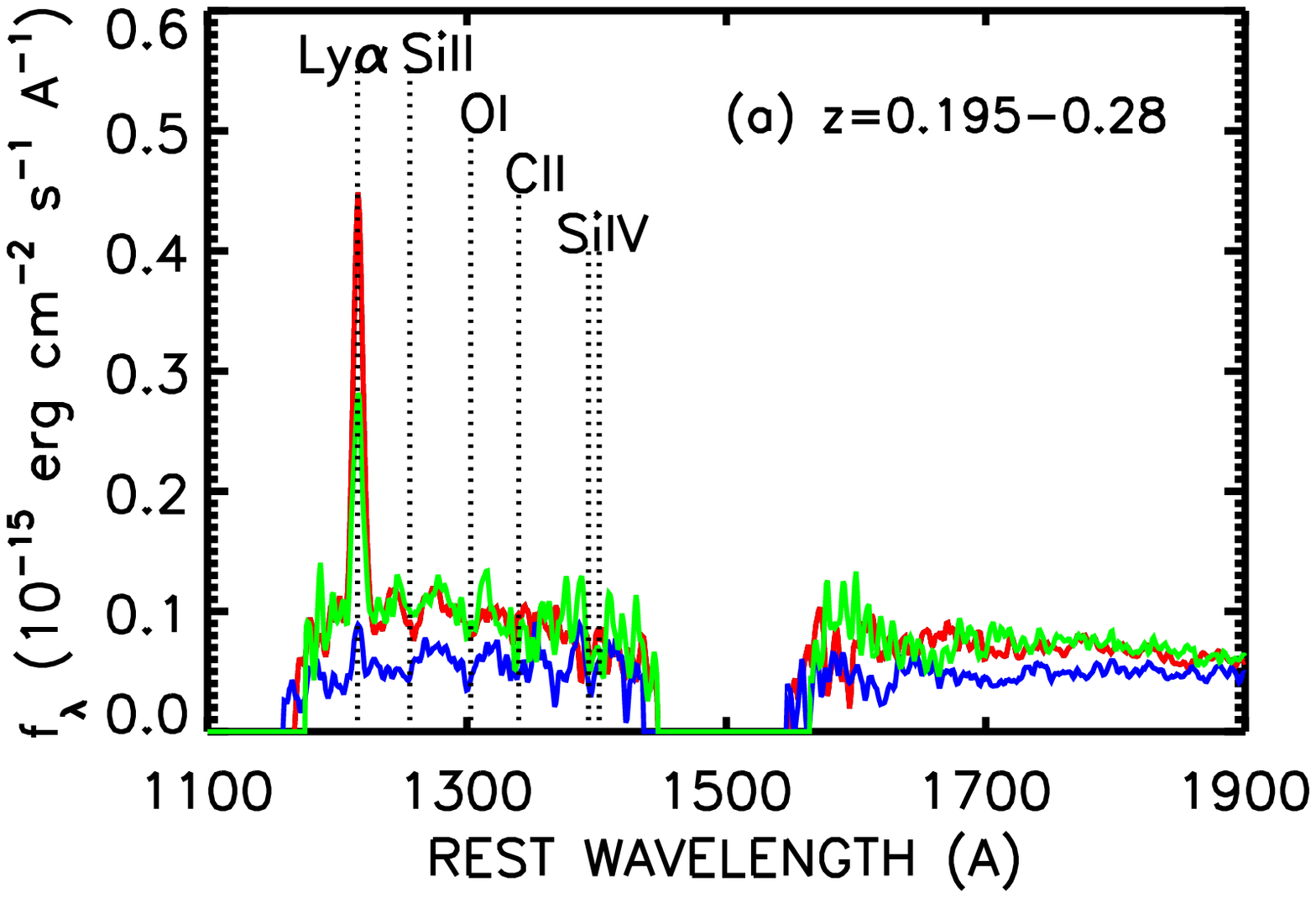}
\includegraphics[width=3.5in,angle=0,scale=1.]{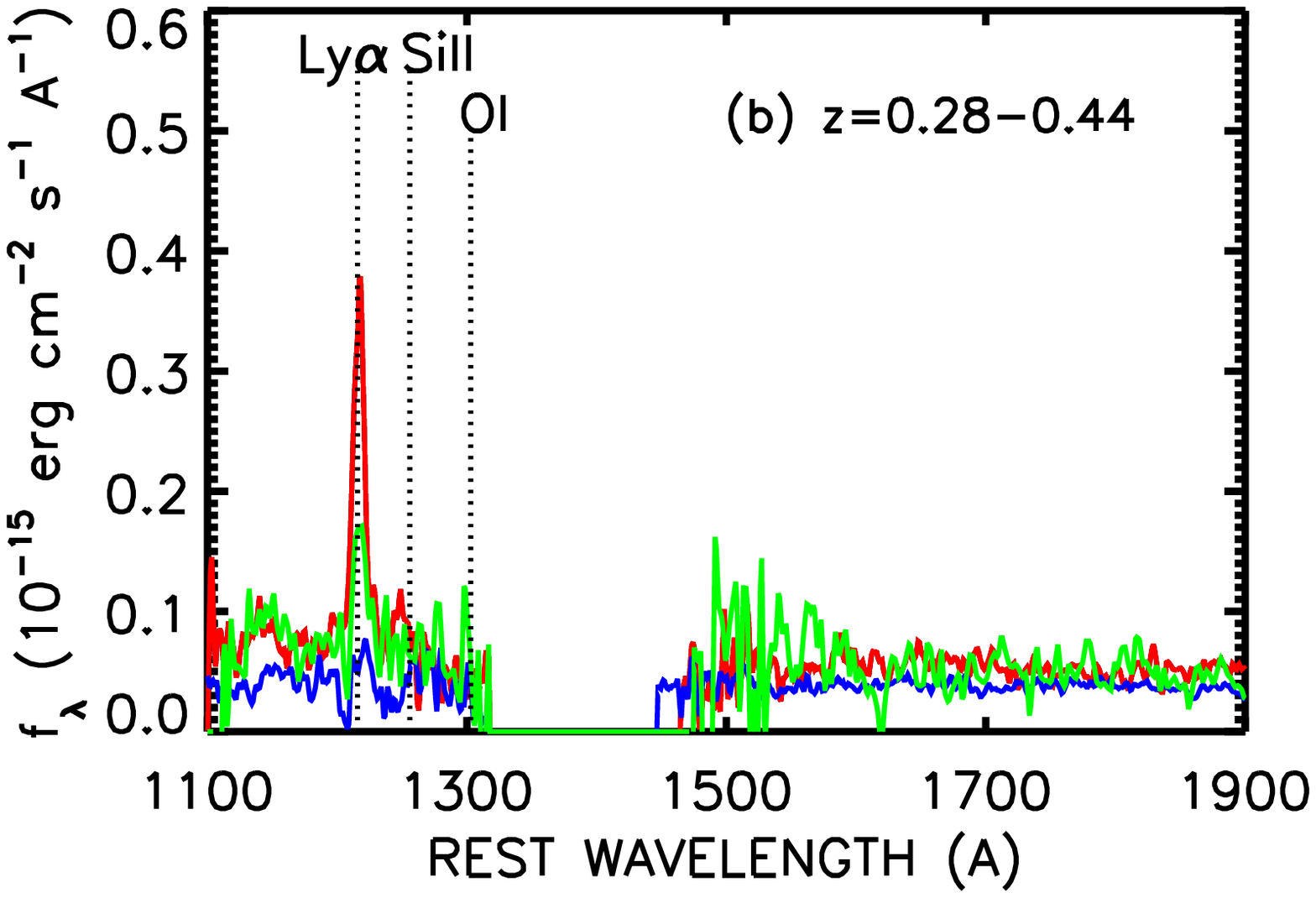}
\caption{Average {\em GALEX\/} spectrum for the objects
with rest-frame EW(Ly$\alpha)>20$~\AA\ 
(red spectrum), detected Ly$\alpha$ with rest-frame
EW(Ly$\alpha)\le20$~\AA\ (green spectrum),
and NUV-continuum selected objects (blue spectrum).
The vertical dotted lines show the strong features in the
spectra. In (a) we show the redshift interval $0.195-0.28$,
and in (b) we show the redshift interval $0.28-0.44$.
\label{stacked_spectra}
}
\end{figure}

With larger samples we could begin to determine the 
incompleteness corrections as a function of EW. 
Here we follow a simpler route of using a stacking analysis
to determine the $L$(Ly$\alpha)/L$(H$\alpha$) ratio in the
sources from the NUV-continuum sample that lie in the
redshift range but do not have identified Ly$\alpha$.
We can then estimate the fraction of Ly$\alpha$ photons 
that escape from NUV-continuum selected galaxies at a
particular redshift using the ratio of the total 
Ly$\alpha$ light to the total H$\alpha$ light
and combine this with the measured values in the
objects with identified Ly$\alpha$ to form an average
appropriate for the total galaxy sample in the redshift range.

To do this, we constructed average rest-frame galaxy spectra.
These are shown in Figure~\ref{stacked_spectra}
for three samples: rest-frame EW(Ly$\alpha)>20$~\AA\ 
(red spectrum), detected Ly$\alpha$ galaxies
with EW(Ly$\alpha)\le20$~\AA\ (green spectrum),
and NUV-continuum selected objects (blue spectrum).
The averaged spectra are constructed using the optically measured
redshifts with an average offset determined from Figure~\ref{wave_scale}
to match the {\em GALEX\/} wavelength calibration. The strong
metal absorption lines produce weak features in the 
low-resolution spectra, which are marked on the figure with
vertical dotted lines. To
simplify the averaging procedure, we have divided the
redshift range into two intervals: (a) $z=0.195-0.28$
and (b) $z=0.28-0.44$. This split has the secondary
advantage of allowing us to see that the results are robust
and reproducible with different samples.

We can see from Figure~\ref{stacked_spectra} that the
NUV-continuum selected galaxies do have weak Ly$\alpha$
emission.  We measured the Ly$\alpha$ flux
and compared it to the average H$\alpha$ flux of
the same sample to determine 
an average Ly$\alpha$/H$\alpha$ ratio for 
these sources. For the full $0.195-0.44$ range
we find a value of 0.23, but both redshift ranges
shown in Figure~\ref{stacked_spectra} give similar
values.  However, for these weaker sources
there may be a substantial correction for the presence
of Ly$\alpha$ absorption in some of the galaxies. A maximum
estimate for this can be obtained by measuring the line
flux relative to the zero level rather than relative to
the continuum. This gives a maximum Ly$\alpha$/H$\alpha$ ratio
for the sources of 0.54. 

If we make a weighted addition of the 5\% of 
LAEs with rest-frame EW(Ly$\alpha)>20$~\AA\
and Ly$\alpha$/H$\alpha$ ratios of 2.6, the 2\% of LAEs with
EW(Ly$\alpha)\le20$~\AA\ and Ly$\alpha$/H$\alpha$
ratios of 1.0, and the 93\% of 
non-LAEs with Ly$\alpha$/H$\alpha$ ratios of 0.23 or 0.54, we 
obtain an average Ly$\alpha$/H$\alpha$ ratio of $0.36-0.65$
at this redshift. Comparing this with the case~B ratio, 
we find that, on average, about $3-8$\% of the Ly$\alpha$ 
photons are escaping from the entire
galaxy population at $z=0.3$. A minimum of a 
quarter of these photons are emerging from the small 
fraction of the identified LAEs alone, and the fraction could be
as high as 40\%.

\begin{figure}
\includegraphics[width=3.5in,angle=0,scale=1.]{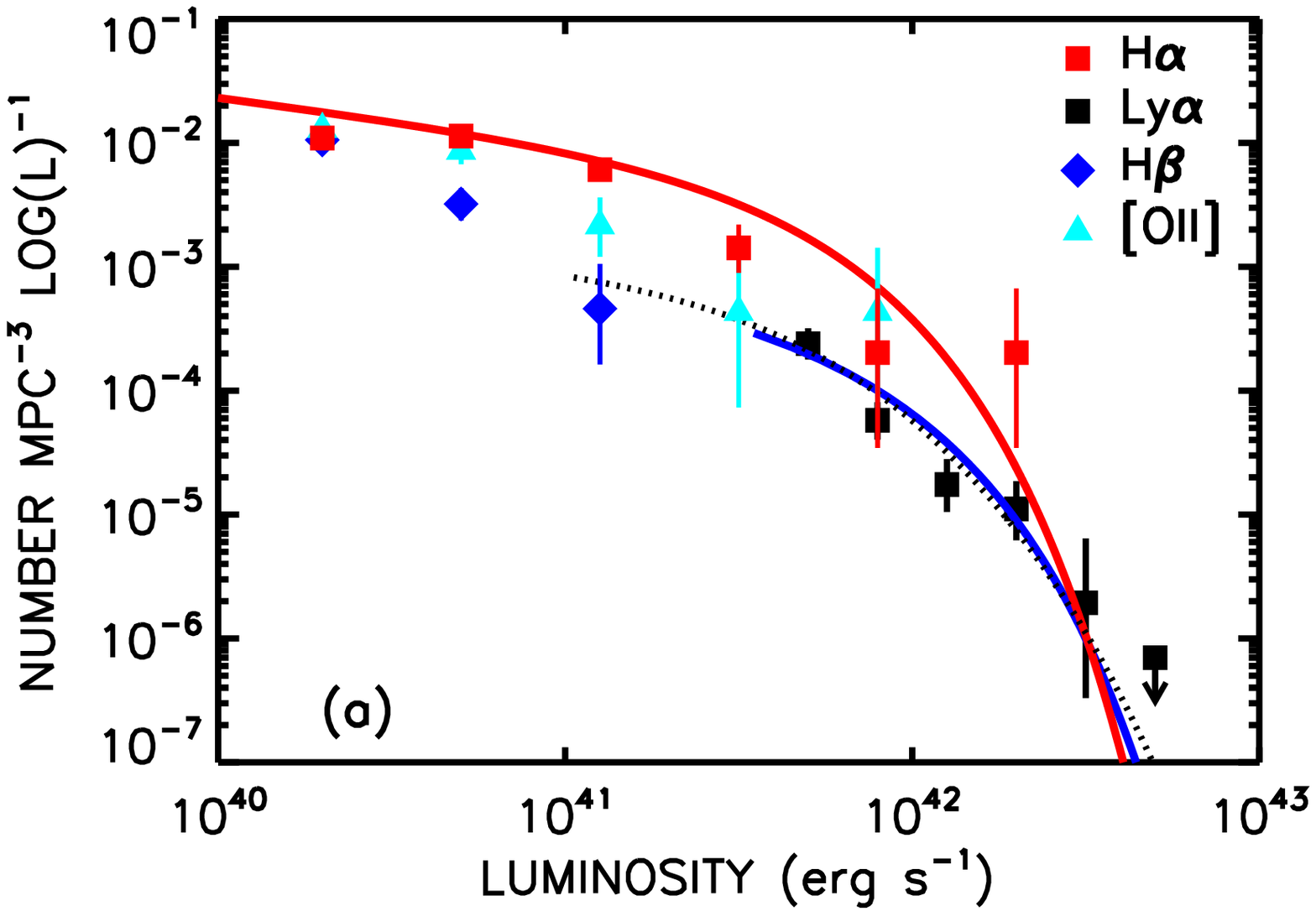}
\includegraphics[width=3.5in,angle=0,scale=1.]{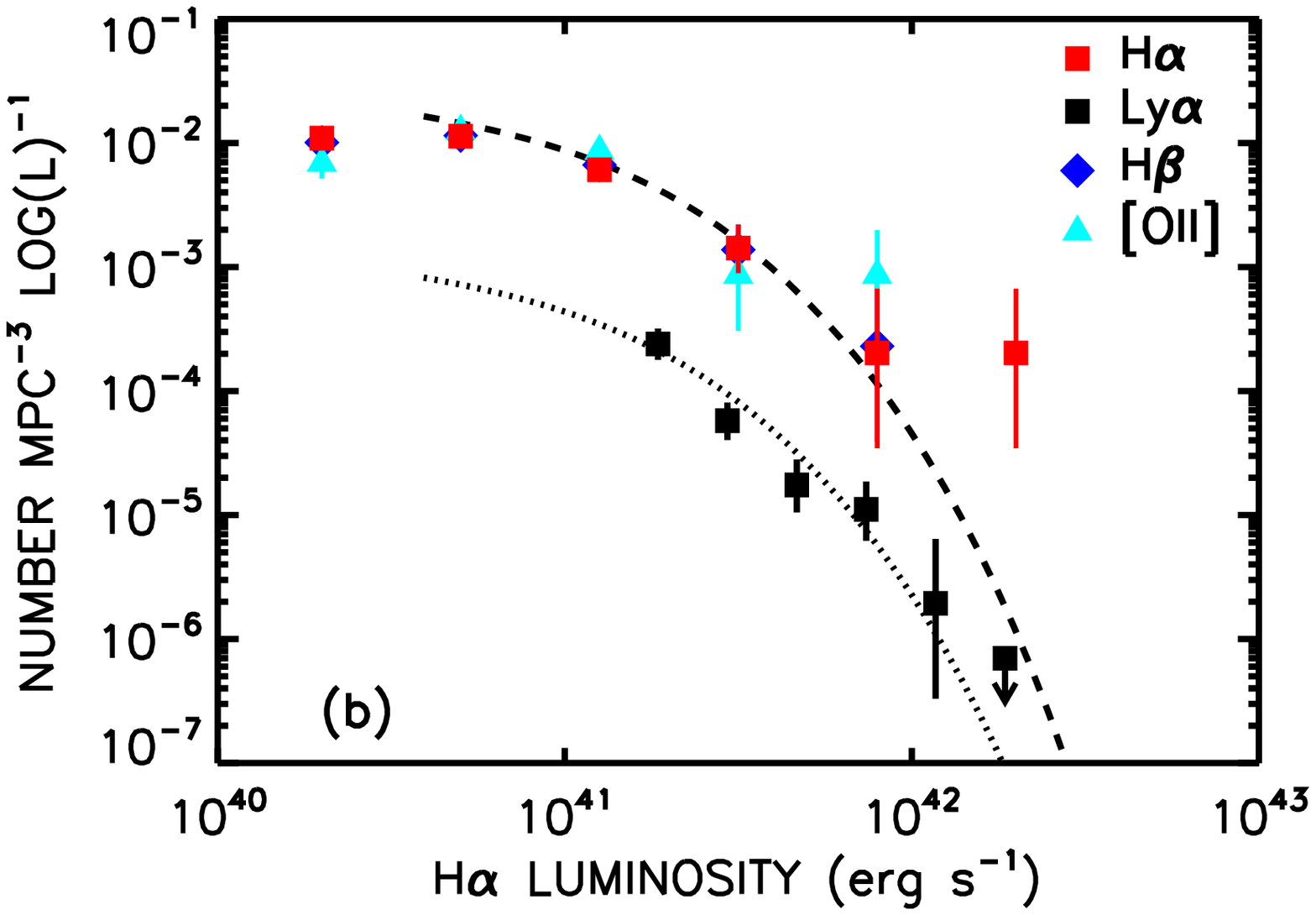}
\caption{(a) Comparison of the $z=0.195-0.44$
H$\alpha$ (red squares), H$\beta$ (blue diamonds), and 
[OII]$\lambda 3727$ (cyan triangles) emission-line 
selected galaxy LFs from the GOODS-N data with the 
LAE LF (black squares and black dotted curve). 
The red solid curve shows the Tresse \& Maddox (1998) 
Schechter (1976) function fit to their H$\alpha$ LF at $z=0.2$ 
scaled to the present geometry. (b) As in (a), but the 
luminosities of the lines have now been scaled to the 
H$\alpha$ luminosity. The black dashed curve shows the 
factor of 20 change in normalization required to bring 
the LAE LF into rough consistency with the other 
emission-line LFs.  
\label{line-comp}
}
\end{figure}

We may also look at the issue of the overall Ly$\alpha$
escape fraction by 
comparing the LFs of the LAEs with the LFs of other 
emission-line selected galaxies.
For the redshift interval $z=0.195-0.44$ we show
in Figure~\ref{line-comp}(a) the LAE LF (black 
squares and dotted curve) compared with the H$\alpha$ 
(red squares), H$\beta$ (blue diamonds), 
and [OII]$\lambda3727$ (cyan triangles) emission-line
selected galaxy LFs
from the GOODS-N sample of Barger et al.\ (2008). 
We also show the $z=0.2$ H$\alpha$ LF of 
Tresse \& Maddox (1998) rescaled to the present geometry 
(red curve), which agrees well with the GOODS-N H$\alpha$ LF.
In Figure~\ref{line-comp}(b) we show the LFs for each of 
the emission lines rescaled to the H$\alpha$ luminosity.
For [OII] and H$\beta$ we made the conversion using 
the star formation rate relations relative to H$\alpha$ from
Cowie \& Barger (2008), while for Ly$\alpha$ we made the
conversion with the ratio of 2.6 derived above.

We can see from Figure~\ref{line-comp}(b) that the
other emission-line LFs give fully consistent LFs when 
placed on the common scale. However, the LAE LF, 
while similar in shape, is much lower in normalization.
This again emphasizes that the LAEs comprise only a 
fraction of the star-forming galaxies at these redshifts. 
The LAE LF can be renormalized 
to match the H$\alpha$ LF by multiplying by a factor of 
roughly 20.  This is shown as the dashed black curve in 
Figure~\ref{line-comp}(b).
Thus, this alternative analysis also suggests that 
5\% of the Ly$\alpha$ light is escaping at this redshift.

\subsection{Ly$\alpha$ versus H$\alpha$ Equivalent Widths}
\label{la_ew_ha}

\begin{figure}
\hskip -0.6cm
\includegraphics[width=3.7in,angle=0,scale=1.]{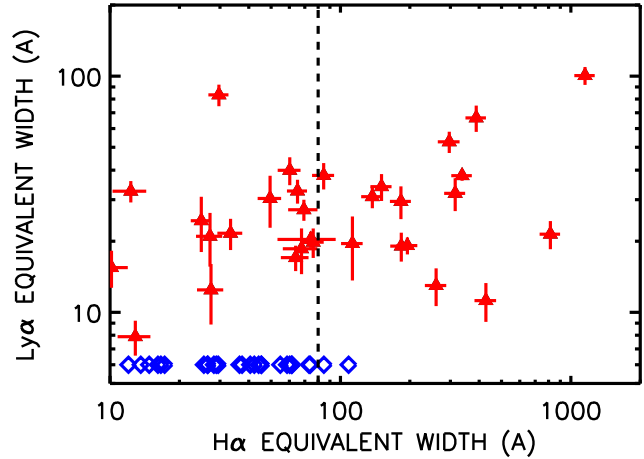}
\caption{Rest-frame EW(Ly$\alpha$) vs. rest-frame 
EW(H$\alpha$) for the {\em GALEX\/} sources that appear to 
be star formers based on their optical spectra and that lie 
in the redshift interval $z=0.195-0.44$.  These include
the optically-confirmed Ly$\alpha$ Galaxies (red solid triangles) 
from Table~12 and the optically-measured NUV-continuum 
selected galaxies 
(blue open diamonds at a nominal value of 6~\AA) from Table~14.
The $\pm1\sigma$ error bars are shown for the EW(Ly$\alpha$) 
of the Ly$\alpha$ Galaxies. In many cases the statistical error
in EW(H$\alpha$) is too small to be seen,
and we have instead shown a 10\% systematic error.  The black 
vertical dashed line shows Kakazu et al.\ (2007)'s definition 
of ultra-strong emission-line galaxies.  This roughly 
matches the EW(H$\alpha$) above which a very large fraction of the 
sources are Ly$\alpha$ Galaxies.
\label{ewha-ew}
}
\end{figure}

One of the most interesting questions is whether there is 
a way to pick out Ly$\alpha$ emission-line
galaxies using only optical spectra.  The best diagnostic 
seems to be the magnitude of the EW in the Balmer lines. 
In Figure~\ref{ewha-ew} we show the EW(Ly$\alpha$) versus  
the EW(H$\alpha$) for the optically-confirmed Ly$\alpha$
Galaxies (red solid triangles) and for the optically-confirmed 
NUV-continuum selected galaxies (blue open diamonds).
Overall, the EW(H$\alpha$) for the Ly$\alpha$ Galaxies 
(median value of 76~\AA) are significantly higher than 
those for the NUV-continuum selected galaxies
(median value of 36~\AA). Roughly half have a rest-frame 
EW(H$\alpha$)~$>80$~\AA.  Kakazu et al.\ (2007) call such 
sources ultra-strong emission-line galaxies (USELs). 
In our optically-identified sample, we see that all but 
two of the USELs are Ly$\alpha$ Galaxies and all
of the very high EW(H$\alpha$) sources are.  Clearly we
have optically observed a much smaller fraction of the
NUV-continuum selected sample than of the candidate
Ly$\alpha$ Galaxy sample, so it is possible that 
we might see more scattering into the very high EW(H$\alpha$) 
region with more observations.  However, the current data 
strongly suggest that a large fraction of USELs are Ly$\alpha$ 
Galaxies.  

As is well known, the EW(H$\alpha$) can give a rough 
estimate of the age of the star formation in a galaxy.
For a Salpeter (1955) IMF and a constant star formation rate, 
the EW(H$\alpha$) would drop smoothly to a value of 
80~\AA\ at about $10^9$~yr (Leitherer et al.\ 1999), 
while an instantaneous starburst would drop below this 
value after about $10^7$~yr.  It is therefore possible 
that the presence of a high EW(Ly$\alpha$) is simply an 
age effect, with the youngest galaxies having the 
strongest Ly$\alpha$ emission.  However, it could also 
be that there are other effects that let the Ly$\alpha$ 
photons out more easily, such as Ly$\alpha$ galaxies 
having lower metallicities or more 
kinematic disturbances than the general population.
We now turn to the measurement of these quantities.

\subsection{Metallicities}
\label{la_metal}

Only a small number of the sources have a detectable 
[OIII]$\lambda4363$ line that we can use to make a 
direct estimate of the O abundance.
We will discuss these at the end of the section.
For our primary analysis of the metallicities
we use the N2$=\log($[NII]$\lambda6584/$H$\alpha$) 
diagnostic ratio (Storchi-Bergmann et al.\ 1994).
We use N2 since the 
R23=(1.3[OIII]$\lambda5007$+[OII]$\lambda3727$)/H$\beta$
diagnostic ratio of Pagel et al.\ (1979)
is multivalued over the metallicity range of interest 
(McGaugh 1991), and our spectral flux calibration is 
not adequate to use the [NII]/[OII] ratio. 
The N2 diagnostic also has the advantage of being the method 
used by Erb et al.\ (2006) to estimate the metallicities of their 
$z\sim2$ galaxy sample.  The downside of the N2 diagnostic is that it is 
highly sensitive to the ionization parameter (e.g., Kewley \& Dopita 2002).
Other drawbacks to the N2 diagnostic include variations in N/O, and
its sensitivity to contamination by a high-[NII]/H$\alpha$ AGN
contribution.

Pettini \& Pagel (2004) showed that locally there is a reasonably 
tight relation between $12+\log$(O/H) and N2 for systems where the
O abundance has been determined with the direct
method; their linear fit gives
$12+\log$(O/H)$=8.90+0.57$N2 over the range N2$=-2.5$ to $-0.5$.
Extrapolating this to higher redshifts requires assuming that
there is no change in the typical ionization parameter, which may well
be incorrect. However, at $z=0.195-0.44$ Cowie \& Barger (2008)
found a narrow range of ionization parameters ($q\sim 2\times10^7$)
that, when combined with the photoionization code-based estimates
of Kobulnicky \& Kewley (2004), gives a broadly similar though
considerably more analytically complex equation
to the Pettini \& Pagel (2004) relation.  Cowie \& Barger (2008) 
also showed that other line diagnostics gave similar 
metallicity-luminosity relations
to that derived from N2 in the redshift interval $z=0.195-0.44$.
The maximum deviation between the relation used in 
Cowie \& Barger (2008) and the Pettini \& Pagel (2004)
linear relation over the range $-2$ to $-0.5$ is 
$-0.26$~dex at $-2$ and $+0.26$~dex at $-0.5$. 
The local data may be slightly better represented by the
Cowie \& Barger (2008) relation over this range, though the 
differences are probably not very meaningful. 
In the following we will use the Pettini \& Pagel (2004) 
linear relation for simplicity and to allow a direct 
comparison with the high-redshift results, but we will 
always show the measured N2 as our primary variable.

\begin{figure}
\includegraphics[width=4.0in,angle=0,scale=1.]{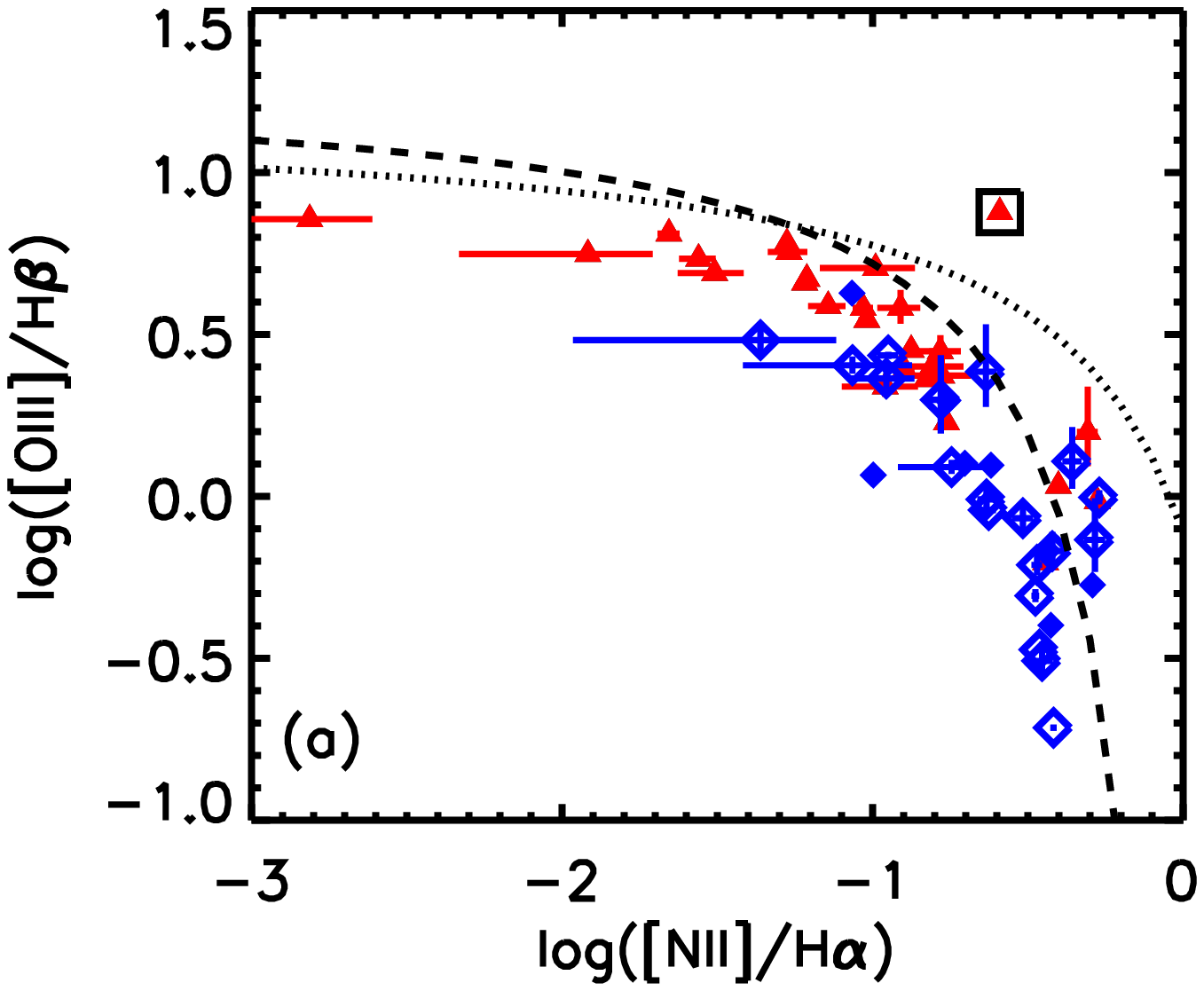}
\includegraphics[width=4.0in,angle=0,scale=1.]{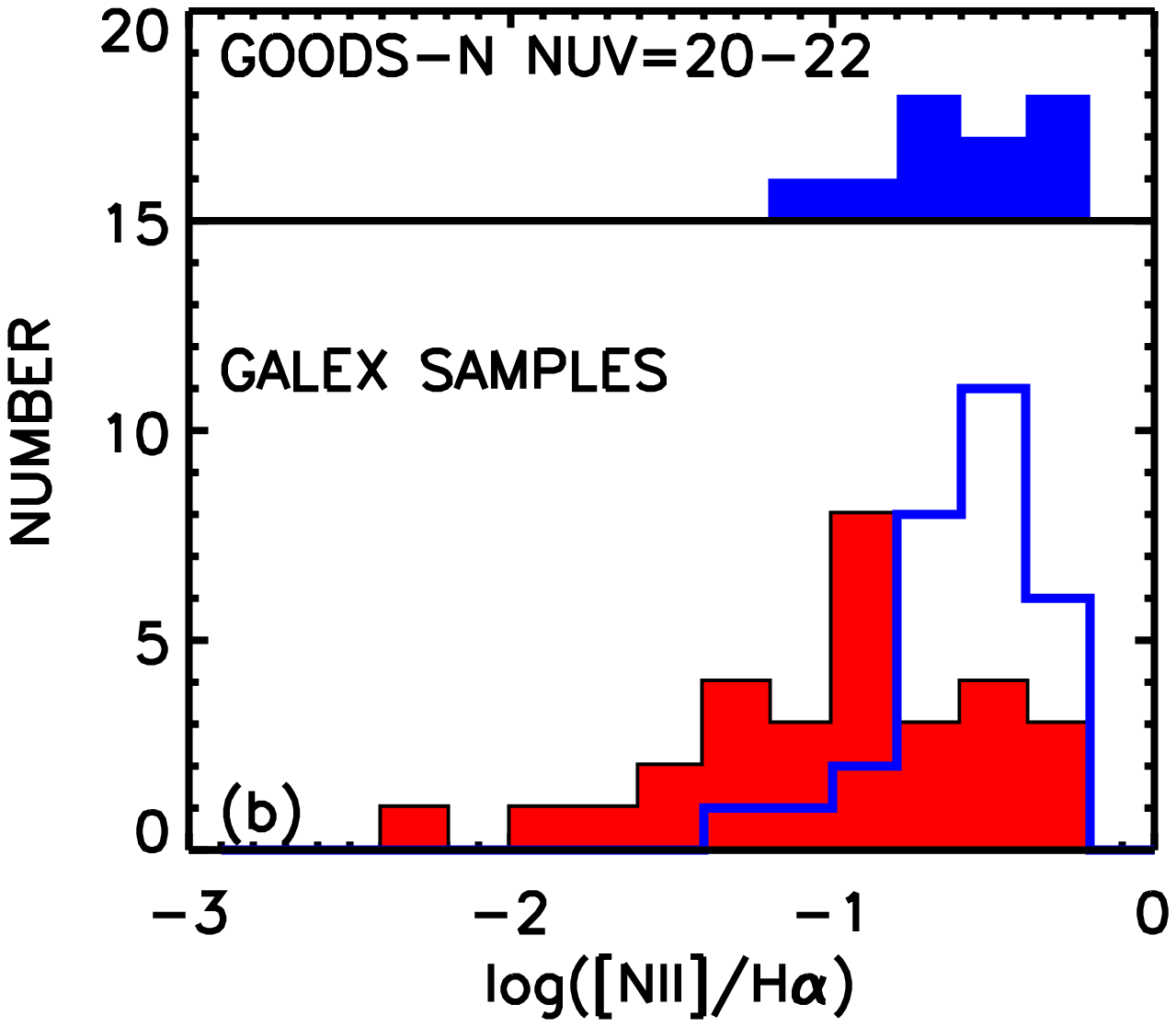}
\caption{(a) BPT diagram for the {\em GALEX\/} sources 
that do not contain broad lines in their optical 
spectra and that lie in the redshift interval $z=0.195-0.44$.
Only objects with significant H$\beta$ line detections ($>3\sigma$)
and where the spectrum covers the [NII]$\lambda6584$ line are 
shown, giving 27 optically-confirmed Ly$\alpha$ Galaxies 
(red solid triangles) and 19 optically-confirmed NUV-continuum selected 
galaxies (blue open diamonds).  The error bars are $\pm1\sigma$. 
We also show the GOODS-N NUV-continuum selected galaxy 
sample with NUV~$=20-22$ and redshifts 
between $z=0.15$ and $z=0.48$ (blue small solid diamonds).
The dotted curve is the maximum starburst 
curve of Kewley et al.\ (2001), and the dashed curve is 
the Kauffmann et al.\ (2003) star former/AGN separator 
from the SDSS data.  The black large open square denotes
the source GALEX1240+6233, 
which would be classified as an AGN on the basis of this 
diagram.  (b) Number of sources vs. $\log$ [NII]/H$\alpha$. 
All sources with high signal-to-noise ($>10\sigma$) H$\alpha$
detections and where the [NII]$\lambda6584$ line is covered
by the spectrum are shown.
The red shaded histogram shows the 30 optically-confirmed 
Ly$\alpha$ Galaxies selected in this way, the blue open histogram shows the 
29 optically-confirmed NUV-continuum selected galaxies, and the blue 
shaded histogram (upper part of diagram) shows the GOODS-N 
NUV-continuum selected galaxies.
\label{bpt}}
\end{figure}

In Figure~\ref{bpt}(a) we show the Baldwin et al.\ (1981; BPT) 
diagnostic diagram of [OIII]$\lambda5007$/H$\beta$ versus 
[NII]$\lambda6584$/H$\alpha$ for the {\em GALEX\/} sources that
are not classified as AGNs based on their UV spectra,
do not show broad lines in their optical spectra,
and lie in the redshift interval $z=0.195-0.44$.
Only spectra with a significantly detected H$\beta$ line
(signal-to-noise greater than three) are included in the diagram.
These include the optically-confirmed Ly$\alpha$ Galaxies 
(red solid triangles) and the optically-confirmed NUV-continuum 
selected galaxies (blue open diamonds).
We also show the GOODS-N NUV-continuum selected 
galaxy sample with NUV$=20-22$ and redshifts 
$z=0.15-0.48$ (blue solid diamonds). 
The diagram separates AGN-dominated sources from 
star-forming galaxies.  The dotted curve is the maximum 
starburst curve of Kewley et al.\ (2001), and the dashed 
curve is the Kauffmann et al.\ (2003) star former/AGN separator 
from the SDSS data.  Sources lying well above these curves 
are AGNs.  Although most of our sources lie along the 
star-forming galaxy track, the diagram suggests that
one of the optically-confirmed Ly$\alpha$ 
Galaxies (GALEX1240+6233) is in fact an AGN.
We have eliminated this object from all other figures,
except the BPT diagrams of Figure~\ref{bpt_morph}.
Finkelstein et al.\ (2009) similarly see a small number 
of BPT-selected AGNs in their analysis of 
the Deharvang et al.\ (2008) Ly$\alpha$ sample.

Finkelstein et al.\ (2009) also classified GALEX1417+5228
as an AGN based on the presence of HeII$\lambda4686$ 
in the spectrum.
Three of the present optical spectra show detectable HeII:
GALEX1001+0233 ($f$(HeII)/$f$(H$\beta)=0.019 \pm 0.005$),
GALEX1240+6233 ($f$(HeII)/$f$(H$\beta)=0.072 \pm 0.015$),
and GALEX1417+5228 ($f$(HeII)/$f$(H$\beta)=0.018 \pm 0.005$).
GALEX1240+6233 is classified as an AGN based on the BPT
diagram, and the strong HeII confirms this interpretation.
However, we do not think that the presence of the weaker HeII lines
necessarily implies that the other two objects are AGNs rather than 
metal-poor star-forming galaxies. Roughly 10\% of blue compact
galaxies have detectable HeII at the level seen in these
galaxies, probably produced by Wolf-Rayet stars or shocks
in the galaxies, and the presence of the line becomes more
common as one moves to more metal-poor galaxies 
(e.g., Thuan \& Izotov 2005).
GALEX1001+0233 and GALEX1417+5228 also have very 
weak or undetected NII$\lambda6584$ which would require them to 
be metal-poor AGNs.
While a very small number of such objects have now been
found (Izotov \& Thuan 2008), they would not be expected to lie 
on the low-metallicity star track in the BPT diagram as the present
objects do (Groves et al.\ 2006). In the case of 
GALEX1001+0233, where the spectrum covers NeV$\lambda3426$, 
there is no sign of this high-excitation line. We therefore classify
GALEX1001+0233 and GALEX1417+5228 as star formers. 

Sources lying at lower values of [NII]$\lambda6584$/H$\alpha$
correspond to lower metallicity galaxies, and we can
see immediately from Figure~\ref{bpt}(a) that the
optically-confirmed Ly$\alpha$ Galaxies 
have lower values of [NII]$\lambda6584$/H$\alpha$ than 
the combined NUV-continuum selected galaxies.  
This may be more clearly seen in 
Figure~\ref{bpt}(b), where we show the distribution of 
$\log($[NII]$\lambda6584$/H$\alpha$) for the 
optically-confirmed Ly$\alpha$ Galaxies (red shaded histogram), 
for the optically-confirmed NUV-continuum selected galaxies 
(blue open histogram), and for the GOODS-N NUV-continuum 
selected galaxies with NUV$=20-22$ 
(upper blue shaded histogram).  While the Ly$\alpha$ Galaxy 
distribution overlaps with the NUV-continuum selected 
galaxy distribution, it clearly extends to lower values, 
and the median [NII]$\lambda6584$/H$\alpha$ is lower.  
A rank sum test gives a $4\times10^{-4}$ probability that the two 
{\em GALEX\/} samples are similar and only a $9\times10^{-5}$
probability that the Ly$\alpha$ Galaxy sample is 
drawn from the combined NUV-continuum selected samples
from both GALEX and the GOODS-N.

\begin{figure}
\includegraphics[width=3.7in,angle=0,scale=1.]{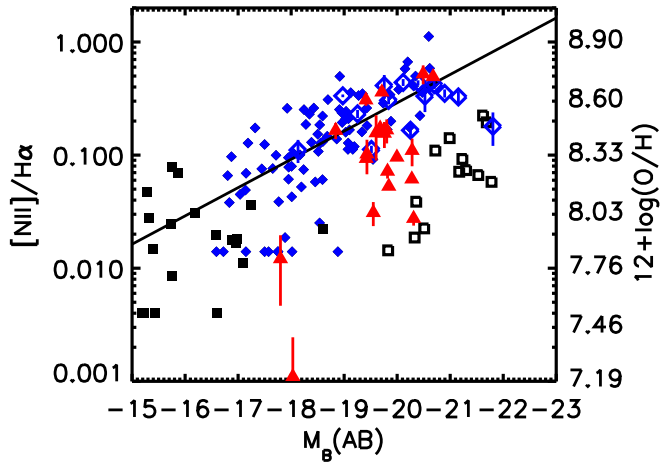}
\caption{[NII]$\lambda6584$/H$\alpha$ ratio vs.
absolute rest-frame $B$ magnitude for the {\em GALEX\/}
sources that appear to be star formers based on their
optical and UV spectra and that lie in the redshift interval
$z=0.195-0.44$. These include the optically-confirmed 
Ly$\alpha$ Galaxies (red solid triangles) and the optically-confirmed
NUV-continuum selected galaxies (blue open diamonds).
We also show the GOODS-N NUV~$<24$ galaxy sample with 
redshifts between $z=0.15$ and $z=0.48$ (blue solid diamonds).
The error bars are $\pm1\sigma$.
The black line shows the linear fit of 
N2~$=\log$([NII]$\lambda6584$/H$\alpha$) relative to 
$M_B$(AB) for all of the blue diamonds.
The metallicity that would be inferred from the Pettini
\& Pagel (2004) calibration is shown on the right-hand axis. 
Finally, we show the strong optical emission-line selected
galaxies in the redshift interval $z=0.2-0.45$ from
the Kakazu et al.\ (2007) and Hu et al.\ (2009) USEL
samples (black solid squares) and from the 
Salzer et al.\ (2009) KISS sample (black open squares).
\label{met_lum}
}
\end{figure}

Given the metallicity-luminosity relation, where lower
luminosity sources also have lower metallicities,
this result could mean that the Ly$\alpha$ Galaxies 
are simply lower luminosity galaxies than the NUV-continuum 
selected galaxies. In order to test this we show
in Figure~\ref{met_lum} [NII]$\lambda6584$/H$\alpha$ versus
absolute rest-frame $B$ magnitude, $M_B$(AB), for the 
optically-confirmed Ly$\alpha$ Galaxies (red solid triangles)
and for the optically-confirmed NUV-continuum selected 
galaxies (blue open diamonds). (We have not attempted
to construct the corresponding metallicity-mass relation
because of the difficulty of computing the mass from
the optical magnitudes in these star formation dominated
objects with strong emission lines.)  The values of $12+\log$(O/H) 
computed from the Pettini \& Pagel (2004) linear relation 
are shown on the right-hand axis of the figure. 
Only galaxies with SDSS magnitudes are shown, and we 
interpolated between
the SDSS model C $g$ and $r$ magnitudes to obtain $M_{B}$(AB).
We also show in the figure the GOODS-N
NUV-continuum selected galaxy sample with NUV~$<24$ 
and redshifts in the interval $z=0.15-0.48$ (blue solid diamonds).
Here the $M_{B}$(AB) magnitudes are computed from the 
AUTO magnitudes in the F606W bandpass of the ACS GOODS-N 
data (Giavalisco et al.\ 2004). As would be expected if 
the photometry is consistent, the {\em GALEX\/} 
optically-confirmed NUV-continuum selected galaxies 
(blue open diamonds) match to the bright end of the 
GOODS-N NUV-continuum selected galaxies (blue solid diamonds). 
The optically-confirmed Ly$\alpha$ Galaxies (red solid triangles) 
lie systematically lower in metallicity at a given luminosity.

All of the blue diamonds are well fit by the relation
\begin{equation}
{\rm N2}  = -0.54 - 0.25( M_{B}{\rm (AB)} + 20) \,,
\label{eqnn2}
\end{equation}
which we show as the black solid line in Figure~\ref{met_lum}.
Translating Equation~\ref{eqnn2} to a 
metallicity-luminosity relation using the 
Pettini \& Pagel (2004) linear relation gives 
\begin{equation}
12+\log({\rm O/H})  = 8.59 - 0.14(M_{B}{\rm (AB)} + 20) \,.
\label{eqnn2b}
\end{equation}
For $M_B{\rm (AB)}=-19$ to $-20$, the median value of 
$12+\log$(O/H) for the optically-confirmed Ly$\alpha$ 
Galaxies lies about 0.2~dex lower than that for the 
combined NUV-continuum selected galaxies.  At brighter 
magnitudes there is less separation, but this is almost 
certainly an effect of the saturation in N2 as the 
metallicity approaches the solar metallicity (see the
right-hand region of Figure~\ref{bpt}(a)).
A similar effect is seen in the very bright {\em GALEX\/} 
sample of Hoopes et al.\ (2007).

We also include in Figure~\ref{met_lum} strong optical
emission-line selected galaxies such as the USELs 
(black solid squares) of Kakazu et al.\ (2007) and 
Hu et al.\ (2009) and the KISS galaxies (black open squares) 
of Salzer et al.\ (2009).  The Cardamone et al.\ (2009)
sample of strong emission-line objects (found
with a green color selection from the SDSS galaxies) lies closer
to the track of the NUV-continuum selected galaxies,
and we do not show these on the figure. The optically-confirmed 
Ly$\alpha$ Galaxies appear to stretch from the values 
of the NUV-continuum selected galaxies down 
to the values of the strong optical emission-line selected 
galaxies.  This appears consistent with the range of 
EW(H$\alpha$) in the optically-confirmed Ly$\alpha$ Galaxies 
(Figure~\ref{ewha-ew}), which shows a substantial but not 
complete overlap with the strong optical emission-line galaxies.

Six of the optically-confirmed Ly$\alpha$ Galaxies have 
significantly detected ($>4\sigma$) [OIII]$\lambda4363$ 
auroral lines. However, one of these (GALEX1240+6233) 
is classified as an AGN based on the BPT diagram.  For the 
five remaining sources we used the `direct' or 
$T_e$ method to determine the metallicity
(e.g., Seaton 1975; Pagel et al.\ 1992; Pilyugin \& Thuan 2005; 
Izotov et al.\ 2006).
To derive $T_e$\oiii\ and the oxygen abundances, we used the
Izotov et al.\ (2006) formulae, which were
developed with the latest atomic data and photoionization models.
All five sources give abundances which are broadly consistent with
the Pettini \& Pagel (2004) N2 determinations.

The lowest metallicity source is GALEX1417+5228.
This extremely high EW(H$\alpha$)
($\sim1400$~\AA\ in the rest frame) source is shown in 
Figure~\ref{whole_spectrum}.  It is the interesting source 
we mentioned in Section~\ref{lae_prop}.
The [NII]$\lambda6584$/H$\alpha$
ratio in Figure~\ref{met_lum} is only at the $1\sigma$
level, and the $1\sigma$ upper limit from the Pettini
\& Pagel (2004) calibration gives $12+\log$(O/H)$<7.4$.
The [OIII]$\lambda4363$ line is exceptionally strong, 
from which we derive $12+\log$(O/H)$~=7.65\pm0.03$. 
Thus, {\em GALEX\/}1417+5228 is near the extremely 
low-metallicity class and below all of the strong
emission-line sources in the KISS sample of
Salzer et al.\ (2009), as can be seen from 
Figure~\ref{met_lum}. The remaining 4 galaxies
(GALEX0332-2811, GALEX0959+0151, GALEX1000+0201, and GALEX1001+0233)
have direct abundances $12+\log$(O/H)=($8.14\pm0.03, 8.27\pm0.03,
7.96\pm0.12$ and $7.94\pm0.03$) compared to the N2 determinations
of $12+\log$(O/H)=($7.96\pm0.02,8.17\pm0.02, 8.20\pm0.02$ and $8.01\pm0.03$).

\begin{figure}
\includegraphics[width=3.7in,angle=0,scale=1.]{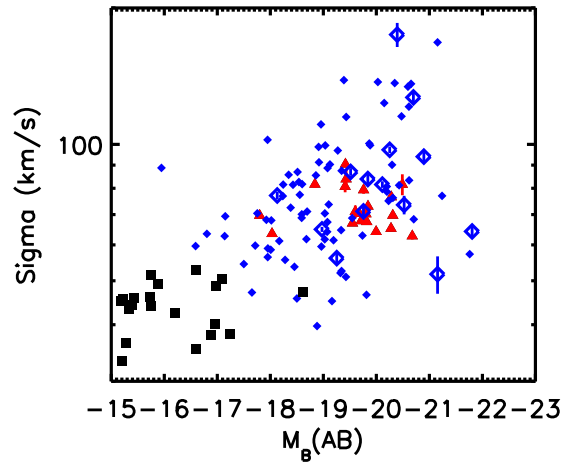}
\caption{Measured velocity dispersion $\sigma$ 
from the H$\alpha$ line widths vs. absolute rest-frame $B$ 
magnitude for the {\em GALEX\/} sources that appear to
be star formers based on their optical spectra and
that lie in the redshift interval $z=0.195-0.44$. 
These include the optically-confirmed Ly$\alpha$
Galaxies (red solid triangles) and the optically-confirmed
NUV-continuum selected galaxies (blue open diamonds).
The error bars are $\pm1\sigma$.
We also show the GOODS-N NUV~$<24$ galaxy sample with
redshifts between $z=0.15$ and $z=0.48$
(this excludes broad-line AGNs and intermediate-type
Seyferts; blue solid diamonds). 
Finally, we show the strong optical emission-line selected
galaxies in the redshift interval $z=0.2-0.45$ 
from the Kakazu et al.\ (2007) and Hu et al.\ (2009) USEL
samples (black solid squares).
\label{nuv_widths}
}
\end{figure}

\begin{figure}
\hskip -0.4cm
\includegraphics[width=4.0in,angle=0,scale=1.]{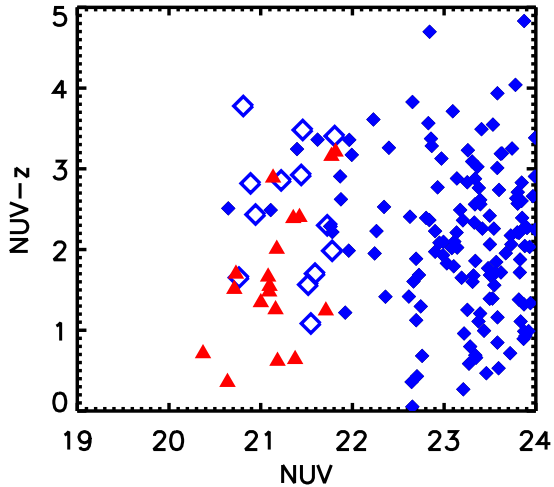}
\caption{NUV$-$Gunn $z$ color vs. NUV magnitude
for the {\em GALEX\/} sources that appear to be star
formers based on their optical spectra and that lie
in the redshift interval $z=0.195-0.44$. These include
the optically-confirmed Ly$\alpha$ Galaxies 
(red solid triangles) and the optically-confirmed NUV-continuum
selected galaxies (blue open diamonds).  We also show
the GOODS-N NUV~$<24$ galaxy sample with redshifts
between $z=0.15$ and $z=0.48$ (blue solid diamonds).
\label{nuv_colors}
}
\end{figure}

\subsection{Line Widths}
\label{ha_line_widths}

There is also a strong correlation between the line widths
measured in H$\alpha$ and luminosity.
In Figure~\ref{nuv_widths} we plot the velocity dispersions
$\sigma$ from the H$\alpha$ line widths versus 
$M_B$(AB). The optically-confirmed Ly$\alpha$ Galaxies 
(red solid triangles) and the optically-confirmed NUV-continuum
selected galaxies (blue open diamonds) show 
consistent values for both $\sigma$ and $M_B$(AB). 
They are also broadly consistent with the bright end of the 
GOODS-N NUV~$<24$ galaxy sample (blue solid diamonds) with 
any differences being attributable to differences in the 
photometry.  The USELs (black solid squares) lie
at the faint end of the distribution. Thus, the 
optically-confirmed Ly$\alpha$ Galaxies are being drawn 
from a population with the same mass to luminosity ratios 
and kinematical structure as the
NUV-continuum selected galaxy population.

\subsection{Colors and Extinctions}
\label{col_ext}

The optically-confirmed Ly$\alpha$ Galaxies also have bluer
colors than the optically-confirmed NUV-continuum selected 
galaxies of the same 
luminosity.  We illustrate this in Figure~\ref{nuv_colors}, 
where we plot observed-frame NUV$-$Gunn $z$ color using
the Gunn $z$ model C magnitudes from the SDSS versus
NUV magnitude for the optically-confirmed Ly$\alpha$ 
Galaxies (red solid triangles) and for the optically-confirmed
NUV-continuum selected galaxies (blue open diamonds).  
While the number of sources in the sample is small, it 
appears that the Ly$\alpha$ Galaxies 
are approximately a magnitude bluer in color than the 
NUV-continuum selected galaxies of the same absolute magnitude.
We also show in the figure the GOODS-N NUV~$<24$ galaxy 
sample (blue solid diamonds)
using the {\em HST\/} ACS F850LP AUTO magnitudes.
The spread in colors matches the distribution
seen in both {\em GALEX\/} samples, and we can see that as we 
move to fainter NUV magnitudes, the fraction of blue sources
increases. This may suggest that Ly$\alpha$ Galaxies
will be more common in sources below the NUV~$\sim22$ limit 
of the {\em GALEX\/} spectroscopic observations.

\begin{figure}
\hskip -0.4cm
\includegraphics[width=4.0in,angle=0,scale=1.]{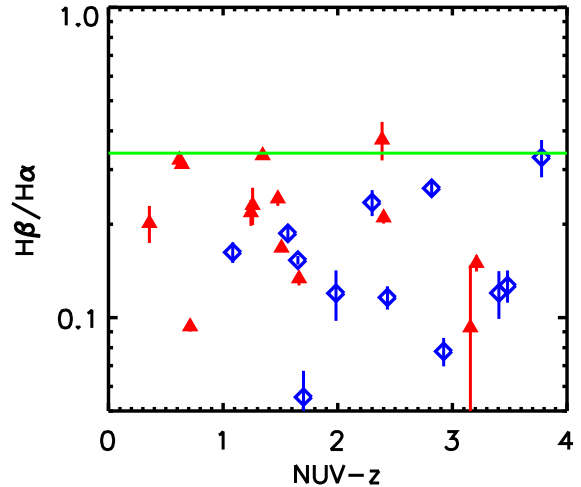}
\caption{Balmer ratio H$\alpha$/H$\beta$
vs. NUV$-$Gunn $z$ color for the {\em GALEX\/} sources that
appear to be star formers based on their optical
spectra and that lie in the redshift interval $z=0.195-0.44$.
These include the optically-confirmed Ly$\alpha$
Galaxies (red solid triangles) and the optically-confirmed
NUV-continuum selected galaxies (blue open diamonds).
Note that we show only sources where the spectra are
of high enough quality to make an accurate measurement
and where rest-frame EW(H$\alpha)<1000$~\AA\ to avoid 
galaxies where the broadband colors
are substantially perturbed by the emission lines.
The green horizontal line shows the expected case~B 
ratio in the absence of extinction.
\label{colext}
}
\end{figure}

The NUV$-$Gunn $z$ color difference appears to be at least 
partly an extinction effect. In Figure~\ref{colext} we plot 
the H$\beta$/H$\alpha$ Balmer ratio versus the NUV$-$Gunn $z$
color using the Gunn $z$ model C magnitudes from the SDSS.  
We show only sources where the spectra are
of high enough quality to make an accurate measurement
and where rest-frame EW(H$\alpha)<1000$~\AA\ to avoid 
galaxies where the broadband colors
are substantially perturbed by the emission lines.
However, the emission lines act in the sense of reducing
the difference between the populations, since in the stronger
H$\alpha$ emission-line galaxies, which are primarily found
in the LAE sample, the line contribution brightens the
$z$-band if it is in the correct redshift interval and
makes the NUV$-$Gunn $z$ colors redder. 
The Balmer line fluxes are calculated from the observed
EWs combined with the continuum fluxes at the emission-line 
wavelengths inferred from the broadband
magnitudes of the galaxies.  For H$\beta$ we have
applied a 1~\AA\ correction to the EWs to allow for 
underlying stellar absorption (Cowie \& Barger 2008). 
The green horizontal line shows the expected case~B ratio in the
absence of extinction.

The bluer optically-confirmed Ly$\alpha$ Galaxies are 
consistent with having only weak extinction within the rather 
substantial systematic uncertainties, while the redder
galaxies, which include the 
optically-confirmed NUV-continuum selected galaxies, 
have a lower average Balmer ratio. 
However, the difference in the Balmer ratios
is not statistically significant with a rank
sum test giving a 7\% probability that the
two samples are drawn from the same distribution.
The median extinctions inferred from the 
two samples are A$_v=1.30$ (0.94, 2.16) for
the LAEs and 2.64 (1.61, 2.85) for the optically-confirmed
NUV-continuum selected galaxies,
where the quanities in brackets give the 68\%
confidence range. The difference in the color 
distributions is marginally significant with
only a 1.7\% probability that the two distributions are the same.  
However, it is clear that larger samples and
preferably more accurate measurements of the 
Balmer fluxes are required to proceed further.

We may also compare with the UV spectral slopes as
measured in the {\em GALEX\/} spectra. As can be seen 
visually in Figure~\ref{stacked_spectra}, there is very little
difference between the average slope of the LAE selected
galaxies and the average slope of the NUV-continuum selected
galaxies. We have measured the slopes of all of the individual 
galaxies in the LAE sample (Table~\ref{tab11}) and in the 
NUV-continuum sample (Table~\ref{tab13}). 
The median $\beta$ of the LAE (NUV-continuum) sample is 
$-1.23\pm0.22$ ($-1.07\pm0.28$),
and there appears to be no significant difference in the distribution
of the slopes in the two samples. The measured continuum slope
translates to an A$_{1600}$ extinction of 1.88
(2.20) for the LAE (NUV-continuum) samples
using the calibration of Meurer et al.\ (1999). (This ignores
any small correction for the wavelength range over which the
index is measured, which should be similar for the two samples.)
Thus, the absolute continuum UV extinction does not appear to
be related to the strength of the Ly$\alpha$ line, which
must have a complex and indirect dependence on the metallicity
and optical colors where there is a significant dependence.

\subsection{Galaxy Morphologies}
\label{secmorph}

\begin{figure}
\epsscale{.2}
\includegraphics[width=3.7in,angle=0,scale=1.]{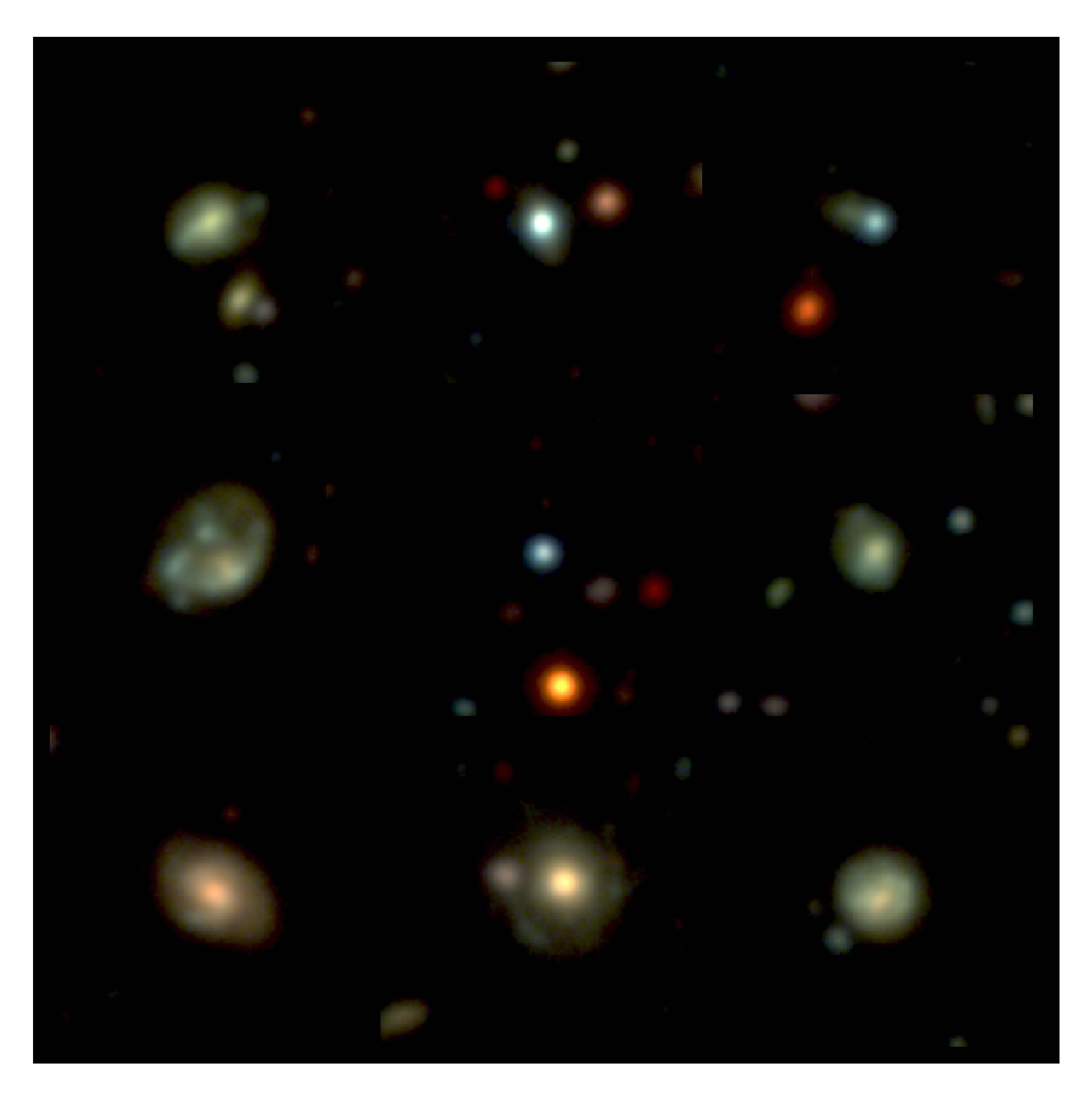}
\includegraphics[width=3.7in,angle=0,scale=1.]{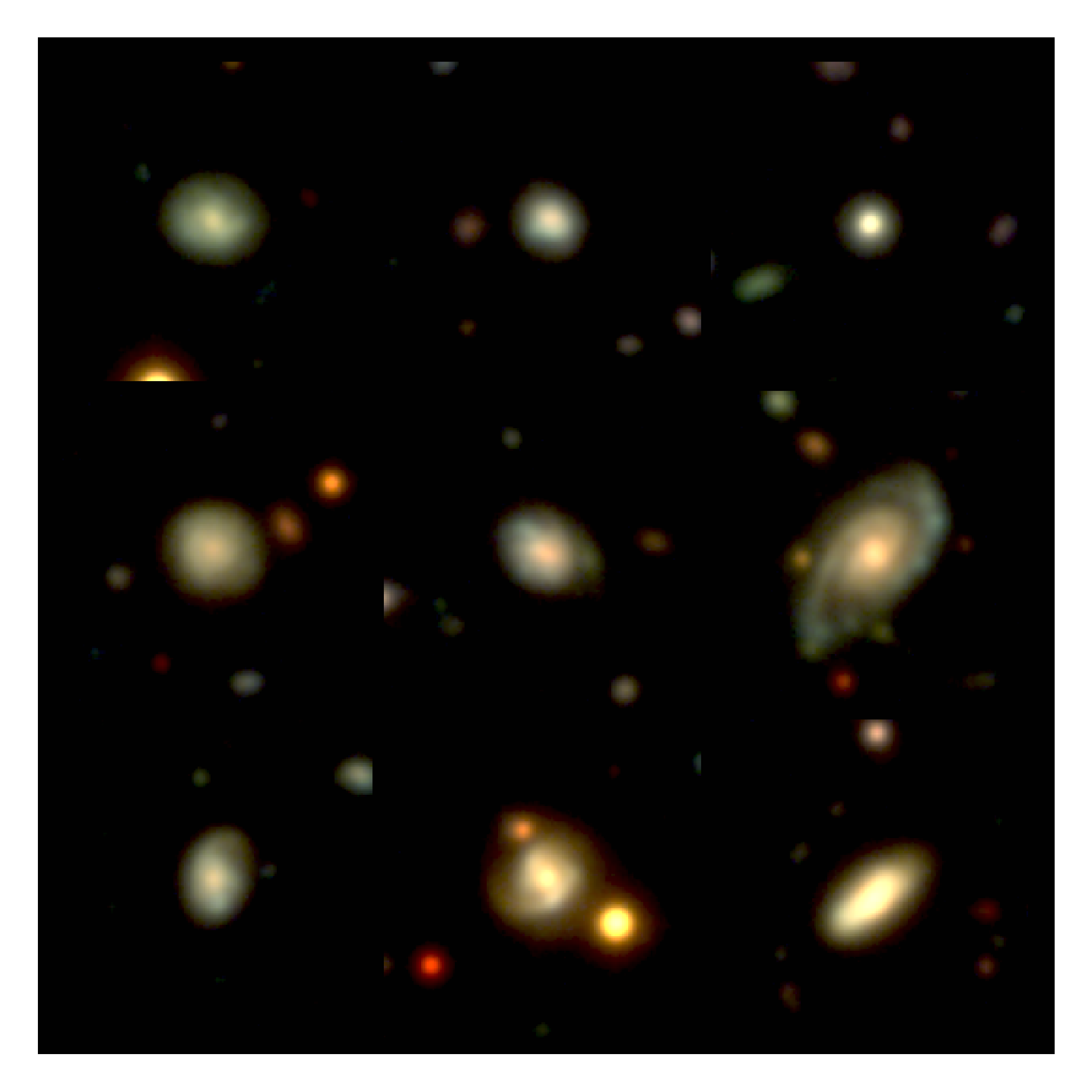}
\caption{The morphologies of some of
the {\em GALEX\/} sources in 
the GROTH~00 field that appear to be star formers based
on their optical spectra and that lie in the redshift
interval $z=0.195-0.44$.  These include sources in
the optically-confirmed Ly$\alpha$ Galaxy sample 
(left panel) and sources in the optically-confirmed
NUV-continuum selected galaxy sample (right panel). 
The blue, green, and red colors
correspond to the $u^*$, $g'$, and $i'$ band images from
the CFHT Legacy Survey deep observations of this field. In the 
right panel nearly all of the galaxies are spirals. 
We classify only the two right-most galaxies in the left
row as compact. Some of the optically-confirmed Ly$\alpha$ 
Galaxies in the left panel are also spirals
(the bottom three galaxies), but a much larger
fraction are mergers (the leftmost galaxy in
the second row) or compacts (the remaining galaxies).
\label{morphs}
}
\end{figure}

\begin{figure}
\hskip -0.4cm
\includegraphics[width=4.0in,angle=0,scale=1.]{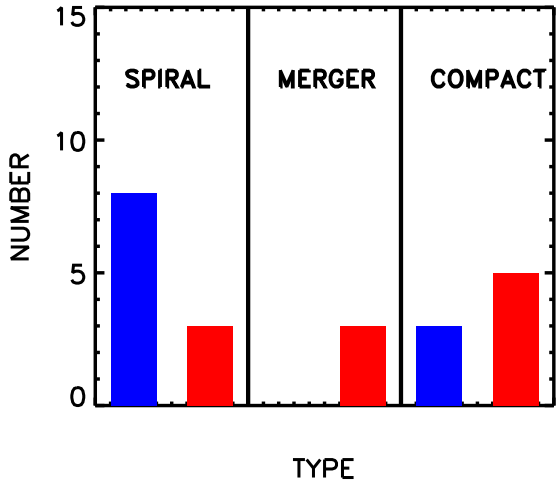}
\caption{Distribution of galaxy types for the
{\em GALEX\/} sources in the GROTH~00 and SIRTFFL~00
fields that appear to be star formers
based on their optical spectra and that lie in the
redshift interval $z=0.195-0.44$.  These include the
optically-confirmed Ly$\alpha$ Galaxies (red bars) and 
the optically-confirmed NUV-continuum selected galaxies
(blue bars).  The NUV-continuum selected galaxies are 
primarily spirals, while the Ly$\alpha$ Galaxies 
contain a much larger fraction of mergers and compact 
galaxies.
\label{morphtype_dist}
}
\end{figure}

We used the deep $i'$-band ground-based data from the
CFHT MegaPipe database of the GROTH~00 and SIRTFFL~00 fields 
to make rough morphological classifications
of the galaxies in our samples.  We divided the galaxies into
three classes: spirals, mergers showing clear signs of major
interactions, and smaller compact or irregular galaxies. 
The classifications were made in a blind fashion without 
reference to the properties of the galaxies to avoid any 
subjective bias. Most of the galaxies in the optically-confirmed
NUV-continuum selected galaxy sample are large, easily recognizable
spirals (lower panel of Figure~\ref{morphs}). This is also 
true for the GOODS-N galaxies with NUV~$=20-22$,
where all but one of the sources (a merger) 
fall into this class.  The optically-confirmed Ly$\alpha$ 
Galaxies are much more heterogeneous (upper panel of 
Figure~\ref{morphs}).  Perhaps somewhat surprisingly, some 
of the Ly$\alpha$ Galaxies are large face-on spiral galaxies 
(see also Finkelstein 2009).
However, as can be seen from Figure~\ref{morphtype_dist}, the
Ly$\alpha$ Galaxy sample (red bars) contains a 
much larger fraction of mergers and compact galaxies than 
the NUV-continuum selected sample (blue bars).

\begin{figure}
\includegraphics[width=3.7in,angle=0,scale=1.]{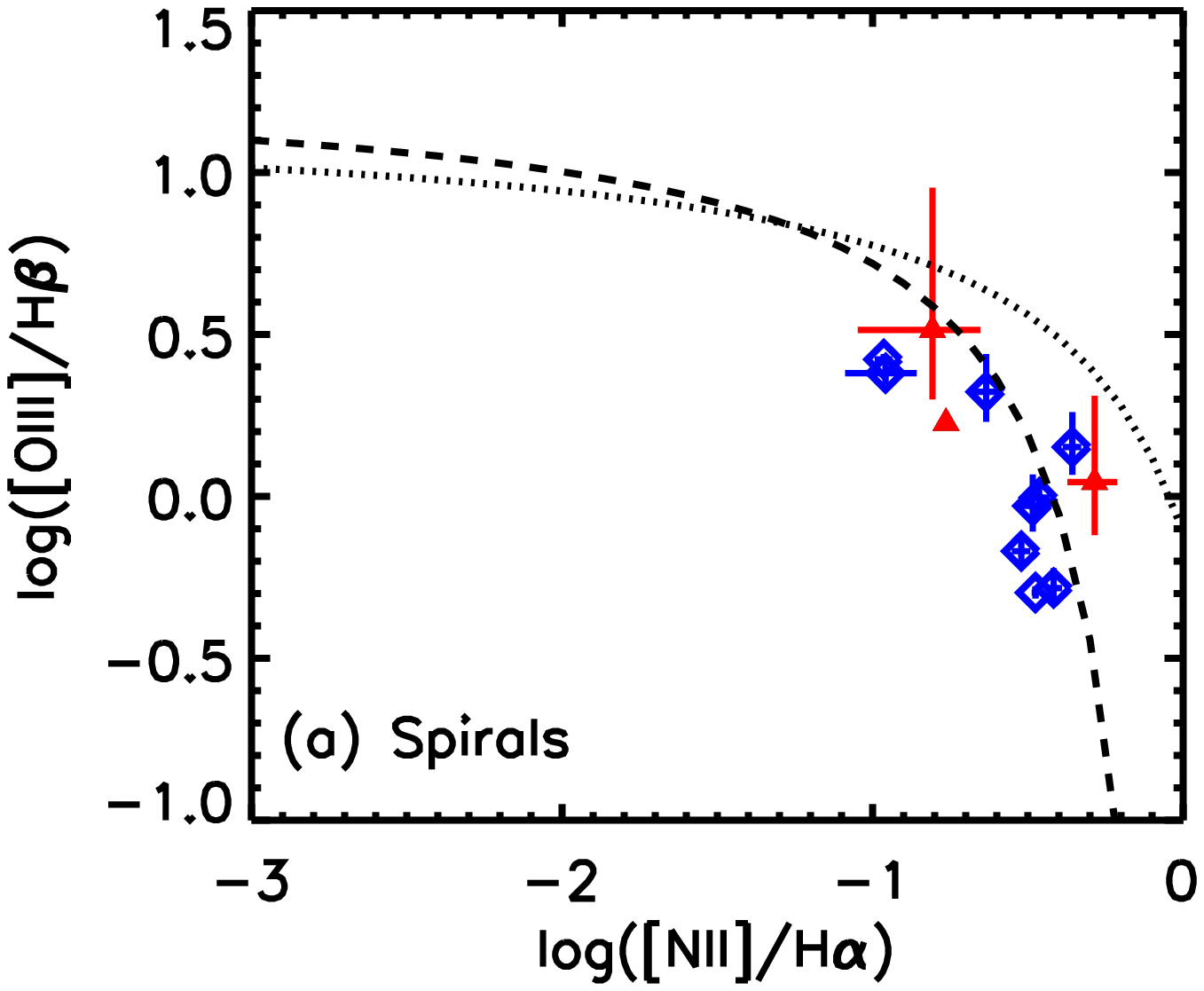}
\includegraphics[width=3.7in,angle=0,scale=1.]{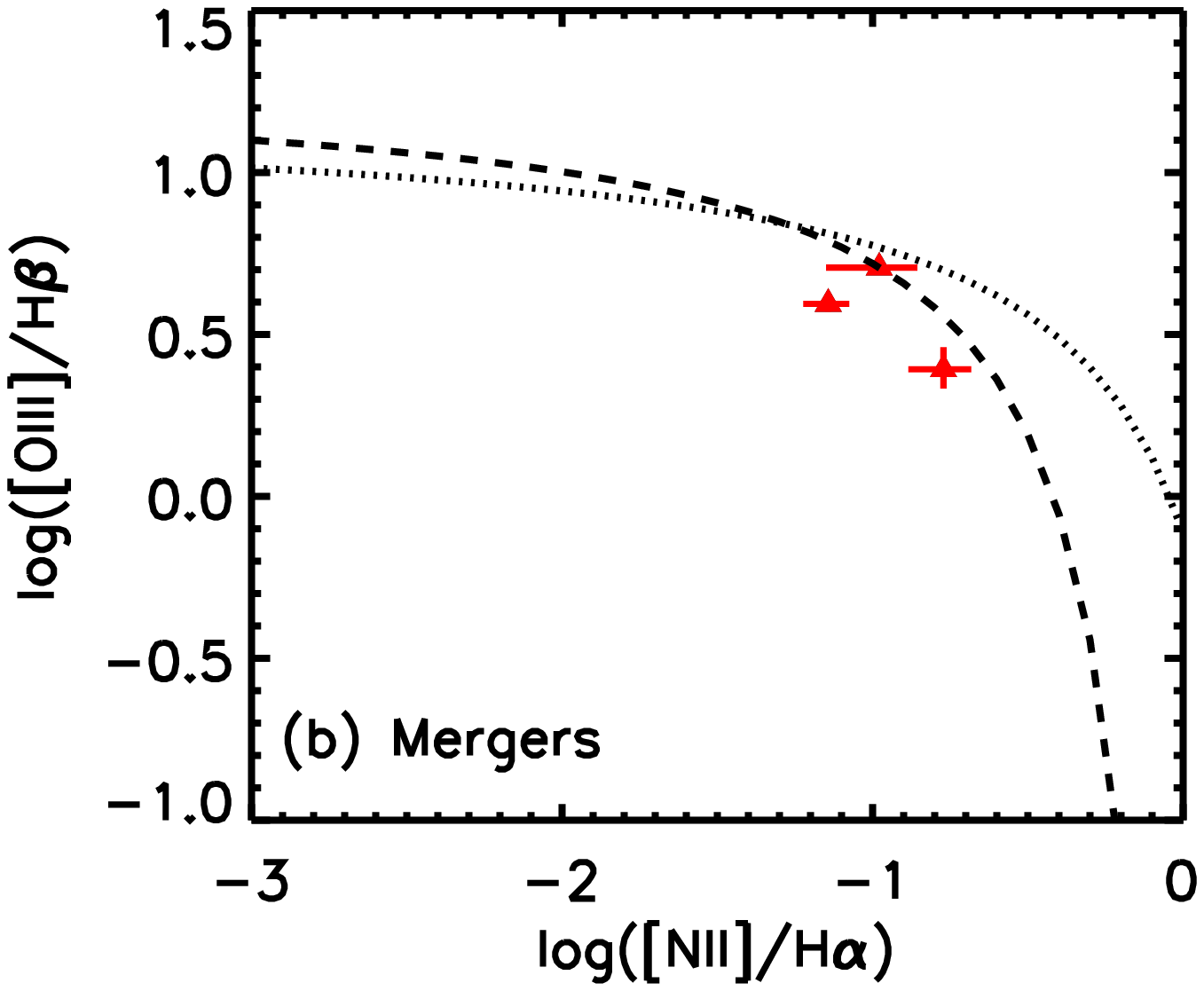}
\includegraphics[width=3.7in,angle=0,scale=1.]{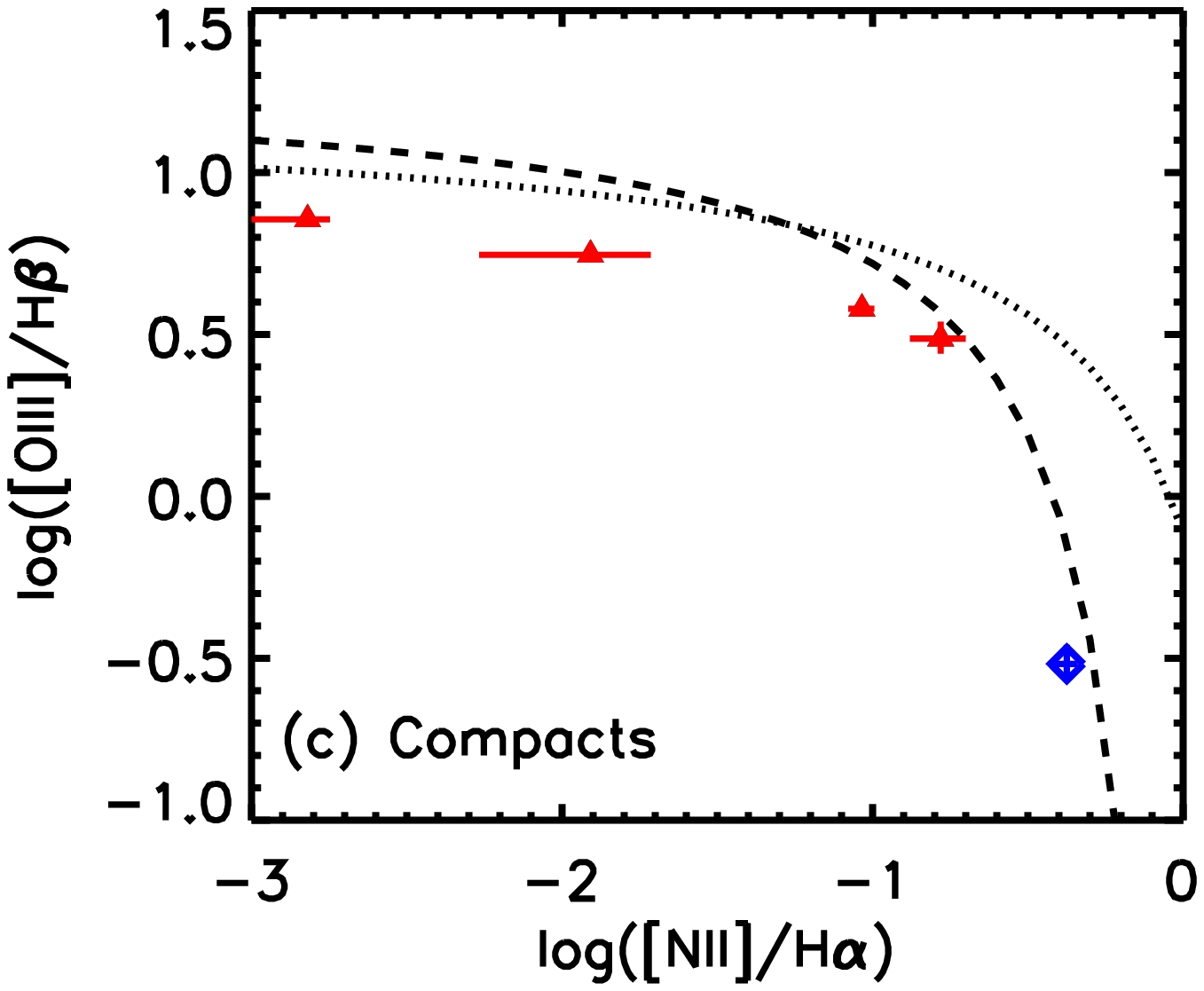}
\caption{
BPT diagrams for the {\em GALEX\/} sources that
appear to be star formers based on their optical
spectra, lie in the redshift interval
$z=0.195-0.44$, and are in the GROTH~00 or
SIRTFFL~00 fields where we have made morphological
classifications.  Only sources with significant
detections ($>3\sigma$) in either of the
H$\beta$ or [OIII]$\lambda5007$ lines are shown.
The sample is divided into optically-confirmed
Ly$\alpha$ Galaxies (red solid triangles) and  
optically-confirmed NUV-continuum selected galaxies 
(blue open diamonds) divided by morphology: 
(a) spirals, (b) mergers, and (c) compacts.
\label{bpt_morph}
}
\end{figure}

In Figure~\ref{bpt_morph} we show the BPT diagrams for
the optically-confirmed Ly$\alpha$ Galaxy sample 
(red solid triangles) and for the optically-confirmed NUV-continuum 
selected sample (blue open diamonds) divided by morphology 
into (a) spirals, (b) mergers, and (c) compacts.
All of the spirals, including the Ly$\alpha$ Galaxies,
have high N2 ratios, showing that
they are near-solar metallicity sources. 
The mergers (all three are Ly$\alpha$ 
Galaxies) appear to have slightly lower metallicities 
than the spirals.  Finally, based on this small 
sample, the compact galaxies appear 
to split between higher metallicity
NUV-continuum selected galaxies and lower metallicity
Ly$\alpha$ Galaxies. Thus, the Ly$\alpha$ Galaxies
appear to be a mixture of normal spirals, merging
galaxies, and low-metallicity compact and irregular galaxies.
It is these latter sources that seem primarily to weight the 
metallicities of the Ly$\alpha$ Galaxies to lower values 
than those of the NUV continuum-selected galaxies.


\section{Conclusions}
\label{seccon}

Perhaps the single most significant conclusion we can draw
from the low-redshift {\em GALEX\/} samples is that 
LAEs are much less common at low redshifts 
than they were in the past (Deharvang et al.\ 2008). 
In this paper we have shown that formally defined LAEs
(rest-frame EW(Ly$\alpha)>20$~\AA) constitute about
5\% of the local NUV-continuum selected population 
at $z=0.3$, as opposed to $20-25$\%
of this population at $z=3$ (Shapley et al.\ 2003). 

\begin{figure}
\hskip -0.4cm
\includegraphics[width=3.7in,angle=0,scale=1.]{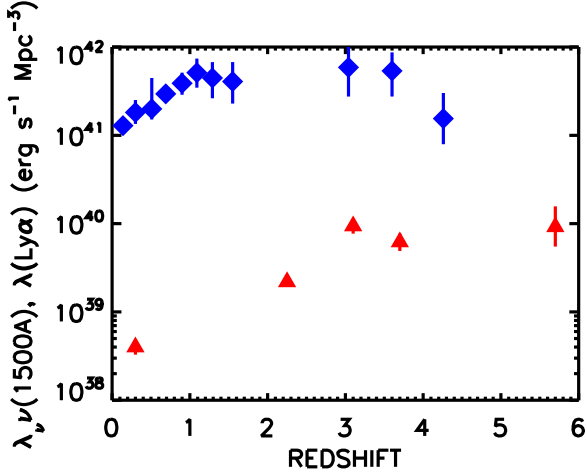}
\caption{Relative evolution of the Ly$\alpha$ 
luminosity densities in the LAEs (red solid triangles) and 
the $\nu L_\nu$(1500~\AA) luminosity densities in 
UV-continuum selected galaxies (blue solid diamonds).  
The Ly$\alpha$ luminosity densities are from the present
work at the lowest redshift and from K.~Nilsson (2009, priv. comm.),
and Ouchi et al.\ (2008) at 
higher redshifts. Gronwall et al. (2007) gives a similar luminosity
density at $z=3.1$. The UV-continuum luminosity densities 
are from Tresse et al.\ (2007). $\pm1$ sigma error bars
are shown for all the points except that of Nilsson.
In some cases they are smaller than the symbol size.
\label{rel_lumden}
}
\end{figure}

This rise is probably best seen by looking at the
relative evolution of the volume emissivities of Ly$\alpha$
in LAEs and the volume emissivities of $\nu L_\nu$(1500~\AA) 
in UV-continuum selected galaxies with redshift.   The
volume emissivities are obtained from the  integrated LF fits
at each redshift.  In Figure~\ref{rel_lumden} we show the 
Ly$\alpha$ luminosity densities in the LAEs (low redshifts, 
this paper; higher redshifts, K.~Nilsson 2009, priv. comm.;
Gronwall et al.\ 2007; Ouchi et al.\ 2008) with red solid 
triangles.  We show the UV-continuum luminosity densities 
from Tresse et al.\ (2007) with blue solid diamonds.  
Both rise rapidly from the present time to redshift three, 
but the rise in the Ly$\alpha$ luminosity densities is much 
larger than the rise in the UV-continuum luminosity densities.
At higher redshifts the UV-continuum luminosity densities 
turn down while the Ly$\alpha$ luminosity densities
remain flat. Thus, as a function of increasing redshift 
over the whole redshift range $z=0-6$
we appear to be seeing an increase in the amount of escaping 
Ly$\alpha$ relative to the escaping UV-continuum from the entire 
UV-continuum selected galaxy population.
We note that we are not making an
extinction correction in either quantity here and that
the comparison is of the light emerging from the galaxy population
in the Ly$\alpha$ line and the UV continuum.

The complexity of the low-redshift population
makes it hard to provide a single explanation
for this evolution. Indeed, the second clear result
from the {\em GALEX\/} data is that low-redshift 
Ly$\alpha$ galaxies are not a monolithic population. 
There is a large fraction of low-metallicity compact
galaxies, as might be expected, and merging
also seems to make it easier to see Ly$\alpha$.
However, there is also a population of more normal 
near-solar metallicity spiral galaxies
contained in the population. 
Furthermore, while on average the low-redshift Ly$\alpha$ galaxies
have lower metallicities than the NUV-continuum selected
galaxies without Ly$\alpha$ emission, the range of 
metallicities in the Ly$\alpha$ galaxies is wide, 
stretching from near-solar metallicities
down to extremely metal-poor galaxies.

\begin{figure}
\hskip -0.4cm
\includegraphics[width=3.7in,angle=0,scale=1.]{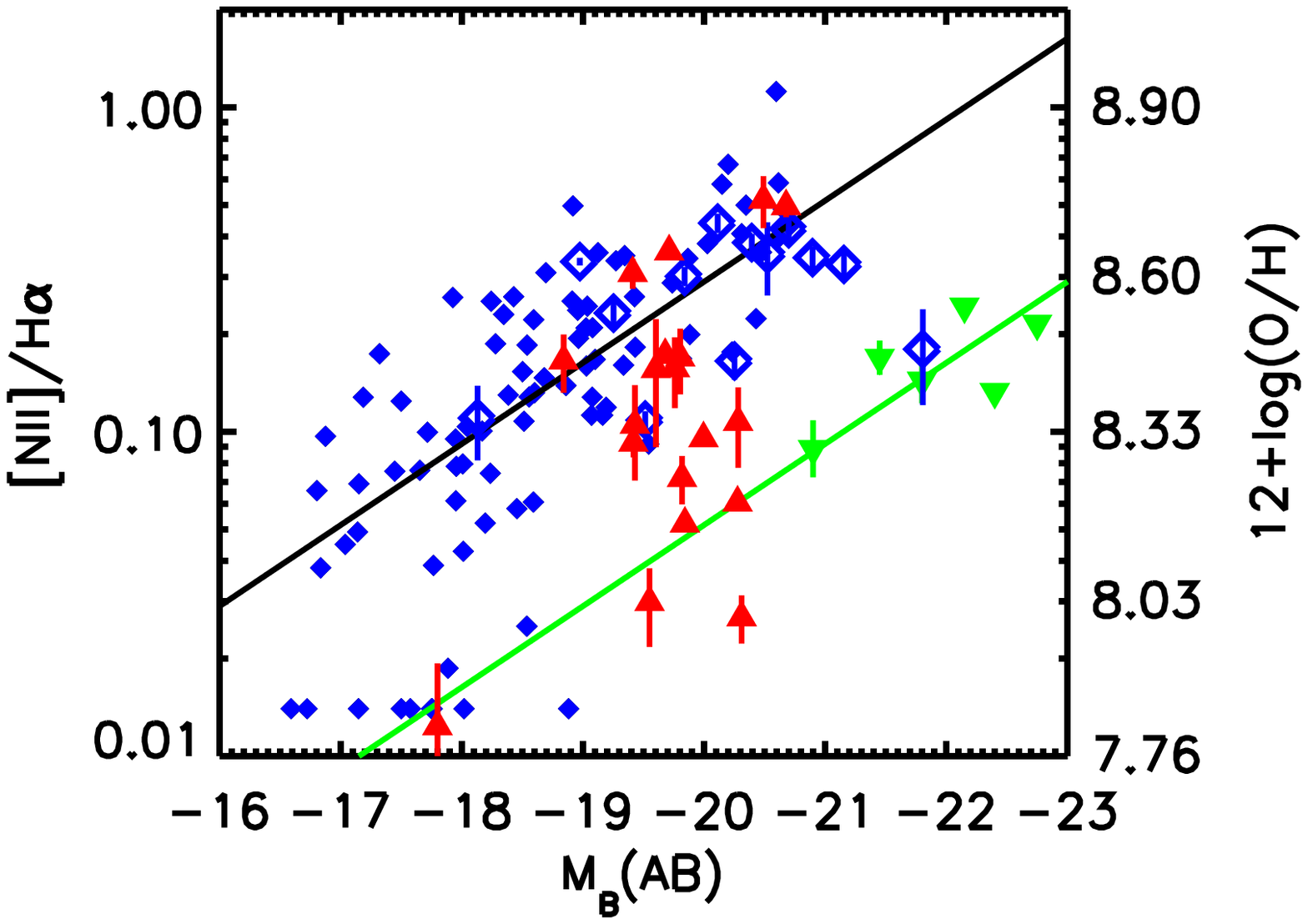}
\caption{[NII]$\lambda6584$/H$\alpha$ ratio vs.
absolute rest-frame $B$ magnitude for the {\em GALEX\/} 
sources that appear to be star formers based on their 
UV and optical spectra and that lie in the redshift interval
$z=0.195-0.44$. These include the optically-confirmed
Ly$\alpha$ Galaxies (red solid triangles) and the 
optically-confirmed NUV-continuum selected galaxies 
(blue open diamonds).  The error bars are $\pm1\sigma$.
We also show the GOODS-N
NUV~$<24$ galaxy sample with redshifts between
$z=0.15$ and $z=0.48$ (blue solid diamonds).
The black line shows the linear fit of 
N2~$=\log$([NII]$\lambda6584$/H$\alpha$) relative to 
$M_B$(AB) for all of the blue diamonds.
The metallicity that would be inferred from the Pettini
\& Pagel (2004) calibration is shown on the right-hand
axis.  Finally, we show the values measured for
the $z\sim2$ LBGs by Erb et al.\ (2006; green inverted
solid triangles).  The green line shows the local 
metallicity-magnitude relation shifted by 3 magnitudes 
to match the metallicities of the $z\sim2$ sources.
\label{nuv_n2ha}
}
\end{figure}

Nevertheless, the overall observed trend of 
an increasing Ly$\alpha$ fraction with increasing redshift
would be expected if the metals and dust content are 
decreasing as we move to high redshift. These factors
increase the probability of seeing Ly$\alpha$ in the
low-redshift sample. The one aspect of this 
that we can test with the present data is the metallicity 
evolution.  We can make a detailed comparison with high-redshift
studies, because Erb et al.\ (2006) also used the N2 
relation in determining the metallicities of their z$\sim2$ 
galaxy sample.
In Figure~\ref{nuv_n2ha} we compare our determinations of N2 
for the optically-confirmed Ly$\alpha$ Galaxies (red solid triangles), 
the optically-confirmed NUV-continuum selected galaxies 
(blue open diamonds), and the GOODS-N NUV-continuum selected 
galaxies (blue solid diamonds) with the Erb et al.\ (2006) 
determinations of N2 for the $z\sim2$ galaxies 
(green inverted solid triangles) versus $M_B$(AB). 
(Note that the Erb et al.\ points differ from the low-redshift
points in that they are binned averages rather 
than individual points corresponding to single galaxies.) 
Consistent with Erb et al.\ (2006), we find that at
the same value of N2, the $z\sim2$ sources are 
3~mag brighter in $M_B$(AB) (green line) than our
$z\sim 0.3$ NUV-continuum selected sample.
In translating this into a metallicity, we must 
remember the dependence of N2 on the ionization parameter.  
A factor of only 3 increase in $q$ between the typical local 
value and $z\sim2$ would increase the inferred metallicity
in the $z\sim2$ galaxies by 0.21~dex and remove much of the 
inferred evolution.  However, in the simplest interpretation,
where we treat this purely as a metallicity effect,
the $z\sim2$ galaxies have relatively low metallicities for their 
luminosities, and they lie in the same metallicity
range as the $z\sim 0.3$ Ly$\alpha$ galaxies.
We can speculate that a higher fraction of low-metallicity
galaxies at $z\sim2$ might equate to a higher fraction of 
LAEs at $z\sim2$, which could be a partial explanation 
for the higher fraction of LAEs observed at high redshifts 
than at low redshifts.

\acknowledgements

We would like to thank the anonymous referee and Jean-Michel
Deharvang for critically reading the first draft of
this manuscript and for providing extensive and extremely
helpful comments. We would also like to thank Kim Nilsson
and Steve Finkelstein for their input, which also improved
the paper.  We are indebted to the staff of the Subaru and Keck
Observatories for their excellent assistance with the
observations, and to Drew Phillips and Greg Wirth for their help in
creating innovative mask designs for the DEIMOS spectrograph. 
We gratefully acknowledge support from NSF
grants AST-0709356 (L.~L.~C.), AST-0708793 (A.~J.~B.), 
and AST-0687850 (E.~M.~H.), the University of 
Wisconsin Research Committee with funds granted by the 
Wisconsin Alumni Research Foundation and the David and 
Lucile Packard Foundation (A.~J.~B.), and a grant from NASA 
through an award issued by JPL 1289080 (E.~M.~H.).
Some of the data presented in this paper were obtained 
from the Multimission Archive at the Space Telescope 
Science Institute (MAST). STScI is operated by the 
Association of Universities for Research in Astronomy, 
Inc., under NASA contract NAS5-26555. Support 
for MAST for non-{\em HST\/} data is provided by the NASA 
Office of Space Science via grant NAG5-7584 and by 
other grants and contracts.
This reseach used the facilities of the Canadian
Astronomy Data Centre operated by the National Research
Council of Canada with the support of the Canadian
Space Agency.
This research has made use of the NASA/IPAC Extragalactic
Database (NED) which is operated by the Jet Propulsion
Laboratory, California Institute of Technology, under
contract with the National Aeronautics and Space
Administration.


\newpage
%
%
\begin{deluxetable}{ccccccccc}
\renewcommand\baselinestretch{1.0}
\tablewidth{0pt}
\tablecaption{{\em GALEX\/} FIELDS}
\scriptsize
\small\addtolength{\tabcolsep}{-3pt}
\tablehead{Name & R.A. (J2000.0) & Decl. (J2000.0) & Exposure & NUV$_{lim}$ & E(B-V) & Gal Lat & Number & Number(ID)  \\ (1) & (2) & (3) & (4)  & (5) & (6) & (7) & (8) & (9)}
\startdata
GROTH 00  &  214.99182 & 52.78173  &  291  &  21.8  &  0.007  &  59.5 & 1152  &  69  \cr
NGPDWS 00  &  219.15610	& 35.17135   &  156  &  21.5  &  0.009 & 66.2  & 738  &  46  \cr
CDFS 00  &  53.12779  &	-27.87137  &  149  &  21.5  &  0.008  & -54.4 & 876  &  39  \cr
COSMOS 00  &  150.11900	& 2.20583 &  140  &  21.5  &  0.018 & 42.1  & 779  &  29  \cr
ELAISS1 00  &  9.63857	& -43.99023  &  84  &  21.2  &  0.007 & -72.9  & 589  &  19  \cr
SIRTFFL 00  &  259.12387 & 59.90915  &  80  &  21.2  &  0.020 & 35.0  & 800  &  22  \cr
LOCK 00  &  162.67843 &	58.73117  &  48  &  20.9  &  0.009 & 52.1  & 390  &  12  \cr
SIRTFFL 01  &  260.41425 & 59.34286  &  34  &  20.7  &  0.029 & 34.4  & 612  &  14  \cr
HDFN 00  &  189.20946 &	62.19772  &  24  &  20.5  &  0.012 & 54.8  & 265 &  10  \cr
\enddata
\label{tab1}
\end{deluxetable}

%
%
\begin{deluxetable}{clccccccc}
\renewcommand\baselinestretch{1.0}
\tablewidth{0pt}
\tablecaption{{\em GALEX\/} SPECTRAL SAMPLE IDENTIFICATIONS: GROTH 00}
\small\addtolength{\tabcolsep}{-3pt}
\tablehead{Number & Name & R.A. (J2000.0) & Decl. (J2000.0) & NUV & FUV & $z$ & Line Width & ID \\ 
(1) & (2) & (3) & (4)  & (5) & (6) & (7) & (8) & (9)}
\startdata
 1 &    GALEX1416+5252    & 214.147766   &  52.871555   &   21.12   &   24.14   &    1.146 &    9391$\pm$ 466 &       AGN  \cr
 2 &    GALEX1416+5259    & 214.201279   &  52.984028   &   20.68   &   21.45   &    1.233 &    6221$\pm$ 177 &       AGN  \cr
 3 &    GALEX1416+5250    & 214.208328   &  52.838863   &   21.49   &   21.97   &    0.288 &    1832$\pm$ 404 &    Galaxy  \cr
 4 &    GALEX1417+5305    & 214.291290   &  53.086723   &   21.29   &   21.82   &    0.268 &    1985$\pm$ 227 &    Galaxy  \cr
 5 &    GALEX1417+5235    & 214.300888   &  52.599110   &   21.24   &   21.76   &    0.262 &    1932$\pm$ 285 &       AGN  \cr
 6 &    GALEX1417+5230    & 214.352707   &  52.507141   &   20.37   &   21.20   &    0.896 &    7618$\pm$ 554 &    Galaxy  \cr
 7 &    GALEX1417+5230    & 214.399918   &  52.508358   &   20.82   &   22.39   &    0.988 &    6585$\pm$  94 &       AGN  \cr
 8 &    GALEX1417+5304    & 214.419083   &  53.079052   &   21.56   &   22.14   &    0.271 &    1587$\pm$ 281 &    Galaxy  \cr
\enddata
\label{tab2}
\end{deluxetable}

%
%
\begin{deluxetable}{clccccccc}
\renewcommand\baselinestretch{1.0}
\tablewidth{0pt}
\tablecaption{{\em GALEX\/} SPECTRAL SAMPLE IDENTIFICATIONS: NGPDWS 00}
\scriptsize
\small\addtolength{\tabcolsep}{-3pt}
\tablehead{Number & Name & R.A. (J2000.0) & Decl. (J2000.0) & NUV & FUV & $z$ & Line Width & ID \\ 
(1) & (2) & (3) & (4)  & (5) & (6) & (7) & (8) & (9)}
\startdata
 1 &    GALEX1434+3512    & 218.502533   &  35.205639   &   21.15   &   22.97   &    1.193 &    8546$\pm$ 511 &       AGN  \cr
 2 &    GALEX1434+3502    & 218.602798   &  35.044998   &   20.10   &   20.19   &    0.280 &    4605$\pm$ 109 &       AGN  \cr
 3 &    GALEX1434+3448    & 218.672165   &  34.808582   &   20.79   &   21.68   &    0.732 &    8811$\pm$ 432 &       AGN  \cr
 4 &    GALEX1434+3529    & 218.686600   &  35.488083   &   20.08   &   20.98   &    0.873 &    9495$\pm$ 283 &       AGN  \cr
 5 &    GALEX1434+3456    & 218.698334   &  34.934277   &   19.82   &   22.97   &    0.927 &    9631$\pm$ 195 &       AGN  \cr
 6 &    GALEX1434+3524    & 218.710663   &  35.409168   &   20.24   &   21.51   &    0.994 &    5629$\pm$  85 &       AGN  \cr
 7 &    GALEX1434+3532    & 218.718338   &  35.547527   &   21.44   &   21.81   &    0.198 &    1781$\pm$ 340 &    Galaxy  \cr
 8 &    GALEX1434+3508    & 218.739197   &  35.141582   &   20.46   &   22.35   &    1.213 &    4443$\pm$ 316 &    Galaxy  \cr
\enddata
\label{tab3}
\end{deluxetable}

%
%
\begin{deluxetable}{clccccccc}
\renewcommand\baselinestretch{1.0}
\tablewidth{0pt}
\tablecaption{{\em GALEX\/} SPECTRAL SAMPLE IDENTIFICATIONS: CDFS 00}
\scriptsize
\small\addtolength{\tabcolsep}{-3pt}
\tablehead{Number & Name & R.A. (J2000.0) & Decl. (J2000.0) & NUV & FUV & $z$ & Line Width & ID \\ 
(1) & (2) & (3) & (4)  & (5) & (6) & (7) & (8) & (9)}
\startdata
 1 &    GALEX0330-2748    &  52.624874   & -27.815527   &   21.21   &   23.94   &    0.855 &    5966$\pm$ 454 &       AGN  \cr
 2 &    GALEX0330-2744    &  52.666042   & -27.747833   &   21.32   &   21.96   &    0.710 &    9635$\pm$1603 &       AGN  \cr
 3 &    GALEX0330-2735    &  52.676041   & -27.585638   &   21.10   &   22.43   &    1.130 &    9888$\pm$ 609 &       AGN  \cr
 4 &    GALEX0330-2759    &  52.696835   & -27.988083   &   21.15   &   22.23   &    0.685 &    8226$\pm$ 959 &       AGN  \cr
 5 &    GALEX0330-2732    &  52.732708   & -27.547722   &   20.41   &   21.18   &    1.013 &    5587$\pm$ 328 &    Galaxy  \cr
 6 &    GALEX0330-2816    &  52.737583   & -28.279501   &   21.11   &   21.55   &    0.285 &    2124$\pm$ 286 &    Galaxy  \cr
 7 &    GALEX0331-2751    &  52.759998   & -27.858416   &   21.50   &   21.97   &    0.340 &    2769$\pm$ 375 &    Galaxy  \cr
 8 &    GALEX0331-2756    &  52.836498   & -27.946917   &   21.30   &   22.24   &    0.685 &    8353$\pm$ 748 &       AGN  \cr
\enddata
\label{tab4}
\end{deluxetable}

%
%
\begin{deluxetable}{clccccccc}
\renewcommand\baselinestretch{1.0}
\tablewidth{0pt}
\tablecaption{{\em GALEX\/} SPECTRAL SAMPLE IDENTIFICATIONS: COSMOS 00}
\scriptsize
\small\addtolength{\tabcolsep}{-3pt}
\tablehead{Number & Name & R.A. (J2000.0) & Decl. (J2000.0) & NUV & FUV & $z$ & Line Width & ID \\ 
(1) & (2) & (3) & (4)  & (5) & (6) & (7) & (8) & (9)}
\startdata
 1 &    GALEX0958+0205    & 149.663834   &   2.085083   &   21.22   &   21.72   &    1.218 &    5888$\pm$ 396 &       AGN  \cr
 2 &    GALEX0958+0213    & 149.739044   &   2.220722   &   20.78   &   23.70   &    1.025 &    6963$\pm$ 191 &       AGN  \cr
 3 &    GALEX0959+0219    & 149.761612   &   2.318472   &   19.99   &   20.55   &    0.349 &    4751$\pm$ 137 &       AGN  \cr
 4 &    GALEX0959+0209    & 149.828262   &   2.164305   &   21.00   &  999.00   &    1.158 &    4174$\pm$ 175 &       AGN  \cr
 5 &    GALEX0959+0149    & 149.873871   &   1.832972   &   21.42   &   21.76   &    0.208 &    1378$\pm$ 320 &    Galaxy  \cr
 6 &    GALEX0959+0212    & 149.912643   &   2.200389   &   21.20   &   22.07   &    0.698 &    5802$\pm$ 843 &       AGN  \cr
 7 &    GALEX0959+0151    & 149.917953   &   1.856028   &   21.10   &   21.47   &    0.254 &    2662$\pm$ 313 &    Galaxy  \cr
 8 &    GALEX1000+0235    & 150.053833   &   2.589667   &   19.39   &   20.66   &    0.706 &    9088$\pm$ 197 &       AGN  \cr
\enddata
\label{tab5}
\end{deluxetable}

%
%
\begin{deluxetable}{clccccccc}
\renewcommand\baselinestretch{1.0}
\tablewidth{0pt}
\tablecaption{{\em GALEX\/} SPECTRAL SAMPLE IDENTIFICATIONS: ELAISS1 00}
\scriptsize
\small\addtolength{\tabcolsep}{-3pt}
\tablehead{Number & Name & R.A. (J2000.0) & Decl. (J2000.0) & NUV & FUV & $z$ & Line Width & ID \\ 
(1) & (2) & (3) & (4)  & (5) & (6) & (7) & (8) & (9)}
\startdata
 1 &    GALEX0036-4348    &   9.061083   & -43.816639   &   21.15   &   22.22   &    0.876 &    8611$\pm$ 598 &       AGN  \cr
 2 &    GALEX0037-4347    &   9.265166   & -43.799919   &   20.68   &   23.65   &    1.240 &    6359$\pm$ 362 &    Galaxy  \cr
 3 &    GALEX0037-4342    &   9.337708   & -43.711136   &   20.61   &   20.98   &    0.225 &    2130$\pm$  97 &       AGN  \cr
 4 &    GALEX0037-4338    &   9.371167   & -43.635670   &   20.52   &   20.85   &    0.300 &    2846$\pm$ 380 &    Galaxy  \cr
 5 &    GALEX0038-4414    &   9.511625   & -44.233448   &   20.99   &   22.21   &    0.878 &    5255$\pm$ 302 &       AGN  \cr
 6 &    GALEX0038-4352    &   9.520541   & -43.874584   &   20.63   &   21.08   &    0.222 &    2348$\pm$ 196 &    Galaxy  \cr
 7 &    GALEX0038-4352    &   9.539542   & -43.878113   &   19.49   &   20.23   &    0.936 &   10019$\pm$ 258 &       AGN  \cr
 8 &    GALEX0038-4414    &   9.559041   & -44.243584   &   21.29   &   21.14   &    0.271 &    2211$\pm$  79 &       AGN  \cr
\enddata
\label{tab6}
\end{deluxetable}

%
%
\begin{deluxetable}{clccccccc}
\renewcommand\baselinestretch{1.0}
\tablewidth{0pt}
\tablecaption{{\em GALEX\/} SPECTRAL SAMPLE IDENTIFICATIONS: SIRTFFL 00}
\scriptsize
\small\addtolength{\tabcolsep}{-3pt}
\tablehead{Number & Name & R.A. (J2000.0) & Decl. (J2000.0) & NUV & FUV & $z$ & Line Width & ID \\ 
(1) & (2) & (3) & (4)  & (5) & (6) & (7) & (8) & (9)}
\startdata
 1 &    GALEX1712+6007    & 258.131622   &  60.122082   &   20.98   &  999.00   &    1.204 &    5908$\pm$ 388 &       AGN  \cr
 2 &    GALEX1712+6005    & 258.144928   &  60.091888   &   19.55   &   20.85   &    0.999 &    9189$\pm$ 215 &       AGN  \cr
 3 &    GALEX1712+5956    & 258.149719   &  59.946888   &   20.37   &   20.99   &    0.225 &    1296$\pm$ 243 &    Galaxy  \cr
 4 &    GALEX1712+6001    & 258.197723   &  60.022778   &   20.80   &   23.63   &    0.786 &    3812$\pm$ 316 &    Galaxy  \cr
 5 &    GALEX1713+6001    & 258.290100   &  60.024193   &   20.26   &   21.42   &    0.818 &    6461$\pm$ 239 &       AGN  \cr
 6 &    GALEX1714+5949    & 258.591766   &  59.833332   &   21.09   &   21.28   &    0.231 &    1946$\pm$ 179 &    Galaxy  \cr
 7 &    GALEX1714+5956    & 258.683746   &  59.947247   &   21.13   &   21.62   &    0.222 &    2059$\pm$ 482 &    Galaxy  \cr
 8 &    GALEX1715+5956    & 258.751678   &  59.936642   &   19.71   &   21.07   &    0.695 &   12588$\pm$1511 &       AGN  \cr
\enddata
\label{tab7}
\end{deluxetable}

%
%
\begin{deluxetable}{clccccccc}
\renewcommand\baselinestretch{1.0}
\tablewidth{0pt}
\tablecaption{{\em GALEX\/} SPECTRAL SAMPLE IDENTIFICATIONS: LOCK 00}
\scriptsize
\small\addtolength{\tabcolsep}{-3pt}
\tablehead{Number & Name & R.A. (J2000.0) & Decl. (J2000.0) & NUV & FUV & $z$ & Line Width & ID \\ 
(1) & (2) & (3) & (4)  & (5) & (6) & (7) & (8) & (9)}
\startdata
 1 &    GALEX1047+5827    & 161.773422   &  58.461693   &   20.33   &   21.13   &    0.248 &    4451$\pm$ 478 &    Galaxy  \cr
 2 &    GALEX1048+5910    & 162.048798   &  59.179695   &   20.60   &   21.46   &    0.711 &    7525$\pm$ 664 &       AGN  \cr
 3 &    GALEX1049+5853    & 162.371796   &  58.893944   &   20.30   &   20.74   &    0.251 &    3975$\pm$ 205 &       AGN  \cr
 4 &    GALEX1050+5819    & 162.500885   &  58.317833   &   18.04   &   18.88   &    0.835 &    8654$\pm$  91 &       AGN  \cr
 5 &    GALEX1051+5838    & 162.771729   &  58.642475   &   20.65   &   21.35   &    0.915 &    4969$\pm$ 782 &       AGN  \cr
 6 &    GALEX1051+5916    & 162.775833   &  59.273808   &   19.13   &   20.70   &    0.775 &    9575$\pm$ 222 &       AGN  \cr
 7 &    GALEX1051+5832    & 162.807800   &  58.544777   &   20.89   &   22.03   &    0.863 &    8187$\pm$ 725 &       AGN  \cr
 8 &    GALEX1051+5913    & 162.921417   &  59.218193   &   19.06   &   19.72   &    0.440 &    7587$\pm$ 359 &       AGN  \cr
 9 &    GALEX1052+5851    & 163.072922   &  58.856194   &   20.57   &   21.62   &    0.970 &    9599$\pm$ 628 &       AGN  \cr
10 &    GALEX1053+5905    & 163.284698   &  59.089584   &   19.15   &   19.53   &    0.434 &    3642$\pm$ 187 &       AGN  \cr
11 &    GALEX1053+5855    & 163.292374   &  58.925808   &   19.11   &   19.98   &    1.180 &    6607$\pm$ 124 &       AGN  \cr
12 &    GALEX1054+5852    & 163.587448   &  58.871887   &   20.42   &   21.62   &    1.184 &    6321$\pm$ 353 &       AGN  \cr
\enddata
\label{tab8}
\end{deluxetable}

%
%
\begin{deluxetable}{clccccccc}
\renewcommand\baselinestretch{1.0}
\tablewidth{0pt}
\tablecaption{{\em GALEX\/} SPECTRAL SAMPLE IDENTIFICATIONS: SIRTFFL 01}
\scriptsize
\small\addtolength{\tabcolsep}{-3pt}
\tablehead{Number & Name & R.A. (J2000.0) & Decl. (J2000.0) & NUV & FUV & $z$ & Line Width & ID \\ 
(1) & (2) & (3) & (4)  & (5) & (6) & (7) & (8) & (9)}
\startdata
 1 &    GALEX1718+5933    & 259.665924   &  59.566555   &   18.49   &   18.95   &    0.396 &    8589$\pm$ 398 &       AGN  \cr
 2 &    GALEX1719+5934    & 259.819031   &  59.580418   &   20.04   &   20.61   &    0.363 &    3107$\pm$ 862 &    Galaxy  \cr
 3 &    GALEX1720+5932    & 260.134705   &  59.535336   &   19.96   &   20.63   &    1.169 &   12141$\pm$ 862 &       AGN  \cr
 4 &    GALEX1720+5910    & 260.234222   &  59.166832   &   20.46   &   21.98   &    1.022 &    5963$\pm$ 525 &    Galaxy  \cr
 5 &    GALEX1721+5924    & 260.269806   &  59.414192   &   19.71   &   20.52   &    0.783 &    9317$\pm$ 490 &       AGN  \cr
 6 &    GALEX1722+5901    & 260.734222   &  59.027248   &   20.56   &   21.73   &    0.933 &    7905$\pm$ 559 &    Galaxy  \cr
 7 &    GALEX1723+5951    & 260.793243   &  59.851692   &   18.86   &   21.48   &    0.996 &   14026$\pm$ 680 &       AGN  \cr
 8 &    GALEX1723+5853    & 260.794189   &  58.884003   &   20.63   &   22.04   &    1.045 &    9044$\pm$ 562 &       AGN  \cr
\enddata
\label{tab9}
\end{deluxetable}

%
%
\begin{deluxetable}{clccccccc}
\renewcommand\baselinestretch{1.0}
\tablewidth{0pt}
\tablecaption{{\em GALEX\/} SPECTRAL SAMPLE IDENTIFICATIONS: HDFN 00}
\scriptsize
\small\addtolength{\tabcolsep}{-3pt}
\tablehead{Number & Name & R.A. (J2000.0) & Decl. (J2000.0) & NUV & FUV & $z$ & Line Width & ID \\ 
(1) & (2) & (3) & (4)  & (5) & (6) & (7) & (8) & (9)}
\startdata
 1 &    GALEX1234+6233    & 188.504761   &  62.554085   &   19.91   &   20.73   &    1.123 &   12033$\pm$ 655 &       AGN  \cr
 2 &    GALEX1234+6226    & 188.561661   &  62.444443   &   17.99   &   18.89   &    0.973 &    4311$\pm$ 212 &       AGN  \cr
 3 &    GALEX1234+6220    & 188.670288   &  62.335999   &   20.37   &   20.85   &    0.398 &    2708$\pm$ 374 &       AGN  \cr
 4 &    GALEX1237+6217    & 189.279007   &  62.284000   &   20.09   &   20.88   &    1.022 &   11390$\pm$ 878 &       AGN  \cr
 5 &    GALEX1237+6211    & 189.498337   &  62.184139   &   19.41   &   20.70   &    0.913 &    8273$\pm$ 227 &       AGN  \cr
 6 &    GALEX1238+6202    & 189.567291   &  62.035610   &   19.30   &   20.21   &    1.001 &   10252$\pm$ 432 &       AGN  \cr
 7 &    GALEX1238+6143    & 189.702667   &  61.731861   &   18.34   &   19.25   &    0.832 &    5570$\pm$  66 &       AGN  \cr
 8 &    GALEX1239+6206    & 189.881638   &  62.105251   &   19.75   &   20.70   &    1.201 &    7959$\pm$ 410 &       AGN  \cr
 9 &    GALEX1240+6233    & 190.041428   &  62.561668   &   20.28   &   20.69   &    0.210 &    2094$\pm$ 209 &    Galaxy  \cr
10 &    GALEX1240+6223    & 190.138916   &  62.390915   &   20.44   &   21.43   &    1.112 &    8787$\pm$1011 &       AGN  \cr
\enddata
\label{tab10}
\end{deluxetable}

%
%
\begin{deluxetable}{cccccccll}
\renewcommand\baselinestretch{1.0}
\tablewidth{0pt}
\tablecaption{DEIMOS OBSERVATIONS: {\em GALEX\/} CANDIDATE Ly$\alpha$ GALAXIES}
\scriptsize
\tablehead{NAME & R.A. (J2000.0) & Decl. (J2000.0) & NUV & $g$ & $z_{galex}$ & $z_{optical}$ & EW(H$\alpha$) &  Optical Type \\ (1) & (2) & (3) & (4)  & (5) & (6) & (7) & (8) & (9) }
\startdata
GALEX0038-4352    &   9.520541   & -43.874580   &  20.63   & \nodata   &     0.2220 &     0.2191 &     15$\pm$   1 &  Galaxy    \cr
GALEX0038-4343    &   9.666203   & -43.722527   &  20.94   & \nodata   &     0.2104 &     0.2072 &    314$\pm$   2 &  Galaxy    \cr
GALEX0331-2811    &  52.962204   & -28.189028   &  20.68   & \nodata   &     0.2162 &     0.2132 &      3$\pm$   0 &  Galaxy    \cr
GALEX0331-2814    &  52.976501   & -28.238640   &  21.49   & \nodata   &     0.2824 &     0.2802 &     38$\pm$   0 &  Galaxy    \cr
GALEX0331-2809    &  52.999294   & -28.164413   &  21.18   & \nodata   &     0.2392 &     0.2362 &    227$\pm$   1 &  Galaxy    \cr
GALEX0332-2801    &  53.049751   & -28.025028   &  21.51   & \nodata   &     0.2191 &     0.2155 &    472$\pm$   1 &  Galaxy    \cr
GALEX0332-2811    &  53.061584   & -28.186525   &  21.18   & \nodata   &     0.2651 &     0.2788 &    234$\pm$   1 &  Galaxy    \cr
GALEX0332-2813    &  53.078033   & -28.222441   &  21.17   & \nodata   &     0.2824 &     0.2788 &    \nodata   1 &  Seyfert 1.8       \cr
\enddata
\label{tab11}
\end{deluxetable}

%
%
\begin{deluxetable}{ccccccc}
\renewcommand\baselinestretch{1.0}
\tablewidth{0pt}
\tablecaption{DEIMOS OBSERVATIONS: {\em GALEX\/} AGNS}
\scriptsize
\tablehead{NAME & R.A. (J2000.0) & Decl. (J2000.0) & NUV & $z_{galex}$ & $z_{optical}$ & Optical Type \\ (1) & (2) & (3) & (4)  & (5) & (6) & (7) }
\startdata
GALEX0332-2810    &  53.184074   & -28.174442   &  21.04   &     1.1519 &     1.1480 &  AGN       \cr
GALEX1000+0152    & 150.139114   &   1.876969   &  21.08   &     0.8323 &     0.8309 &  AGN       \cr
GALEX1000+0202    & 150.145584   &   2.043078   &  21.25   &     1.1778 &     1.1764 &  AGN       \cr
GALEX1417+5230    & 214.399780   &  52.508324   &  20.82   &     0.9878 &     0.9851 &  AGN       \cr
GALEX1418+5237    & 214.519180   &  52.629162   &  21.32   &     1.0900 &     1.0885 &  AGN       \cr
GALEX1418+5310    & 214.657745   &  53.171547   &  21.01   &     0.8957 &     0.8922 &  AGN       \cr
GALEX1420+5240    & 215.042847   &  52.674995   &  21.17   &     0.5526 &     0.5483 &  Galaxy    \cr
GALEX1436+3453    & 219.131790   &  34.890221   &  20.33   &     1.0677 &     1.0652 &  AGN       \cr
GALEX1436+3450    & 219.152878   &  34.842777   &  20.66   &     0.2830 &     0.2810 &  AGN       \cr
GALEX1437+3522    & 219.311295   &  35.381779   &  19.33   &     0.6299 &     0.6236 &  AGN       \cr
GALEX1437+3448    & 219.345505   &  34.801777   &  20.98   &     0.9497 &     0.9476 &  AGN       \cr
GALEX1437+3445    & 219.370041   &  34.763195   &  19.22   &     0.6284 &     0.6224 &  AGN       \cr
GALEX1712+6007    & 258.131622   &  60.122139   &  20.98   &     1.2037 &     1.2020 &  AGN       \cr
GALEX1712+6005    & 258.145172   &  60.091831   &  19.55   &     0.9993 &     0.9973 &  AGN       \cr
GALEX1715+5956    & 258.751862   &  59.936611   &  19.71   &     0.6950 &     0.7000 &  AGN       \cr
\enddata
\label{tab12}
\end{deluxetable}

%
%
\begin{deluxetable}{ccccccll}
\renewcommand\baselinestretch{1.0}
\tablewidth{0pt}
\tablecaption{DEIMOS OBSERVATIONS: {\em GALEX\/} SPECTRA WITHOUT STRONG Ly$\alpha$ EMISSION}
\scriptsize
\tablehead{NAME & R.A. (J2000.0) & Decl. (J2000.0) & NUV & $g$ & $z_{optical}$ & EW(H$\alpha$) &  Optical Type \\ (1) & (2) & (3) & (4)  & (5) & (6) & (7) & (8) }
\startdata
GALEX0331-2810    &  52.932369   & -28.173195   &  21.21   & \nodata   &     0.2169 &     21$\pm$ 1 &  Galaxy    \cr
GALEX0331-2811    &  52.948586   & -28.193085   &  21.46   & \nodata   &     0.3814 &     81$\pm$ 8 &  Galaxy    \cr
GALEX0331-2810    &  52.949966   & -28.168806   &  21.19   & \nodata   &     0.3798 &     56$\pm$ 1 &  Galaxy    \cr
GALEX0331-2811    &  52.952209   & -28.184692   &  20.67   & \nodata   &     0.2157 &    103$\pm$ 1 &  Galaxy    \cr
GALEX0331-2812    &  52.953033   & -28.208719   &  20.79   & \nodata   &     0.2136 &     31$\pm$ 1 &  Galaxy    \cr
GALEX0331-2813    &  52.958126   & -28.221001   &  21.27   & \nodata   &     0.2730 &     93$\pm$ 3 &  Galaxy    \cr
GALEX0331-2813    &  52.968121   & -28.231945   &  21.04   & \nodata   &     0.2793 &     37$\pm$ 1 &  Galaxy    \cr
GALEX0331-2814    &  52.978458   & -28.235889   &  21.26   & \nodata   &     0.3160 &    81$\pm$ 1 &  Galaxy    \cr
\enddata
\label{tab13}
\end{deluxetable}

%
%
\begin{deluxetable}{cccccccc}
\renewcommand\baselinestretch{1.0}
\tablewidth{0pt}
\tablecaption{{\em GALEX\/} X-ray Properties}
\scriptsize
\tablehead{NAME & R.A. (J2000.0) & Decl. (J2000.0) & $f_{2-8~{\rm keV}}$ & $f_{0.5-2~{\rm keV}}$ & $\log(L_{2-8~{\rm keV}})$ & z$_{galex}$ &  class  \\ (1) & (2) & (3) & (4)  & (5) & (6) & (7) & (8) }
\startdata
GALEX0332-2740  &        53.1104  &       -27.6766  &     1.54e-14  & 
   2.20e-14  &         43.91   &         1.0396  &  AGN  \cr 
GALEX0332-2741  &        53.1125  &       -27.6847  &     6.55e-14  & 
   5.97e-14  &         44.19   &        0.7373  &  AGN  \cr 
GALEX0331-2756  &        52.8365  &       -27.9469  &     1.70e-14  & 
   1.03e-14  &         43.53   &        0.6850  &  AGN  \cr 
GALEX0332-2739  &        53.1588  &       -27.6625  &     2.55e-14  & 
   2.34e-14  &         43.91   &        0.8381  &  AGN  \cr 
GALEX0333-2756  &        53.3705  &       -27.9447  &     4.98e-14  & 
   3.91e-14  &         44.20   &        0.8381 &  AGN  \cr 
GALEX1048+5910  &        162.049  &        59.1797  &     2.23e-14  & 
   2.58e-14  &         43.68   &        0.7110  &  AGN  \cr 
GALEX1049+5853  &        162.372  &        58.8940  &     8.69e-14  & 
   9.08e-14  &         43.24   &        0.2506  &  AGN  \cr 
GALEX1237+6217  &        189.279  &        62.2840  &     1.89e-14  & 
   1.57e-14  &         43.98   &         1.019  &  AGN  \cr 
GALEX1237+6211  &        189.498  &        62.1840  &     9.11e-14  & 
   5.64e-14  &         44.55   &        0.9129  &  AGN  \cr 
GALEX0332-2753  &        53.2362  &       -27.8879  &    -4.67e-16  & 
   2.05e-16  &        $<41.0$   &        0.3745  &  Galaxy  \cr 
GALEX0332-2801   &       53.0497   &      -28.0250   &  \nodata & 
 \nodata  &    $<40.5$ &  0.2191   &      Galaxy \cr
\enddata
\label{tab14}
\end{deluxetable}
%
%
\pagestyle{empty}
\begin{deluxetable}{lcccccccc}
\renewcommand\baselinestretch{1.0}
\tablewidth{0pt}
\tablecaption{LOW-REDSHIFT {\em GALEX\/} LAE SAMPLE}
\scriptsize
\small\addtolength{\tabcolsep}{-4pt}
\tablehead{NAME & R.A. (J2000.0) & Decl. (J2000.0) & NUV & FUV & $z_{galex}$ & $\log L({\rm Ly}\alpha)$ & EW({\rm Ly}$\alpha$) &  $z_{optical}$ \\ (1) & (2) & (3) & (4)  & (5) & (6) & (7) & (8) & (9) }
\startdata
GALEX1718+5949    & 259.748260   &  59.825752   &  20.91   &  23.31   &   0.3717 &   42.24$\pm$0.08 &    115$\pm$49 &    \nodata \cr
GALEX1417+5228    & 214.430923   &  52.468307   &  21.90   &  21.79   &   0.2104 &   42.01$\pm$0.01 &    101$\pm$ 9 &    0.2067 SF \cr
GALEX0331-2814    &  52.976460   & -28.238554   &  21.49   &  21.84   &   0.2824 &   42.15$\pm$0.02 &     84$\pm$ 9 &    0.2802 SF \cr
GALEX1436+3525    & 219.193008   &  35.417641   &  21.22   &  22.20   &   0.3697 &   42.32$\pm$0.05 &     81$\pm$16 &    0.3689 AGN \cr
GALEX0334-2815    &  53.541370   & -28.255472   &  21.57   &  22.77   &   0.3450 &   41.93$\pm$0.08 &     75$\pm$27 &    0.3348 AGN \cr
GALEX1422+5252    & 215.533752   &  52.873722   &  21.30   &  22.52   &   0.3055 &   42.01$\pm$0.02 &     71$\pm$ 8 &    0.3017 AGN \cr
GALEX0332-2801    &  53.049873   & -28.025000   &  21.51   &  21.71   &   0.2191 &   41.76$\pm$0.02 &     67$\pm$ 9 &    0.2155 SF \cr
GALEX1240+6233    & 190.041428   &  62.561668   &  20.28   &  20.69   &   0.2104 &   42.01$\pm$0.04 &     61$\pm$10 &    0.2068 BPT AGN \cr
\enddata
\label{tab15}
\end{deluxetable}

%
%
\pagestyle{empty}
\begin{deluxetable}{ccccccccc}
\renewcommand\baselinestretch{1.0}
\tablewidth{0pt}
\tablecaption{MODERATE-REDSHIFT {\em GALEX\/} LAE SAMPLE}
\scriptsize
\small\addtolength{\tabcolsep}{-4pt}
\tablehead{NAME & R.A. (J2000.0) & Decl. (J2000.0) & NUV & FUV & $z_{galex}$ & $\log L({\rm Ly}\alpha)$ & EW({\rm Ly}$\alpha$) &  $z_{optical}$ \\ (1) & (2) & (3) & (4)  & (5) & (6) & (7) & (8) & (9) }
\startdata
GALEX1421+5221    & 215.421539   &  52.359886   &  21.64   &  25.20   &   0.9475 &   43.21$\pm$0.02 &     52$\pm$ 4 &    \nodata \cr
GALEX1712+6001    & 258.197723   &  60.022778   &  20.80   &  23.63   &   0.7863 &   43.19$\pm$0.03 &     37$\pm$ 4 &    0.7780 SF \cr
GALEX1437+3541    & 219.281967   &  35.692280   &  20.93   &  22.79   &   0.6640 &   42.91$\pm$0.05 &     30$\pm$ 5 &    \nodata \cr
GALEX1420+5313    & 215.237045   &  53.218750   &  21.26   &  23.87   &   0.7488 &   42.84$\pm$0.04 &     27$\pm$ 3 &    \nodata \cr
\enddata
\label{tab16}
\end{deluxetable}

\end{document}